\documentclass[12pt,titlepage,letterpaper]{utarticle}

\usepackage{amsmath}
\usepackage{amsfonts}
\usepackage{amssymb}
\usepackage{amsthm}
\usepackage{mathtools}
\usepackage{graphicx}
\usepackage{color}
\usepackage{xparse}
\usepackage[utf8x]{inputenc}
\usepackage{tocloft}
\usepackage[titletoc,title]{appendix}
\usepackage{hyperref}
\usepackage{longtable}

\definecolor{aqua}{rgb}{0, 1.0, 1.0}
\definecolor{fuschia}{rgb}{1.0, 0, 1.0}
\definecolor{gray}{rgb}{0.502, 0.502, 0.502}
\definecolor{lime}{rgb}{0, 1.0, 0}
\definecolor{maroon}{rgb}{0.502, 0, 0}
\definecolor{navy}{rgb}{0, 0, 0.502}
\definecolor{olive}{rgb}{0.502, 0.502, 0}
\definecolor{purple}{rgb}{0.502, 0, 0.502}
\definecolor{silver}{rgb}{0.753, 0.753, 0.753}
\definecolor{teal}{rgb}{0, 0.502, 0.502}

\makeatletter
\newdimen\itex@wd%
\newdimen\itex@dp%
\newdimen\itex@thd%
\def\itexspace#1#2#3{\itex@wd=#3em%
\itex@wd=0.1\itex@wd%
\itex@dp=#2ex%
\itex@dp=0.1\itex@dp%
\itex@thd=#1ex%
\itex@thd=0.1\itex@thd%
\advance\itex@thd\the\itex@dp%
\makebox[\the\itex@wd]{\rule[-\the\itex@dp]{0cm}{\the\itex@thd}}}
\makeatother

\makeatletter
\newif\if@sup
\newtoks\@sups
\def\append@sup#1{\edef\act{\noexpand\@sups={\the\@sups #1}}\act}%
\def\reset@sup{\@supfalse\@sups={}}%
\def\mk@scripts#1#2{\if #2/ \if@sup ^{\the\@sups}\fi \else%
  \ifx #1_ \if@sup ^{\the\@sups}\reset@sup \fi {}_{#2}%
  \else \append@sup#2 \@suptrue \fi%
  \expandafter\mk@scripts\fi}
\def\tensor#1#2{\reset@sup#1\mk@scripts#2_/}
\def\multiscripts#1#2#3{\reset@sup{}\mk@scripts#1_/#2%
  \reset@sup\mk@scripts#3_/}
\makeatother

\makeatletter
\newbox\slashbox \setbox\slashbox=\hbox{$/$}
\def\itex@pslash#1{\setbox\@tempboxa=\hbox{$#1$}
  \@tempdima=0.5\wd\slashbox \advance\@tempdima 0.5\wd\@tempboxa
  \copy\slashbox \kern-\@tempdima \box\@tempboxa}
\def\slash{\protect\itex@pslash}
\makeatother

\def\clap#1{\hbox to 0pt{\hss#1\hss}}

\def\mathrlap{\mathpalette\mathrlapinternal}

\def\mathrlapinternal#1#2{\rlap{$\mathsurround=0pt#1{#2}$}}

\let\oldroot\root
\def\root#1#2{\oldroot #1 \of{#2}}
\renewcommand{\sqrt}[2][]{\oldroot #1 \of{#2}}

\DeclareSymbolFont{symbolsC}{U}{txsyc}{m}{n}
\SetSymbolFont{symbolsC}{bold}{U}{txsyc}{bx}{n}
\DeclareFontSubstitution{U}{txsyc}{m}{n}

\DeclareSymbolFont{stmry}{U}{stmry}{m}{n}
\SetSymbolFont{stmry}{bold}{U}{stmry}{b}{n}

\DeclareFontFamily{OMX}{MnSymbolE}{}
\DeclareSymbolFont{mnomx}{OMX}{MnSymbolE}{m}{n}
\SetSymbolFont{mnomx}{bold}{OMX}{MnSymbolE}{b}{n}
\DeclareFontShape{OMX}{MnSymbolE}{m}{n}{
    <-6>  MnSymbolE5
   <6-7>  MnSymbolE6
   <7-8>  MnSymbolE7
   <8-9>  MnSymbolE8
   <9-10> MnSymbolE9
  <10-12> MnSymbolE10
  <12->   MnSymbolE12}{}

\makeatletter
\def\re@DeclareMathSymbol#1#2#3#4{%
    \let#1=\undefined
    \DeclareMathSymbol{#1}{#2}{#3}{#4}}
\re@DeclareMathSymbol{\neArrow}{\mathrel}{symbolsC}{116}
\re@DeclareMathSymbol{\neArr}{\mathrel}{symbolsC}{116}
\re@DeclareMathSymbol{\seArrow}{\mathrel}{symbolsC}{117}
\re@DeclareMathSymbol{\seArr}{\mathrel}{symbolsC}{117}
\re@DeclareMathSymbol{\nwArrow}{\mathrel}{symbolsC}{118}
\re@DeclareMathSymbol{\nwArr}{\mathrel}{symbolsC}{118}
\re@DeclareMathSymbol{\swArrow}{\mathrel}{symbolsC}{119}
\re@DeclareMathSymbol{\swArr}{\mathrel}{symbolsC}{119}
\re@DeclareMathSymbol{\nequiv}{\mathrel}{symbolsC}{46}
\re@DeclareMathSymbol{\Perp}{\mathrel}{symbolsC}{121}
\re@DeclareMathSymbol{\Vbar}{\mathrel}{symbolsC}{121}
\re@DeclareMathSymbol{\sslash}{\mathrel}{stmry}{12}
\re@DeclareMathSymbol{\bigsqcap}{\mathop}{stmry}{"64}
\re@DeclareMathSymbol{\biginterleave}{\mathop}{stmry}{"6}
\re@DeclareMathSymbol{\invamp}{\mathrel}{symbolsC}{77}
\re@DeclareMathSymbol{\parr}{\mathrel}{symbolsC}{77}
\makeatother

\makeatletter
\def\Decl@Mn@Delim#1#2#3#4{%
  \if\relax\noexpand#1%
    \let#1\undefined
  \fi
  \DeclareMathDelimiter{#1}{#2}{#3}{#4}{#3}{#4}}
\def\Decl@Mn@Open#1#2#3{\Decl@Mn@Delim{#1}{\mathopen}{#2}{#3}}
\def\Decl@Mn@Close#1#2#3{\Decl@Mn@Delim{#1}{\mathclose}{#2}{#3}}
\Decl@Mn@Open{\llangle}{mnomx}{'164}
\Decl@Mn@Close{\rrangle}{mnomx}{'171}
\Decl@Mn@Open{\lmoustache}{mnomx}{'245}
\Decl@Mn@Close{\rmoustache}{mnomx}{'244}
\makeatother

\makeatletter
\DeclareRobustCommand\widecheck[1]{{\mathpalette\@widecheck{#1}}}
\def\@widecheck#1#2{%
    \setbox\z@\hbox{\m@th$#1#2$}%
    \setbox\tw@\hbox{\m@th$#1%
       \widehat{%
          \vrule\@width\z@\@height\ht\z@
          \vrule\@height\z@\@width\wd\z@}$}%
    \dp\tw@-\ht\z@
    \@tempdima\ht\z@ \advance\@tempdima2\ht\tw@ \divide\@tempdima\thr@@
    \setbox\tw@\hbox{%
       \raise\@tempdima\hbox{\scalebox{1}[-1]{\lower\@tempdima\box
\tw@}}}%
    {\ooalign{\box\tw@ \cr \box\z@}}}
\makeatother

\makeatletter
\NewDocumentCommand\mathraisebox{moom}{%
\IfNoValueTF{#2}{\def\@temp##1##2{\raisebox{#1}{$\m@th##1##2$}}}{%
\IfNoValueTF{#3}{\def\@temp##1##2{\raisebox{#1}[#2]{$\m@th##1##2$}}%
}{\def\@temp##1##2{\raisebox{#1}[#2][#3]{$\m@th##1##2$}}}}%
\mathpalette\@temp{#4}}
\makeatletter

\makeatletter
\def\udots{\mathinner{\mkern2mu\raise\p@\hbox{.}
\mkern2mu\raise4\p@\hbox{.}\mkern1mu
\raise7\p@\vbox{\kern7\p@\hbox{.}}\mkern1mu}}
\makeatother

\newcommand{\gt}{>}
\newcommand{\lt}{<}

\renewcommand{\scriptsize}{\scriptstyle}

\theoremstyle{plain}

\theoremstyle{definition}

\theoremstyle{remark}

\numberwithin{equation}{section}

\def\Tr{\mathop{\mathrm{Tr}}}

\allowdisplaybreaks

\makeatletter
 \newcommand{\htarget}[1]{\raisebox{\ht\strutbox}{\hypertarget{#1}{}}}
\makeatother

\begin{document}


\preprint{
UTTG--05--14\\
TCC--005--14\\
ICTP--SAIFR/2014--001\\
}

\title{Tinkertoys for the $E_6$ Theory}

\author{Oscar Chacaltana
    \address{
    ICTP South American Institute for\\ Fundamental Research,\\
    Instituto de F\'isica Te\'orica,\\Universidade Estadual Paulista,\\
    01140-070 S\~{a}o Paulo, SP, Brazil\\
    {~}\\
    \email{chacaltana@ift.unesp.br}\\
    },
    Jacques Distler ${}^\mathrm{b}$ and Anderson Trimm
     \address{
      Theory Group and\\
      Texas Cosmology Center\\
      Department of Physics,\\
      University of Texas at Austin,\\
      Austin, TX 78712, USA \\
      {~}\\
      \email{distler@golem.ph.utexas.edu}\\
      \email{atrimm@physics.utexas.edu}
      }
}
\date{March 19, 2014}

\Abstract{
Compactifying the 6-dimensional (2,0) superconformal field theory, of type ADE, on a Riemann surface, $C$, with codimension-2 defect operators at points on $C$, yields a 4-dimensional $\mathcal{N}=2$ superconformal field theory. An outstanding problem is to classify the 4D theories one obtains, in this way, and to understand their properties. In this paper, we turn our attention to the $E_6$ (2,0) theory, which (unlike the A- and D-series) has no realization in terms of M5-branes. Classifying the 4D theories amounts to classifying all of the 3-punctured spheres (``fixtures"), and the cylinders that connect them, that can occur in a pants-decomposition of $C$. We find 904 fixtures: 19 corresponding to free hypermultiplets, 825 corresponding to isolated interacting SCFTs (with no known Lagrangian description) and 60 ``mixed fixtures", corresponding to a combination of free hypermultiplets and an interacting SCFT. Of the 825 interacting fixtures, we list only the 139 ``interesting" ones. As an application, we study the strong coupling limits of the Lagrangian field theories: $E_6$ with 4 hypermultiplets in the $27$ and $F_4$ with 3 hypermultiplets in the $26$.
}

\maketitle

\tocloftpagestyle{empty}
\tableofcontents
\vfill
\newpage
\setcounter{page}{1}

\section{Introduction}\label{introduction}

In \cite{Gaiotto:2009we,Gaiotto:2009hg}, Gaiotto and collaborators introduced a construction for a class of 4D $\mathcal{N}=2$ superconformal theories, realizing them as compactifications of the 6D (2,0) theories on a punctured Riemann surface. Because of its six-dimensional origin, this class of theories (sometimes called class ``S'') is endowed with a powerful set of tools \cite{Gaiotto:2009we,Gaiotto:2009hg,Chacaltana:2010ks,Chacaltana:2011ze,Nanopoulos:2009uw,Nanopoulos:2010ga,Benini:2010uu,Gaiotto:2011xs,Chacaltana:2012zy,Chacaltana:2012ch,Chacaltana:2013oka} for studying its physical properties. For a theory in this class, it is trivial to write its low-energy Seiberg-Witten solution and study $\mathcal{N}=2$ S-duality \cite{Argyres:2007cn}, as well as compute central charges, the global symmetry group, graded dimensions of the Coulomb branch, etc.

Equally interesting, the understanding of the theories in this class expands our knowledge of what 4D $\mathcal{N}=2$ SCFTs actually exist. The set of $\mathcal{N}=2$ SCFTs includes ordinary gauge theories (recently classified in \cite{Bhardwaj:2013qia}), some of which are realized from six dimensions, but also isolated interacting fixed points of the renormalization group with no known Lagrangian description. Gaiotto's construction generates infinitely many new examples of these interacting theories. Thus, while it seems unlikely that this construction covers the full gamut of 4D $\mathcal{N}=2$ SCFTs, it becomes interesting to ask what subset it \emph{does} cover.

In \cite{Chacaltana:2010ks}, we started a systematic classification of the 4D $\mathcal{N}=2$ SCFTs, arising from the $A_{N-1}$ series of (2,0) theories. The classification is possible because the various pieces of the construction can themselves be classified. The (2,0) theories are classified by a choice of simply-laced Lie algebra $\mathfrak{j}$ \cite{Witten:1995zh}; (a class of) punctures on the Riemann surface are labeled by nilpotent orbits in $\mathfrak{j}$ \cite{Gukov:2006jk, Gukov:2008sn}, and degenerating Riemann surfaces can be decomposed into a collection of three-punctured spheres connected by cylinders. The set of theories becomes larger if we allow for the (2,0) theory to be ``twisted'' by an outer-automorphism of $\mathfrak{j}$, when traversing a nontrivial cycle of the punctured Riemann surface, $C$. In particular, a twist when circling a puncture introduces a new class of defects, called ``twisted punctures,'' which are classified by nilpotent orbits in non-simply-laced Lie algebras obtained by dividing $\mathfrak{j}$ by the action of the outer automorphism \cite{Tachikawa:2010vg, Chacaltana:2012zy}. In \cite{Chacaltana:2011ze}, we extended the classification of \cite{Chacaltana:2010ks} to the (untwisted) $D_N$ series, and in \cite{Chacaltana:2012ch} and \cite{Chacaltana:2013oka} we incorporated $\mathbb{Z}_2$ outer automorphism twists in the $A_{2N-1}$ series and in the $D_N$ series, respectively.

In this paper, we extend our classification program to the (2,0) theory of type $E_6$. We leave the study of this theory in the presence of $\mathbb{Z}_2$ outer automorphism twists for another publication. There is no known construction of the $E_6$ theory as a low-energy theory of a stack of M5 branes, as was the case for the A- and D-series. Rather, the only known construction is as a compactification of IIB string theory on a K3 manifold at an $E_6$ singularity \cite{Witten:1995zh}. Still, computations are possible because the the 4D $\mathcal{N}=2$ compactification of the $E_6$ theory is controlled by a Hitchin system \cite{Gaiotto:2009hg} with gauge group $E_6$.

As a byproduct, we realize $E_6$ gauge theory with matter in the $4(27)$, as well as $F_4$ gauge theory with matter in the $3(26)$, as compactifications of the $E_6$ (2,0) theory on a 4-punctured sphere. The Seiberg-Witten solution to the $E_6$ gauge theory, with $N_f\leq 3$ $27$s, appeared first in \cite{Terashima:1998iz}. Our solution to the superconformal $F_4$ gauge theory is new.

\section{The $E_6$ Theory}\label{the__theory}
\subsection{The Hitchin system}\label{the_hitchin_system}

The Coulomb branch of the 4D $\mathcal{N}=2$ theories obtained from the compactification of the 6D (2,0) theory of type $E_6$ on a Riemann surface $C$ is described by the Hitchin equations on $C$ with complexified gauge group $E_6$ \cite{Gaiotto:2009hg}. We may also include codimension-two defects of the (2,0) theory localized at points on $C$; we refer to these as ``punctures''. A class of punctures is classified by nilpotent orbits (or, equivalently, by embeddings of $\mathfrak{sl}(2)$) in the complexified Lie algebra $\mathfrak{e}_6$ \cite{Chacaltana:2012zy}. One of the main points of the construction is that a number of physical properties of the 4D theories can be computed directly from geometric properties of the nilpotent orbits that label the punctures on $C$, without any detailed knowledge of the (2,0) theory.

A puncture labeled by a nilpotent orbit $\mathcal{O}$, and located at $z=0$ on $C$, corresponds to a local boundary condition for the Higgs field,

\begin{equation}
\Phi(z) = \frac{X}{z} + \dots
\label{higgsfield}\end{equation}
where $\Phi$ is a holomorphic 1-form on $C$ that takes values in $\mathfrak{e}_6$ and transforms in the adjoint representation of the gauge group, $X$ is a representative of the nilpotent orbit $d(\mathcal{O})$ in $\mathfrak{e}_6$, and $\dots$ represents a generic regular function of $z$ taking values in $\mathfrak{e}_6$. Here, $d(\mathcal{O})$ is the image of $\mathcal{O}$ under the Lusztig-Spaltenstein map $d$ \cite{Benini:2010uu,Chacaltana:2011ze,Chacaltana:2012zy}. Representatives of all nilpotent orbits in $\mathfrak{e}_6$ can be found in \cite{LawtherTesterman}, and a diagram specifying the action of $d$, as well as other properties of the $\mathfrak{e}_6$ orbits, are collected in Appendix C of \cite{Chacaltana:2012zy} (taken from \cite{McGovern,Spaltenstein}) When $d$ is not injective, we distinguish different punctures with the same $d(\mathcal{O})$ by their Sommers-Achar group $\mathcal{C}(\mathcal{O})$ \cite{Chacaltana:2012zy}, which is a discrete subgroup of $E_6$, imposing gauge invariance of $\Phi$ under the action of $\mathcal{C}(\mathcal{O})$.

As in our previous papers, we call $\mathcal{O}$, which labels the puncture, the \emph{Nahm pole}, and $d(\mathcal{O})$, which appears in the Hitchin system boundary condition, the \emph{Hitchin pole}. The physical properties of a puncture labeled by $\mathcal{O}$ will be directly related to geometric properties of the orbits $\mathcal{O}$ and $d(\mathcal{O})$, and the discrete group $\mathcal{C}(\mathcal{O})$.

Unlike classical Lie algebras, there is no natural parameterization of the nilpotent orbits of exceptional Lie algebras in terms of partitions or Young diagrams. Instead, the notation due to Bala and Carter is standard in the representation theory literature. This notation has been briefly discussed in previous works \cite{Gukov:2008sn,Tachikawa:2011yr,Chacaltana:2012zy}, but, for completeness, we review it in Appendix \ref{appendix_bala_carter_notation}, and discuss how to extract relevant information from it.

\subsection{$k$-differentials}\label{differentials}

The low-energy solution of the 4D $\mathcal{N}=2$ theory is encoded in the Seiberg-Witten curve, which is given by the spectral curve of the Hitchin system, i.e., by the characteristic polynomial for the Higgs field $\Phi$, in representation $R$ of $\mathfrak{e}_6$:

\begin{displaymath}
\Sigma_R: {\det}_{R}(\Phi-\lambda I) = \lambda^{d} + \lambda^{d-2} s_2 + \lambda^{d-3} s_3 +\dots + s_{d} = 0,
\end{displaymath}
where $d=\dim{R}$ and the $\lambda^{d-1}$ is zero because $\Tr(\Phi)=0$. Different choices of $R$ will yield different curves $\Sigma_R$. However, as discussed in \cite{Martinec:1995by}, the physical information that one can extract from them is the same.

For a choice of $R$, let $s_k$ be the coefficient of $\lambda^{\dim(R)-k}$, for $k=0,1,2,\dots,\dim(R)$. ($s_0$ and $s_1$ are trivial -- they are 1 and 0, respectively.) The $s_k(z)$ are holomorphic $k$-differentials on $C$ (with poles at the punctures), and can be expressed as polynomials in the trace invariants $P_k=\Tr(\Phi^k)$. Notice that both the $s_k$ and the $P_k$ are dependent on the representation $R$.

On the other hand, we are actually interested in the Casimirs of $\Phi$, which are the independent $k$-differentials providing the gauge-invariant information contained in $\Phi$. For a Lie algebra $\mathfrak{g}$, the number of Casimirs is equal to the rank of $\mathfrak{g}$, and their scaling dimensions are the exponents (minus 1) of $\mathfrak{g}$. Unlike the $s_k$ or the $P_k$, the Casimirs encode the non-redundant gauge-invariant information in $\Phi$.

In our previous papers \cite{Chacaltana:2010ks,Chacaltana:2011ze,Chacaltana:2012ch,Chacaltana:2013oka}, the Lie algebra was of classical type, and $R$ was always chosen to be the smallest non-trivial representation (the fundamental for $A_{N-1}$, or the vector for $D_N$). In such cases, the coefficients $s_k$ directly provide a basis for the Casimirs of $\Phi$. For example, for $\mathfrak{g}=A_{N-1}$, we have $N-1$ Casimirs, of dimensions $2,3,4,\dots, N$. These dimensions match precisely those of the non-trivial coefficients $s_k$ if $R$ is chosen to be the fundamental representation. Thus, the $s_k$ can be taken to be the Casimirs of $A_{N-1}$. Similarly, for $\mathfrak{g}=D_N$, the $N$ Casimirs have degrees $2,4,6,\dots, 2N-2;N$. If $R$ is the vector representation, the $s_{k}$ with $k$ odd vanish, and the non-trivial coefficients are $s_2,s_4,\dots, s_{2N-2},s_{2N}$. Here, $s_{2N}$ is the square of the Pfaffian, $s_{2N}=\tilde{s}^2$, and so $\tilde{s}$ has dimension $N$. Thus, as before, the $s_2,s_4,\dots, s_{2N-2};\tilde{s}$ provide a basis of Casimirs of $D_N$.

But if for $A_{N-1}$ and $D_N$ we had chosen $R$ to be, say, the adjoint representation, then, for large enough $N$, the $s_{k}$ would not have given directly the Casimirs, but instead a lot of redundant information. For example, for $A_5$, there are five Casimirs, with dimensions $2,3,4,5,6$. However, we have 34 non-trivial coefficients $s_k$, with dimensions $2,3,4,\dots, 35$. These $s_k$ are polynomials in the five Casimirs.

For $\mathfrak{j}=\mathfrak{e}_6$, the Casimirs have degrees 2,5,6,8,9,12. In our computations, we have chosen $R$ to be the \emph{adjoint} representation of $\mathfrak{e}_6$, as it is readily available in the form of structure constants; we used those from the computer algebra system GAP 4 \cite{GAP4}. Instead of trying to compute the 78 coefficients $s_k$, we focus directly on the trace invariants $P_k$ for values of $k$ only as large as needed to extract the Casimirs. For the adjoint representation of $\mathfrak{e}_6$, the $P_k$ vanish for $k$ odd, and are non-trivial for $k$ even, except for $P_4=\tfrac{1}{32}(P_2)^2$. Also, as we will see below, to extract the Casimirs, we only need to consider the $P_k$ for $k=2,6,8,10,12,14$. From the $P_k$, one can construct a less-redundant basis,

\begin{align*}
\phi_2 =& \tfrac{1}{48}P_2\\
\phi_6 =& \tfrac{1}{24}\left(P_6 -\tfrac{7}{4608}(P_2)^3\right)\\
\phi_8 =& \tfrac{1}{30}\left(P_8 -\tfrac{2}{9}P_6P_2 +\tfrac{155}{663552}(P_2)^4\right)\\
\phi_{10} =& -\tfrac{1}{105}\left(P_{10}-\tfrac{17}{96} P_8P_2+\tfrac{77}{6912}P_6(P_2)^2-\tfrac{427}{63700992}(P_2)^5\right)\\
\phi_{12} =& \tfrac{1}{155}\left(P_{12} - \tfrac{107}{504} P_{10}P_2+ \tfrac{515}{32256}P_8(P_2)^2- \tfrac{41}{108} (P_6)^2+ \tfrac{295}{497664}P_6(P_2)^3-\tfrac{5669}{9172942848} (P_2)^6\right)\\
\phi_{14} =& \tfrac{1}{4389}\left(P_{14}-\tfrac{3479}{14880}P_{12}P_2 + \tfrac{61391}{3214080}P_{10}(P_2)^2- \tfrac{539}{2160}P_8P_6- \tfrac{139733}{617103360}P_8(P_2)^3\right.\\
& \left.+ \tfrac{165781}{4821120}(P_6)^2P_2- \tfrac{3488947}{44431441920}P_6(P_2)^4+\tfrac{19596907}{409480168734720}(P_2)^7 \right)
\end{align*}

This basis was constructed so that it reduces the constraints in our punctures to a minimum. In particular, the pole coefficients for the minimal puncture have no redundancies; that is, the $\phi_k$ are such that it not be possible to reduce their pole orders in $z$ further by a change of basis, for $z$ a local coordinate centered at the minimal puncture. The $\phi_k$ basis also makes apparent how the Casimirs of degree 5 and 9 appear. Specifically, $\phi_{10}$ and $\phi_{14}$ factor,

\begin{displaymath}
\begin{aligned}
\phi_{10} &\equiv (\phi_5)^2,\\
\phi_{14} &\equiv \phi_5\phi_9
\end{aligned}
\end{displaymath}
These relations define the odd-degree differentials $\phi_5$ and $\phi_9$ (up to a sign, which flips under the $\mathbb{Z}_2$ outer automorphism of $E_6$). So, we can declare the $k$-differentials $\{\phi_2,\phi_5,\phi_6,\phi_8,\phi_9,\phi_{12}\}$ to be our basis of $\mathfrak{e}_6$ Casimirs. In the following, by the $\phi_k$, we will refer to the Casimirs, and ignore the auxiliary differentials $\phi_{10}$ and $\phi_{14}$.

As for the Seiberg-Witten curve, to write it explicitly, we need to know how the 78 coefficients $s_k$ depend on the six Casimirs $\phi_k$. Instead, it is much simpler to write down the (representation independent) Seiberg-Witten geometry, given by an ALE fibration over $C$, and which equivalently describes the low-energy solution of 4D $\mathcal{N}=2$ theories, but directly using the Casimirs \cite{Klemm:1996bj,Lerche:1996an}. Let us briefly review that construction.

\subsection{ALE geometry}\label{ale_geometry}

The 4D $\mathcal{N}=2$ SCFT constructed from the compactification of a 6D (2,0) theory of type $J$ (where $J$ is of A-D-E type) on the Riemann surface $C$ can also be obtained, in a dual manner, from IIB string theory on a non-compact Calabi-Yau threefold, locally given by an ALE fibration over $C$ of type $J$ \cite{Klemm:1996bj,Lerche:1996an}. For $\mathfrak{e}_6$, the threefold is realized as the hypersurface

\begin{displaymath}
\begin{aligned}
  X_{\vec{u}} =& \bigl\{ 0 = w^2 + x^3+ y^{4} + \epsilon_2(z) x y^2 + \epsilon_5(z) x y +\epsilon_6(z) y^2 + \epsilon_8(z) x +\epsilon_9(z) y +\epsilon_{12}(z)\bigr\}\\
  & \subset \text{tot}(K_C^6\oplus K_C^4\oplus K_C^3)
\end{aligned}
\end{displaymath}
where the $\epsilon_k(z)$ are $k$-differentials on $C$ \cite{Tachikawa:2011yr} (in the ``Katz-Morrison basis'' \cite{KatzMorrison}), related to our $\phi_k(z)$ by

\begin{displaymath}
\begin{aligned}
\epsilon_2 &= \tfrac{1}{2} \phi_2\\
\epsilon_5 &= \tfrac{1}{6}\phi_5\\
\epsilon_6 &= \tfrac{1}{72} (-3 \phi_2^3 + 2 \phi_6)\\
\epsilon_8 &= \tfrac{1}{144} (-3 \phi_2^4 + 4 \phi_2 \phi_6 - \phi_8)\\
\epsilon_9 &= \tfrac{1}{72} (-\phi_2^2 \phi_5 + 4 \phi_9)\\
\epsilon_{12} &=\tfrac{1}{5184}(4 \phi_{12} + 6 \phi_2^6 - 12 \phi_2^3\phi_6 + 4 \phi_6^2 + 
  3 \phi_2^2 \phi_8)
\end{aligned}
\end{displaymath}
The $\phi_k(z)$, in turn, depend on the Coulomb branch parameters, $\vec{u}$, as we determine below.

The Seiberg-Witten solution is obtained by computing the periods of the holomorphic 3-form, $\Omega$, over a symplectic basis of (rational) 3-cycles on $X_{\vec{u}}$ which are locally of the form of a 2-sphere in the fiber times a curve on $C$. In the conformal case (which will be our focus in this paper), many of these cycles will necessarily be noncompact (the curve on $C$ being a open curve, stretching between punctures). But, precisely for the parabolic case (where the Higgs field $\Phi(z)$ has simple poles at the punctures, with nilpotent residues), the singularity is integrable, and the requisite periods of $\Omega$ are finite.

In our realization of $F_4$ gauge theory in \S\ref{F4_2}, the differentials $\phi_5(z)$ and $\phi_9(z)$ vanish identically. In this case, the Calabi-Yau, $X_{\vec{u}}$, has a holomorphic involution, $y\to-y$, under which $\Omega\to -\Omega$. The 3-cycles which give the Seiberg-Witten solution are the anti-invariant cycles and the periods of $\Omega$ over those cycles are finite, despite the slightly singular nature of $X_{\vec{u}}$ itself.

\subsection{Puncture properties}\label{puncture_properties}

We describe below how to compute the properties of a puncture. There is a systematic way to compute every property of the puncture, except for the constraints, so it is easiest to compute the other properties first, and use them to guess the constraints. Below, let $\mathcal{O}$ be the Nahm nilpotent orbit that labels a given puncture, and $\mathfrak{su}(2)_\mathcal{O}$ the associated $\mathfrak{su}(2)$ embedding in $\mathfrak{e}_6$.

\subsubsection{Flavour groups}\label{flavour_groups}

The Lie algebra $\mathfrak{f}$ of the flavour group $F=F_{\mathcal{O}}$ is the centralizer of $\mathfrak{su}(2)_{\mathcal{O}}$ in $\mathfrak{e}_6$. A list of the centralizers for each $\mathcal{O}$ can be found in Table 14 of \cite{Chacaltana:2012zy}, taken originally from \cite{Carter}.

The levels of the simple, nonabelian factors $\mathfrak{f}_i$ in $\mathfrak{f}$ follow from the decomposition of the adjoint of $\mathfrak{e}_6$ under $\mathfrak{su}(2)\times \mathfrak{f}$. These decompositions can be deduced from the Bala-Carter label for $\mathcal{O}$, and are summarized in \hyperlink{decomptable}{the table} in Appendix \ref{appendix_bala_carter_notation}.

Let the decomposition of the 78 be
\begin{displaymath}
\mathfrak{e}_6 = \bigoplus_n V_n\otimes R_{n,i}
\end{displaymath}
where $V_n$ is the $n$-dimensional irrep of $\mathfrak{su}(2)$ (denoted by ``$n$'' in \hyperlink{decomptable}{the table}) and $R_{n,i}$ is the corresponding (reducible) representation of $\mathfrak{f}_i$. Let $l_{n,i}$ be the index of $R_{n,i}$. Then, the level of $\mathfrak{f}_i$ is $k_i = \sum_n l_{n,i}$.

For example, consider the $3A_1$ puncture, which has $\mathfrak{f}=\mathfrak{su}(3)\times \mathfrak{su}(2)$. From the table in Appendix \ref{appendix_bala_carter_notation}, we have, for $\mathfrak{f}_1=\mathfrak{su}(3)$,
\begin{displaymath}
R_1=8+3(1),\quad R_2=2(8),\quad R_3=8+1,\quad R_4=2(1),
\end{displaymath}
and so the level is $k_{\mathfrak{su}(3)}= 4l_8=24$. Similarly, for $\mathfrak{f}_2=\mathfrak{su}(2)$, we have
\begin{displaymath}
R_1=3+8(1),\quad R_2=8(2),\quad R_3=9(1),\quad R_4=2,
\end{displaymath}
and thus the level is $k_{\mathfrak{su}(2)}= l_3+9l_2=4+9\times1=13$.

\subsubsection{$\delta n_h$ and $\delta n_v$}\label{nh_and_nv}

The effective number of hyper- and vector multiplets, $\delta n_h$ and $\delta n_v$, can be computed using the formulas in eq. (3.19) of \cite{Chacaltana:2012zy}. Basically, given $\mathcal{O}$, one needs to know how $\mathfrak{e}_6$ decomposes into eigenspaces of the Cartan element of $\mathfrak{su}(2)_\mathcal{O}$.

Here, let us recast those formulas in terms of the \emph{weighted Dynkin diagram} for $\mathcal{O}$, which can be found in Table 14 of \cite{Chacaltana:2012zy}. Let $\vec{x}$ be the six-dimensional vector consisting of the labels of the weighted Dynkin diagram for $\mathcal{O}$. Now, for each root $\alpha$ of $E_6$, let $\vec{k}$ be a six-dimensional vector consisting of the (integer) components of $\alpha$ in any basis of simple roots.  The ``Weyl vector'' is $\vec{W}=\tfrac{1}{2}\sum_{\vec{k}\geq 0} \vec{k}$, where the sum is over positive roots. Let $n_0$ and $n_{1/2}$ be the number of roots $\alpha$ that satisfy $(\vec{x}/2)\cdot \vec{k}=0$ and $(\vec{x}/2)\cdot\vec{k}=1/2$, respectively. (The dot product is Euclidean.) In this notation, the formulas in eq. (3.19) of \cite{Chacaltana:2012zy} are:
\begin{equation}
\begin{aligned}
n_h(\vec{x}) &=8\left(\tfrac{1}{12}h^\vee(E_6)\,\text{dim}(E_6)-\frac{1}{2}\vec{W}\cdot\vec{x}\right)+\frac{1}{2}n_{1/2}\\
n_v(\vec{x}) &=8\left(\tfrac{1}{12}h^\vee(E_6)\,\text{dim}(E_6)-\frac{1}{2}\vec{W}\cdot\vec{x}\right)-\frac{1}{2}n_{0}
\end{aligned}
\end{equation}
where $h^\vee(E_6)=12$ denotes the dual Coxeter number of $E_6$.

For example, for $\vec{x}=\vec{0}$, corresponding to the maximal puncture, one has that the adjoint of $E_6$ decomposes trivially into singlets of $\mathfrak{su}(2)$, $78\to 78(1)$, so $n_0(\vec{0})=\dim(E_6)-\text{rank}(E_6)$, and $n_{1/2}(\vec{0})=0$. Thus,
$$
\begin{aligned}
n_h(\vec{0}) &=\tfrac{2}{3}h^\vee(E_6)\,\text{dim}(E_6) = 624\\
n_v(\vec{0}) &=\tfrac{2}{3}h^\vee(E_6)\,\text{dim}(E_6)-\frac{1}{2}(\dim(E_6)-\text{rank}(E_6)) = 588
\end{aligned}
$$
As a self-consistency check, recall that the complex dimension $\dim_{\mathbb{C}}(\mathcal{O})$ of the orbit $\mathcal{O}$ (seen as a manifold) is related linearly to the difference $n_h-n_v$. Specifically, $n_h-n_v=C - \frac{1}{2}\dim_{\mathbb{C}}(\mathcal{O})$, where $C=n_h(\vec{0})-n_v(\vec{0})=\frac{1}{2}(\dim(E_6)-\text{rank}(E_6))$. In other words,
\begin{equation}
\dim_{\mathbb{C}}(\mathcal{O}) = \dim(E_6) - \text{rank}(E_6) - (n_{1/2} + n_{0}),
\label{dimOfromnhnv}\end{equation}
The dimensions of the nilpotent orbits of $E_6$ are listed in Table 14 of \cite{Chacaltana:2012zy}.

For a non-trivial example, consider the puncture $2A_1$, which has weighted Dynkin diagram
\begin{displaymath}
 \includegraphics[width=84pt]{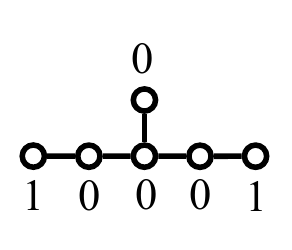}
\end{displaymath}
that is, $\vec{x}=(1,0,0,0,1;0)$. One finds $\vec{W}=(8,15,21,15,8;11)$, $n_{1/2}=16$, $n_0=24$. Thus, $n_h(2A_1)=568$ and $n_v(2A_1)=548$, and one indeed checks \eqref{dimOfromnhnv} for $\dim_{\mathbb{C}}(2A_1)=32$.

\subsubsection{Pole structures}\label{pole_structures}

The ``pole structure'' is the set of leading pole orders $\{p_2,p_5,p_6,p_8,p_9,p_{12}\}$ in the expansion of the Casimirs $\phi_k$ in a coordinate $z$ centered at the puncture, $\phi_k(z)\sim 1/z^{p_k}$.

To compute the pole structure, we need a representative of the Hitchin nilpotent orbit $d(\mathcal{O})$. A table of representatives of all nilpotent orbits of $E_6$ can be found in Table 2 of \cite{LawtherTesterman}. In this table, a nilpotent representative is given by a sum of weighted Dynkin diagrams, and each weighted Dynkin diagram represents an element in the root vector space of $\mathfrak{e}_6$ for a positive root $\alpha$, where $\alpha$ is such that its components in a basis of simple roots are given by the labels of the Dynkin diagram. The nilpotent representative is the sum of these root-vector space elements. This procedure is most easily understood in terms of an example.

Take, for instance, $\mathcal{O}=D_4(a_1)$. The Hitchin orbit, given by the Spaltenstein dual, is the same, $d(\mathcal{O})=\mathcal{O}=D_4(a_1)$. This orbit has a nilpotent representative $X$ given by a sum of five elements \cite{LawtherTesterman},
\begin{displaymath}
 \includegraphics[width=439pt]{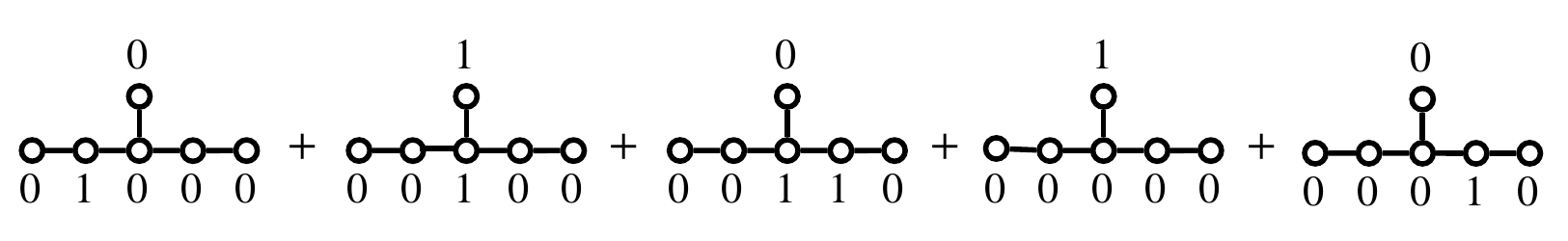}
\end{displaymath}
The five summands above represent arbitrary non-zero elements $X_{\alpha_i}$ ($i=1,\dots,5$) in the root vector spaces for the positive roots
$$
\begin{aligned}
\alpha_1 &=s_2, & \qquad \alpha_4 &=s_6,\\
\alpha_2 &=s_3+s_6, & \qquad \qquad \alpha_5 &=s_4,\\
\alpha_3 &=s_3+s_4,& & 
\end{aligned}
$$
respectively, where $\{s_1,\dots,s_5;s_6\}$ is a basis of simple roots of $E_6$. So,  $X=X_{\alpha_1}+\dots+X_{\alpha_5}$. Fortunately, GAP4 provides a Chevalley basis for the adjoint representation of  $\mathfrak{e}_6$, so it is trivial to find elements $X_{\alpha_i}$. Once we know $X$, we compute $\Phi(z)$ using $X$ as the residue in \eqref{higgsfield}, then the Casimir $k$-differentials $\phi_k$ as in \S\ref{differentials}, and we finally find the pole structure $\{1,3,4,6,6,9\}$ for the $D_4(a_1)$ puncture. (Actually, there are \emph{three} orbits, $D_4(a_1)$, $A_3+A_1$ and $2A_2+A_1$, that map under Spaltenstein to $D_4(a_1)$, so we have three punctures with the same pole structure. However, the other properties of these punctures are different.)

\subsubsection{Constraints}\label{constraints}

The constraints for some $E_6$ punctures are, in some cases, much less obvious than those in the $A_{N-1}$ and $D_N$ series. The guiding quantities to find constraints are $\delta n_v$ and the (complex) dimension of the Hitchin nilpotent orbit, $d$. These are, respectively, the graded and ungraded local contributions to the Coulomb branch.

Let us be specific. Let $z$ be a local coordinate on $C$ centered at the puncture, and let $c^{(k)}_l$ be the coefficient of $z^{-l}$ in the expansion of $\phi_k=\phi_k(z)$ in $z$. Recall that, in the notation of our previous papers, a ``c-constraint'' is a polynomial relation among coefficients $c^{(k)}_l$ (of homogeneous bi-degree in both $k$ and $l$). On the other hand, an ``a-constraint'' is a relation that defines a new quantity, $a^{(k)}_l$, of dimension $k$, in terms of the $c^{(k)}_l$. Only the $c^{(k)}_l$ with $l\gt 0$ parameterize the Hitchin nilpotent orbit \cite{Gukov:2008sn}. In the absence of constraints, all the $c^{(k)}_l$ with $0\lt l\leq p_k$ are independent, so their total number, $\sum{p_k}$, should be equal to the dimension of the Hitchin nilpotent orbit. Thus, if there are no constraints, $\sum{p_k}=d$. A c-constraint reduces the total number of independent parameters by one, whereas an a-constraint does not affect this number. So, one should have:

\begin{displaymath}
\sum{p_k} - (\text{number of c-constraints}) = d
\end{displaymath}
Hence, $d$ tells us how many c-constraints exist. On the other hand, the graded sum of the parameters, that is, the result of adding $(2k-1)$ for each parameter of degree $k$ (in the presence of ``a"-constraints, $k$ is not restricted to the degrees of the Casimirs), should be equal to $n_v$. An a-constraint replaces a parameter of a certain degree $k$ by another one of a different degree $k'<k$. So, to get precisely $n_v$, one must take into account all a-constraints and c-constraints.

\subsubsection{Puncture collisions}\label{puncture_collisions}

Suppose we have two punctures on a plane, so the Higgs field has two simple poles with residues $X_1$ and $X_2$. Near each puncture, the Higgs field $\Phi$ looks like eq. \eqref{higgsfield}. In the limit where the two punctures collide, the Higgs field has one simple pole with residue $X=X_1+X_2$ (by the residue theorem applied to the sphere that bubbles off), which corresponds to a new puncture. Generically, $X$ will be mass deformed. The mass deformations are interpreted as VEVs of the scalars in the gauge multiplet associate to the factor in the gauge group which becomes weakly coupled in the collision limit. One can also study this degeneration by computing the Casimirs $\phi_k$ from the Higgs field before taking the collision limit.

Alternatively, one can bypass the Higgs field, and study the collision directly with the $\phi_k$, by writing a generic $k$-differential with poles at the positions of the two punctures (given by their pole structures), and imposing at each pole the constraints of the corresponding puncture. Then, taking the collision limit, the pole structure and constraints of the resulting puncture on the plane arise naturally.

As an example, let us see that the collision of two $D_5$ punctures on a plane produces an $Sp(2)$ gauge group, gauged off an $A_3$ puncture. Let us write generic Casimirs for the collision, taking the $D_5$ punctures to be at $z=0$ and $z=x$:

\begin{align*}
\phi_2(z) &=\frac{u_2+z v_2 + z(z-x)P_2(z)}{z(z-x)}\\
\phi_5(z)  &=\frac{u_5+zv_5+z(z-x)w_5+z^2(z-x)P_5(z)}{z^2(z-x)^2}\\
\phi_6(z)  &=\frac{u_6+zv_6+z(z-x)w_6+z^2(z-x)P_6(z)}{z^3(z-x)^3}\\
\phi_8(z)  &=\frac{u_8+zv_8+z(z-x)w_8+z^2(z-x)y_8+z^2(z-x)^2 P_8(z)}{z^4(z-x)^4}\\
\phi_9(z)  &=\frac{u_9+zv_9+z(z-x)w_9+z^2(z-x)P_9(z)}{z^4(z-x)^4}\\
\phi_{12}(z)  &=\frac{u_{12}+zv_{12}+z(z-x)w_{12}+z^2(z-x)y_{12}+z^2(z-x)^2P_{12}(z)}{z^6(z-x)^6},\\
\end{align*}
where $P_2(z), P_5(z),\dots, P_{12}(z)$ denote regular functions in $z$. To solve the constraints at each $D_5$ puncture, we introduce new parameters $s_4$ and $t_4$ of dimension four, and write:

\begin{displaymath}
\begin{aligned}
u_6 = &\frac{3s_4u_2}{4},\qquad &v_6=&\frac{3}{2}(t_4u_2+s_4v_2+t_4v_2 x),\\
u_8 = &3s_4^2,\qquad &v_8=& 3(2s_4t_4+t_4^2 x),\\
u_9 =&-\frac{s_4u_5}{4},\qquad &v_9 =& -\frac{1}{4} (t_4u_5+s_4v_5+t_4v_5x),\\
u_{12}=&\frac{3s_4^3}{2},\qquad &v_{12}=&\frac{3}{2}t_4(3s_4^2+3s_4 t_4 x+t_4^2 x^2),\\
w_{12}=&\frac{3}{4}(3s_4t_4^2+s_4w_8+2t_4^3 x),\qquad & y_{12}=&-\frac{3}{4}(t_4^3-t_4w_8-s_4y_8-t_4y_8 x)
\end{aligned}
\end{displaymath}
In the collision limit, $x\to 0$, the new puncture appears at $z=0$. The expansion in $z$ of the Casimirs in this limit is:

\begin{displaymath}
\begin{aligned}
\phi_2(z) =& \frac{u_2}{z^2}+\frac{v_2}{z}+\dots\\
\phi_5(z) =& \frac{u_5}{z^4}+\dots\\
\phi_6(z) =& \frac{3s_4u_2}{2z^6}+\frac{3(t_4u_2+s_4v_2)}{2z^5}+\frac{w_6}{z^4}+\dots\\
\phi_8(z) =&\frac{3s_4^2}{z^8}+\frac{6s_4t_4}{z^7}+\frac{w_8}{z^6}+\dots\\
\phi_9(z) =&-\frac{s_4u_5}{4z^{8}}-\frac{(t_4u_5+s_4v_5)}{4z^7}+\frac{w_9}{z^6}+\dots\\
\phi_{12}(z) =&\frac{3s_4^3}{2z^{12}}+\frac{9s_4^2t_4}{2z^{11}}+\frac{3(3s_4t_4^2+s_4w_8)}{4z^{10}}-\frac{3(t_4^3-t_4w_8-s_4y_8)}{4z^{9}}+\dots,\\
\end{aligned}
\end{displaymath}
where the $\dots$ indicate less singular terms in $z$. So, $u_2$ and $s_4$ can be interpreted as the VEVs of Coulomb branch parameters (of degree two and four) of the gauge group (which, with a little more work, can be checked to be $Sp(2)$). In the limit $u_2,s_4 \to 0$, we obtain the Casimirs for the massless puncture, with pole orders $\{1,4,4,6,7,9\}$, and with constraints

\begin{displaymath}
\begin{aligned}
c^{(9)}_{7} =& \frac{1}{2}\tilde{t}_4 u_5\\
c^{(12)}_{9} =& 6\tilde{t}_4^3-\frac{3}{2}w_8\tilde{t}_4,
\end{aligned}
\end{displaymath}
where $\tilde{t}_4\equiv -t_4/2$. Thus, we get precisely the pole structure and constraints of the $A_3$ puncture.

\subsection{Global symmetries and the superconformal index}\label{global_symmetries_and_the_superconformal_index}

\subsubsection{Cataloguing fixtures using the superconformal index}\label{cataloging_fixtures_using_the_superconformal_index}

For the $E_6$ theory, we find 880 fixtures with three regular punctures which correspond to interacting SCFTs, possibly with additional decoupled hypermultiplets. Each of these SCFTs has a manifest global symmetry group, which is given by the product of the flavor symmetry groups of the three punctures. This global symmetry group may, in general, become enhanced to a larger group.

To determine the global symmetry group and number of free hypermultiplets for each of these fixtures, we use the superconformal index \cite{Kinney:2005ej, Gadde:2009kb, Gadde:2011ik, Gadde:2011uv, Lemos:2012ph}. The superconformal index of $E$-type class $\mathcal{S}$ theories has not yet been systematically studied. However, since the methods used for $A$- and $D$-type theories generalize to any root system, we assume the superconformal index\footnote{In what follows we will consider the Hall-Littlewood limit of the index \cite{Gadde:2011uv}, which depends on one superconformal fugacity, $\tau$.}  for a fixture in the $E_6$ theory takes the usual form

\begin{equation}
\mathcal{I}(\mathbf{a}_i,\tau)=\mathcal{A}(\tau)\sum_{\lambda}\frac{\prod_{i=1}^3\mathcal{K}(\mathbf{a}_i)P^\lambda(\mathbf{a}_i|\tau)}{P^\lambda(\mathbf{a}_\text{triv}|\tau)}
\label{SCI}\end{equation}
where

\begin{itemize}%
\item The sum is over $\lambda$ labeling the highest weights of finite-dimensional irreducible representations of $\mathfrak{e}_6$.

\item The $P^\lambda(\mathbf{a}_i|\tau)$ are Hall-Littlewood polynomials, defined for a general root system by

\end{itemize}
\begin{displaymath}
\begin{split} P^\lambda&=W^{-1}(\tau) \sum_{w \in W}w\left(e^\lambda \prod_{\alpha \in R^+}\frac{1-\tau^2e^{-\alpha}}{1-e^{-\alpha}}\right), \\ W(\tau)& = \sqrt{\sum_{\stackrel{w \in W}{w\lambda=\lambda}}\tau^{2\ell(w)}} \end{split}
\end{displaymath}
where $R^+$ denotes the set of positive roots, $W$ the Weyl group, and $\ell(w)$ the length of the Weyl group element $w$.

\begin{itemize}%
\item $\mathbf{a}_i \equiv \{e^\alpha\}_{\alpha \in R^+}$ denotes a set of flavor fugacities dual to the Cartan subalgebra of the flavor symmetry of the $i^\text{th}$ puncture.  $\mathbf{a}_\text{triv}$ denotes the set of fugacities dual to the Cartan of the embedded $\mathfrak{su}(2)$ of the trivial puncture.

\item The $\mathcal{K}$-factors are discussed in \cite{Gadde:2011uv, Lemos:2012ph, Gaiotto:2012uq, Gaiotto:2012xa}. We will not need their detailed form for our purposes.

\item $\mathcal{A}(\tau)$ is an overall, flavor fugacity independent normalization.

\end{itemize}
Consider a fixture corresponding to an interacting SCFT, with global symmetry $G_\text{global}$, plus free hypermultiplets transforming in a representation $R$ of a flavor symmetry $F$. Let $G_\text{fixt}\equiv G_\text{global} \times F$ denote the global symmetry of the fixture. As discussed in \cite{Gaiotto:2012uq}, the number of free hypers in the fixture and the global symmetry of the fixture can be read off from the first two non-trivial terms in the Taylor expansion of the index. Schematically, this is given by

\begin{equation}
\mathcal{I} = 1 + \chi^{R}_\text{F} \tau + \chi^{adj}_{G_\text{fixt}}\tau^2+\dots
\label{schematic}\end{equation}
where $\chi^{R}_\text{F}$ is the Weyl character of $R$ and $\chi^{adj}_{G_\text{fixt}}$ is the character of the adjoint representation of $G_\text{fixt}$, where both of these representations are viewed as reducible representations of the manifest symmetry algebra. By Taylor expanding $\mathcal{I}_\text{free}=PE[\tau \chi^{R}_\text{F}]$ (where $PE$ denotes the Plethystic exponential) and removing the contribution of the free hypermultiplets in \eqref{schematic}, we arrive at

\begin{displaymath}
\begin{split}
\mathcal{I}_\text{SCFT}&=\mathcal{I}/\mathcal{I}_\text{free} \\
&=1+ \chi^{adj}_{G_\text{global}}\tau^2+\dots
\end{split}
\end{displaymath}
from which we can read off the global symmetry of the interacting SCFT.
\subsubsection{Computing the expansion of the index}\label{computing_the_expansion_of_the_index}

In \eqref{SCI} the term in the sum coming from the trivial representation of $\mathfrak{e}_6$ gives, to second order in $\tau$, \cite{Gaiotto:2012uq}

\begin{displaymath}
\mathcal{I}=1+\chi^{adj}_{G_\text{manifest}}\tau^2+\cdots
\end{displaymath}
encoding the manifest global symmetry group. The global symmetry group of the fixture is enhanced if there are terms contributing at order $\tau^2$ coming from the sum over $\lambda \gt 0$.

To order $\tau^2$, \eqref{SCI} simplifies to \footnote{Since the theories considered here are all ``good'' or ``ugly'' (in the sense of \cite{Gaiotto:2011xs}), the lowest possible contribution from the sum over $\lambda \gt 0$ is at order $\tau$ (see \cite{Gaiotto:2012uq} for a discussion of the superconformal index in the context of the good/ugly/bad trichotomy of 4d $\mathcal{N}=2$ theories). From \eqref{expansion}, we see that $\mathcal{A}(\tau)$ and $\mathcal{K}(\mathbf{a}_i)$ are both 1 + $\mathcal{O}(\tau^2)$, so we can set them both to one in the order $\tau^2$ approximation. We have also used the fact that $P^\lambda=\chi^\lambda + \mathcal{O}(\tau^2)$.} 

\begin{equation}
\mathcal{I}=1+\chi^{adj}_{G_\text{manifest}}\tau^2 + [\sum_{\lambda \gt 0}\frac{\prod_{i=1}^3\chi^\lambda(\mathbf{a_i}|\tau)}{\chi^\lambda(\mathbf{a}_\text{triv}|\tau)}]_{\mathcal{O}(\tau^2)}
\label{expansion}\end{equation}
To compute \eqref{expansion}, we consider each $\mathfrak{e}_6$ representation in the sum to be a reducible representation of $\mathfrak{su}(2) \times \mathfrak{f}$ and plug in the corresponding character expansion, where the embedded $\mathfrak{su}(2)$ has fugacity $\tau$. The decomposition of any $\mathfrak{e}_6$ representation in terms of $\mathfrak{su}(2) \times \mathfrak{f}$ representations can be obtained using the projection matrices listed in Appendix \ref{projection_matrices}.

Of the 881 fixtures involving three regular punctures, we find that 1 is a free-field fixture, 60 are mixed fixtures and another 134 are interacting fixtures with an enhanced global symmetry group. We list these in the tables below. For the remaining 686 interacting fixtures, the global symmetry group is the manifest one.

As an example, consider the fixture
$$
\includegraphics[width=92pt]{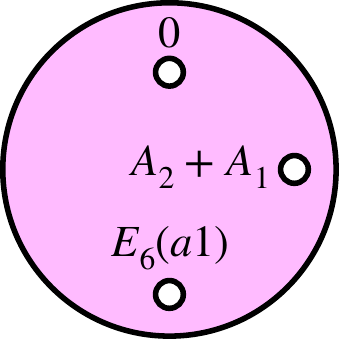}\quad.
$$
The manifest global symmetry is $(E_6)_{24} \times SU(3)_{12} \times U(1)$. The contributions at order $\tau^2$ come from the sum over the $27$, $\overline{27}$, $78$, $351$, $\overline{351}$, $351'$, $\overline{351}'$, and $650$ of $\mathfrak{e}_6$. The expansion of the superconformal index is given by \footnote{For simplicity, we write the dimension to stand for the character of the corresponding representation. The subscript is the $U(1)$ weight.}

\begin{displaymath}
\begin{split}
\mathcal{I} &= 1 + \{(27,1)_1 + (\overline{27},1)_{-1}\} \tau + \{(1,1)_0 + (78,1)_0 +\\
&(650,1)_0 + (27,1)_{-2} + (351',1)_{-2} + (\overline{27},1)_2 +\\
&(\overline{351}',1)_2+ (78,1)_0+(1,8)_0+(27,\overline{3})_0+(\overline{27},3)_0 \}\tau^2+\dots \end{split}
\end{displaymath}
Due to the order $\tau$ term, this is a mixed fixture, with $27$ free hypermultiplets transforming in the fundamental representation of $E_6$. The index of these free hypers is given by

\begin{displaymath}
\begin{split}
\mathcal{I}_\text{free}&=PE[\tau\{(27,1)_1 + (\overline{27},1)_{-1}\}] \\
&=1+\{(27,1)_1 + (\overline{27},1)_{-1}\}\tau +\\
& \{(1,1)_0+(78,1)_0+(650,1)_0+(27,1)_{-2} + (351',1)_{-2}+(\overline{27},1)_2+(\overline{351}',1)_2\}\tau^2+\dots
\end{split}
\end{displaymath}
The index of the underlying SCFT is then

\begin{displaymath}
\begin{split}
\mathcal{I}_\text{SCFT}&=\mathcal{I}/\mathcal{I}_\text{free} \\
&=1 + \{(78,1)_0+(1,8)_0+(27,\overline{3})_0+(\overline{27},3)_0\}\tau^2
\end{split}
\end{displaymath}
We recognize the coefficient of $\tau^2$ as the character of the adjoint representation of $E_8$. Computing the other numerical invariants of the fixture, we find that this is the $(E_8)_{12}$ theory of Minahan and Nemeschansky \cite{Minahan:1996cj} with 27 additional free hypermultiplets.

\subsection{Levels of enhanced global symmetry groups}\label{levels_of_enhanced_global_symmetry_groups}

Since the superconformal index gives the branching rule for the adjoint representation of $G_\text{global}$ under the subgroup $G_\text{manifest}$, it most cases it is straightforward to determine the level of each factor in $G_\text{global}$ from those of $G_\text{manifest}$: If $H_{k'}$ is a subgroup of $G_k$, then $k$ is given by \cite{Argyres:2007cn}

\begin{displaymath}
k=\frac{k'}{I_{\mathrlap{H \hookrightarrow G}}\ \ \ \ \ \ }
\end{displaymath}
where $I_{H \hookrightarrow G}$ is the index of the embedding of $H$ in $G$.

There are two cases which require a little more work. The first is when a manifest $U(1)$ becomes enhanced to $SU(2)$. Since we do not know how to assign a level to a $U(1)$ flavor symmetry (which would require a precise understanding of how the generator is normalized), we cannot immediately determine the level of the enhanced $SU(2)$ from the index.

The second case is when some factor $H_k$ in $G_\text{manifest}$ is embedded diagonally as

\begin{displaymath}
H_k \hookrightarrow H_{k_1} \times H_{k_2}.
\end{displaymath}
Since the only embedding of $H$ in itself has index one, in this case, all we know is that $k_1+k_2=k$.

If any of these remain as factors in $G_\text{global}$ (that is, if they do not combine with some other factor, with known level, to enhance $G_\text{global}$), we cannot determine their levels from the index, and must determine them using an $S$-duality. To do so, we look for a 4-punctured sphere for which the SCFT appears in some degeneration, with $H_{k_i}$ in the centralizer of subgroup of $G_\text{global}$ being weakly gauged.

Unfortunately, there are a few such fixtures for which no puncture can be gauged (some of these can still be gauged in the twisted sector, which will be discussed in \S\ref{a_detour_through_the_twisted_sector}). For these, we do not have a way to determine the levels. In the end, there are two interacting fixtures whose levels we cannot completely determine.

\section{Tinkertoys}\label{tinkertoys}

\subsection{Regular punctures}\label{regular_punctures}

The pole structure $\{p_2,p_5,p_6,p_8,p_9,p_{12}\}$ of a puncture at $z=0$ will be the leading pole orders in $z$ of the differentials $\phi_k(z)$ for $k=2,5,6,8,9,12$. Notice that in some cases there are constraints, not just on the coefficient of this leading singularity, but also on \emph{subleading} terms in the Laurent expansion of the $k$-differentials.

{\footnotesize
\renewcommand{\arraystretch}{2.25}

\begin{longtable}{|c|c|c|c|c|c|}
\hline
\mbox{\shortstack{Nahm\\B-C label}}&\mbox{\shortstack{Hitchin\\B-C label}}&Pole structure&Constraints&Flavour group&$(\delta n_h,\delta n_v)$\\
\hline
\endhead
0&$E_{6}$&$\{1,4,5,7,8,11\}$&-&${(E_6)}_{24}$&$(624,588)$\\
\hline
$A_1$&$E_{6}(a_{1})$&$\{1,4,5,7,8,10\}$&-&${SU(6)}_{18}$&$(590,565)$\\
\hline
$2A_1$&$D_{5}$&$\{1,4,5,7,7,10\}$&-&${Spin(7)}_{16}\times U(1)$&$(568,548)$\\
\hline
$3A_1\,(\underline{ns})$&$(E_{6}(a_{3}),\mathbb{Z}_2)$&$\{1,4,5,6,7,10\}$&-&${SU(3)}_{24}\times {SU(2)}_{13}$&$(549,533)$\\
\hline
$A_2$&$E_{6}(a_{3})$&$\{1,4,5,6,7,10\}$&$\begin{gathered}\\ c^{(12)}_{10}=-\bigl(c^{(6)}_5\bigr)^2+\bigl(a^{(6)}_5\bigr)^2\\ {}\end{gathered}$&${SU(3)}_{12}^2$&$(536,521)$\\
\hline
$A_2+A_1$&$D_{5}(a_{1})$&$\{1,4,5,6,7,9\}$&-&${SU(3)}_{12}\times U(1)$&$(523,510)$\\
\hline
$2A_2$&$D_{4}$&$\{1,3,5,6,6,9\}$&-&${(G_2)}_{12}$&$(496,484)$\\
\hline
$A_2+2A_1$&$A_{4}+A_{1}$&$\{1,4,4,6,7,9\}$&-&${SU(2)}_{54}\times U(1)$&$(510,499)$\\
\hline
$A_3$&$A_{4}$&$\{1,4,4,6,7,9\}$&$\begin{gathered}\\ c^{(9)}_7=\tfrac{1}{2}c^{(5)}_4a^{(4)}_3\\c^{(12)}_9=\begin{array}[t]{l}6\bigl(a^{(4)}_3\bigr)^3\\-\tfrac{3}{2}c^{(8)}_6a^{(4)}_3\end{array}\\ {}\end{gathered}$&${Sp(2)}_{10}\times U(1)$&$(476,466)$\\
\hline
$2A_2+A_1\,(\underline{ns})$&$(D_{4}(a_1),S_3)$&$\{1,3,4,6,6,9\}$&-&${SU(2)}_{26}$&$(482,473)$\\
\hline
$A_3+A_1\,(\underline{ns})$&$(D_{4}(a_1),\mathbb{Z}_2)$&$\{1,3,4,6,6,9\}$&$\begin{gathered}\\ c^{(12)}_9= a^{(4)}_3\Bigl(\tfrac{16}{9}\bigl(a^{(4)}_3\bigr)^2-c^{(8)}_6\Bigr)\\ {}\end{gathered}$&${SU(2)}_9\times U(1)$&$(465,457)$\\
\hline
$D_4(a_1)$&$D_{4}(a_{1})$&$\{1,3,4,6,6,9\}$&$\begin{gathered}\\ c^{(8)}_6=\tfrac{4}{3}\Bigl(\bigl(a^{(4)}_3\bigr)^2+3\bigl(a'^{(4)}_3\bigr)^2\Bigr)\\c^{(12)}_9=\tfrac{4}{9}a^{(4)}_3\Bigl(\bigl(a^{(4)}_3\bigr)^2 - 9\bigl( a'^{(4)}_3\bigr)^2\Bigr)\\ {}\end{gathered}$&$U(1)^2$&$(456,449)$\\
\hline
$A_4$&$A_{3}$&$\{1,3,4,6,6,9\}$&$\begin{gathered}\\ c^{(8)}_6=3\bigl(a^{(4)}_3\bigr)^2\\c^{(9)}_6=\tfrac{1}{4}c^{(5)}_3a^{(4)}_3\\c^{(12)}_9=-\tfrac{3}{2}\bigl(a^{(4)}_3\bigr)^3\\c^{(12)}_8=-\tfrac{3}{4}a^{(4)}_3c^{(8)}_5\\ {}\end{gathered}$&${SU(2)}_8\times U(1)$&$(408,402)$\\
\hline
$D_4$&$2A_{2}$&$\{1,3,4,5,6,8\}$&$\begin{gathered}
c^{(8)}_5=-\begin{array}[t]{l}4c^{(6)}_4c^{(2)}_1+4c^{(5)}_3a^{(3)}_2\\-2\bigl(a^{(3)}_2\bigr)^2c^{(2)}_1\end{array}\\
c^{(9)}_6=-\tfrac{1}{12}a^{(3)}_2\Bigl(c^{(6)}_4+\tfrac{1}{2}\bigl(a^{(3)}_2\bigr)^2\Bigr)\\
c^{(12)}_8=-\begin{array}[t]{l}
\Bigl(c^{(6)}_4+\tfrac{1}{2}\bigl(a^{(3)}_2\bigr)^2\Bigr)\cdot\\
\cdot\Bigl(c^{(6)}_4-\tfrac{3}{2}\bigl(a^{(3)}_2\bigr)^2\Bigr)\end{array}\\
c^{(12)}_7=-12c^{(9)}_5a^{(3)}_2-2c^{(6)}_4c^{(6)}_3\\-c^{(6)}_3\bigl(a^{(3)}_2\bigr)^2\\{}\end{gathered}$&${SU(3)}_{12}$&$(368,362)$\\
\hline
$A_4+A_1$&$A_{2}+2A_{1}$&$\{1,3,4,5,5,7\}$&-&$U(1)$&$(400,395)$\\
\hline
$D_5(a_1)$&$A_2+A_1$&$\{1,3,4,5,5,7\}$&$\begin{gathered}\\ c^{(6)}_4=-\tfrac{1}{8}\left(a^{(3)}_2\right)^2\\c^{(8)}_5=2c^{(5)}_3a^{(3)}_2\\c^{(12)}_7=-6c^{(9)}_5a^{(3)}_2\\ {}\end{gathered}$&$U(1)$&$(355,351)$\\
\hline
$A_5\,(\underline{ns})$&$(A_{2},\mathbb{Z}_2)$&$\{1,2,4,4,4,6\}$&-&${SU(2)}_7$&$(335,331)$\\
\hline
$E_6(a_3)$&$A_{2}$&$\{1,2,4,4,4,6\}$&$\begin{gathered}\\ c^{(6)}_4=\left(a^{(3)}_2\right)^2\\ {} \end{gathered}$&none&$(328,325)$\\
\hline
$D_5$&$2A_{1}$&$\{1,2,3,4,4,6\}$&$\begin{gathered}\\ c^{(6)}_3 = \tfrac{3}{2}c^{(2)}_1 a^{(4)}_2\\ c^{(8)}_4 = 3\bigl(a^{(4)}_2\bigr)^2\\c^{(9)}_4=-\tfrac{1}{4}a^{(4)}_2 c^{(5)}_2\\c^{(12)}_6=\tfrac{3}{2}\bigl(a^{(4)}_2\bigr)^3\\c^{(12)}_5=\tfrac{3}{4}c^{(8)}_3 a^{(4)}_2\\ {} \end{gathered}$&$U(1)$&$(240,238)$\\
\hline
$E_6(a_1)$&$A_{1}$&$\{1,1,2,2,2,3\}$&-&none&$(168,167)$\\
\hline
\end{longtable}

}

Note that there is a special piece, consisting of three punctures: $2A_2+A_1$, $A_3+A_1$ and the special puncture $D_4(a_1)$. For $2A_2+A_1$, the Sommers-Achar group is the nonabelian group, $S_3$. It acts on $a^{(4)},a'^{(4)}$ as

\begin{displaymath}
\begin{pmatrix}a^{(4)}\\ a'^{(4)}\end{pmatrix}\to
\gamma\,
\begin{pmatrix}a^{(4)}\\ a'^{(4)}\end{pmatrix}
\end{displaymath}
for

\begin{displaymath}
\gamma\in\left\{
\begin{pmatrix}1&0\\0&1\end{pmatrix},
\begin{pmatrix}1&0\\0&-1\end{pmatrix},
\tfrac{1}{2}\begin{pmatrix}-1&-3\\1&-1\end{pmatrix},
\tfrac{1}{2}\begin{pmatrix}-1&3\\-1&-1\end{pmatrix},
\tfrac{1}{2}\begin{pmatrix}-1&-3\\-1&1\end{pmatrix},
\tfrac{1}{2}\begin{pmatrix}-1&3\\1&1\end{pmatrix}
\right\}
\end{displaymath}
For $A_3+A_1$, the Sommers-Achar group is the $\mathbb{Z}_2$ subgroup of $S_3$, generated by $a'^{(4)}\to - a'^{(4)}$. For $D_4(a_1)$, the Sommers-Achar group is of course trivial, so that both $a^{(4)},a'^{(4)}$ survive as Coulomb branch parameters.

\subsection{Free-field fixtures}\label{freefield_fixtures}

We denote a 3-punctured sphere, in the tables below, by listing the Bala-Carter labels of the three punctures. For the free-field fixtures, one of the punctures is an irregular puncture\footnote{Or, in the case of fixture 13, a full puncture, corresponding to the trivial orbit, $0$.} (in the sense used in our previous papers), which we denote\footnote{For brevity, we will often omit the level, $k$, when denoting an irregular puncture.} by the pair, $(\mathcal{O}, G_k)$, where $\mathcal{O}$ is the regular puncture obtained as the OPE of the two regular punctures which collide, and this fixture is attached to the rest of the surface via a cylinder

\begin{displaymath}
(\mathcal{O}, G_k) \xleftrightarrow{\quad G\quad} \mathcal{O}
\end{displaymath}
with gauge group $G\subset F\subset E_6$. Here, $F$ is the flavour symmetry group of the puncture, $\mathcal{O}$, and the levels are such that $G$ has vanishing $\beta$-function. 

\bigskip
{
\renewcommand{\arraystretch}{1.25}

\begin{longtable}{|c|l|c|l|}
\hline
\#&Fixture&$n_h$&Representation\\
\hline
\endhead
1&$\begin{matrix}E_6(a_1)\\ E_6(a_1)\end{matrix}\quad (A_5,{SU(2)}_1)$&1&$\tfrac{1}{2}(2)$\\
\hline
2&$\begin{matrix}E_6(a_1)\\ D_5\end{matrix}\quad (A_4,{SU(2)}_0)$&0&empty\\
\hline
3&$\begin{matrix}E_6(a_1)\\ E_6(a_3)\end{matrix}\quad(2A_2,{SU(3)}_0)$&0&empty\\
\hline
4&$\begin{matrix}E_6(a_1)\\ A_5\end{matrix}\quad (2A_2,{(G_2)}_4)$&7&$\tfrac{1}{2}(2,7)$\\
\hline
5&$\begin{matrix}E_6(a_1)\\ D_5(a_1)\end{matrix}\quad(A_2+A_1,{SU(3)}_0)$&0&empty\\
\hline
6&$\begin{matrix}E_6(a_1)\\ A_4+A_1\end{matrix}\quad (2A_1,{(G_2)}_0)$&0&empty\\
\hline
7&$\begin{matrix}E_6(a_1)\\ D_4\end{matrix}\quad (A_2,{SU(3)}_0)$&0&empty\\
\hline
8&$\begin{matrix}E_6(a_1)\\ A_4\end{matrix}\quad (2A_1,{Spin(7)}_4)$&8&$\tfrac{1}{2}(2,8)$\\
\hline
9&$\begin{matrix}E_6(a_1)\\ D_4(a_1)\end{matrix}\quad (0,{Spin(8)}_0)$&0&empty\\
\hline
10&$\begin{matrix}E_6(a_1)\\ A_3+A_1\end{matrix}\quad (0,{Spin(9)}_4)$&9&$\tfrac{1}{2}(2,9)$\\
\hline
11&$\begin{matrix}E_6(a_1)\\ 2A_2+A_1\end{matrix}\quad (0,{(F_4)}_{12})$&26&$\tfrac{1}{2}(2,26)$\\
\hline
12&$\begin{matrix}E_6(a_1)\\ A_3\end{matrix}\quad (0,{Spin(10)}_8)$&20&$\tfrac{1}{2}(4,10)$\\
\hline
13&$\begin{matrix}E_6(a_1)\\ A_2+2A_1\end{matrix}\quad 0$&54&$(2,27)$\\
\hline
14&$\begin{matrix}D_5\\ D_5\end{matrix}\quad (A_3,{Sp(2)}_2)$&4&$1(4)$\\
\hline
15&$\begin{matrix}D_5\\ E_6(a3)\end{matrix}\quad (2A_1,{SU(4)}_0)$&0&empty\\
\hline
16&$\begin{matrix}D_5\\ A_5\end{matrix}\quad (2A_1,{Spin(7)}_4)$&7&$\tfrac{1}{2}(2,7)$\\
\hline
17&$\begin{matrix}D_5\\ D_5(a_1)\end{matrix}\quad (A_1,{SU(5)}_2)$&5&$1(5)$\\
\hline
18&$\begin{matrix}D_5\\ A_4+A_1\end{matrix}\quad (0,{Spin(10)}_8)$&16&$1(16)$\\
\hline
19&$\begin{matrix}D_5\\ D_4\end{matrix}\quad (A_1,{SU(6)}_6)$&18&$3(6)$\\
\hline
\end{longtable}

}

\subsection{Interacting fixtures with one irregular puncture}\label{interacting_fixtures_with_one_irregular_puncture}

In the tables below, $n_d$ is the number of Coulomb branch parameters of degree $d$. The total Coulomb branch dimension is $\sum_d n_d$ and the effective number of vector multiplets is $n_v = \sum_d (2d-1) n_d$.

\bigskip
\begin{longtable}{|l|c|c|l|}
\hline
\endhead
Fixture&$\scriptsize(n_2,n_3,n_4,n_5,n_6,n_8,n_9,n_{12})$&$(n_h,n_v)$&Theory\\
\hline 
$\begin{matrix}E_6(a_1)\\ 2A_2\end{matrix}\quad (0,{(F_4)}_{12})$&$(0,0,0,0,1,0,0,0)$&$(40,11)$&${(E_8)}_{12}$ SCFT\\
\hline
$\begin{matrix}D_5\\ A_4\end{matrix}\quad (0,{Spin(10)}_8)$&$(0,0,1,0,0,0,0,0)$&$(24,7)$&${(E_7)}_{8}$ SCFT\\
\hline
$\begin{matrix}E_6(a_3)\\ E_6(a_3)\end{matrix}\quad (0,{(F_4)}_{12})$&$(0,2,0,0,0,0,0,0)$&$(32,10)$&${\left[{(E_6)}_{6}\, \text{SCFT}\right]}^2$\\
\hline
$\begin{matrix}E_6(a_3)\\ A_5\end{matrix}\quad (0,{(F_4)}_{12})$&$(0,1,0,0,1,0,0,0)$&$(39,16)$&${(E_6)}_{12}\times {SU(2)}_7$ SCFT\\
\hline
$\begin{matrix}A_5\\ A_5\end{matrix}\quad (0,{(F_4)}_{12})$&$(0,0,0,0,2,0,0,0)$&$(46,22)$&${(F_4)}_{12}\times {SU(2)}_7^2$ SCFT\\
\hline
\end{longtable}

\bigskip
The ${(E_6)}_{12}\times {SU(2)}_7$ and ${(F_4)}_{12}\times {SU(2)}_7^2$ first appeared in \cite{Chacaltana:2011ze}, as fixtures in the untwisted $D_4$ theory.

\subsection{Interacting fixtures with enhanced global symmetry}\label{interacting_fixtures_with_enhanced_global_symmetry}

{\footnotesize
\renewcommand{\arraystretch}{1.25}

\begin{longtable}{|c|l|c|c|c|}
\hline
\#&Fixture&{ $\scriptsize(n_2,n_3,n_4,n_5,n_6,n_8,n_9,n_{12})$}&$(n_h,n_v)$&$G_k$\\
\hline
\endhead
1&$\begin{matrix}E_6(a_1)\\A_2\end{matrix}\quad 0$&$(0, 0, 0, 0, 2, 0, 0, 0)$&$(80, 22)$&$[(E_8)_{12} \text{ SCFT}]^2$\\
\hline\htarget{IntFixture2}
2&$\begin{matrix}E_6(a_1)\\3A_1\end{matrix}\quad 0$&$(0, 0, 0, 0, 1, 0, 0, 1)$&$(93, 34)$&$(E_8)_{24} \times SU(2)_{13}$\\
\hline
3&$\begin{matrix}E_6(a_1)\\2A_1\end{matrix}\quad 0$&$(0, 0, 0, 0, 1, 1, 0, 1)$&$(112, 49)$&$(E_7)_{24} \times Spin(7)_{16}$\\
\hline
4&$\begin{matrix}E_6(a_1)\\A_1\end{matrix}\quad A_1$&$(0, 0, 0, 0, 1, 1, 1, 0)$&$(100, 43)$&$SU(12)_{18}$\\
\hline
5&$\begin{matrix}D_5\\D_4(a_1)\end{matrix}\quad 0$&$(0, 0, 3, 0, 0, 0, 0, 0)$&$(72, 21)$&$[(E_7)_8 \text{ SCFT}]^3$\\
\hline\htarget{IntFixture6}
6&$\begin{matrix}D_5\\A_3+A_1\end{matrix}\quad 0$&$(0, 0, 2, 0, 0, 1, 0, 0)$&$(81, 29)$&${\begin{gathered} [{(E_7)}_8\,\text{SCFT}]\\ \times [{(E_7)}_{16} \times {SU(2)}_9\,\text{SCFT}]\end{gathered}}$\\
\hline\htarget{IntFixture7}
7&$\begin{matrix}D_5\\2A_2+A_1\end{matrix}\quad 0$&$(0, 0, 1, 0, 0, 1, 0, 1)$&$(98, 45)$&$(E_7)_{24} \times SU(2)_{26}$\\
\hline
8&$\begin{matrix}D_5\\A_3\end{matrix}\quad 0$&$(0, 0, 2, 1, 0, 1, 0, 0)$&$(92, 38)$&${\begin{gathered} [{(E_7)}_8\,\text{SCFT}]\\ \times [(E_6)_{16} \times Sp(2)_{10} \times U(1)\,\text{SCFT}]\end{gathered}}$\\
\hline
9&$\begin{matrix}D_5\\2A_2\end{matrix}\quad 0$&$(0, 0, 1, 0, 1, 1, 0, 1)$&$(112, 56)$&$(E_7)_{24} \times (G_2)_{12}$\\
\hline
10&$\begin{matrix}D_5\\A_2+2A_1\end{matrix}\quad A_1$&$(0, 0, 1, 1, 0, 1, 1, 0)$&$(92, 48)$&$SU(8)_{18} \times SU(2)_{36} \times U(1)$\\
\hline
11&$\begin{matrix}D_5\\A_2+A_1\end{matrix}\quad A_1$&$(0, 0, 1, 1, 1, 1, 1, 0)$&$(105, 59)$&$SU(7)_{18} \times SU(3)_{12} \times U(1)^2$\\
\hline
12&$\begin{matrix}D_5\\A_2\end{matrix}\quad 2A_1$&$(0, 0, 1, 1, 2, 1, 0, 0)$&$(96, 53)$&$Spin(8)_{16} \times {SU(4)}_{12}^2 \times U(1)$\\
\hline
13&$\begin{matrix}D_5\\A_2\end{matrix}\quad A_1$&$(0, 0, 1, 1, 2, 1, 1, 0)$&$(118, 70)$&$SU(6)_{18} \times {SU(3)}_{12}^2 \times U(1)^2$\\
\hline
14&$\begin{matrix}D_5\\3A_1\end{matrix}\quad 3A_1$&$(0, 0, 1, 1, 1, 0, 0, 1)$&$(90, 50)$&$SU(6)_{24} \times Sp(2)_{13}$\\
\hline
15&$\begin{matrix}D_5\\3A_1\end{matrix}\quad 2A_1$&$(0, 0, 1, 1, 1, 1, 0, 1)$&$(109, 65)$&$Spin(7)_{16} \times SU(4)_{24} \times SU(2)_{13} \times U(1)$\\
\hline
16&$\begin{matrix}D_5\\2A_1\end{matrix}\quad 2A_1$&$(0, 0, 1, 1, 1, 2, 0, 1)$&$(128, 80)$&${Spin(7)}_{16}^2 \times SU(2)_{24} \times U(1)^2$\\
\hline
17&$\begin{matrix}E_6(a_3)\\A_4+A_1\end{matrix}\quad 0$&$(0, 1, 0, 0, 1, 1, 0, 1)$&$(104, 54)$&$(E_7)_{24}$\\
\hline
18&$\begin{matrix}E_6(a_3)\\D_4\end{matrix}\quad 0$&$(0, 2, 0, 0, 1, 0, 0, 0)$&$(72, 21)$&$[(E_8)_{12} \text{ SCFT}] \times [(E_6)_6 \text{ SCFT}]^2$\\
\hline
19&$\begin{matrix}E_6(a_3)\\A_4\end{matrix}\quad 0$&$(0, 1, 1, 0, 1, 1, 0, 1)$&$(112, 61)$&$(E_7)_{24} \times SU(2)_8$\\
\hline
20&$\begin{matrix}E_6(a_3)\\D_4(a_1)\end{matrix}\quad A_2$&$(0, 1, 2, 0, 2, 0, 0, 0)$&$(72, 41)$&${Spin(8)}_{12}^2 \times U(1)^2$\\
\hline
21&$\begin{matrix}E_6(a_3)\\D_4(a_1)\end{matrix}\quad 3A_1$&$(0, 1, 2, 0, 1, 0, 0, 1)$&$(85, 53)$&$Spin(8)_{24} \times SU(2)_{13}$\\
\hline
22&$\begin{matrix}E_6(a_3)\\D_4(a_1)\end{matrix}\quad 2A_1$&$(0, 1, 2, 0, 1, 1, 0, 1)$&$(104, 68)$&$Spin(7)_{16} \times {SU(2)}_{24}^3$\\
\hline
23&$\begin{matrix}E_6(a_3)\\A_3+A_1\end{matrix}\quad A_2$&$(0, 1, 1, 0, 2, 1, 0, 0)$&$(81, 49)$&${Spin(7)}_{12}^2 \times SU(2)_9 \times U(1)$\\
\hline
24&$\begin{matrix}E_6(a_3)\\A_3+A_1\end{matrix}\quad 3A_1$&$(0, 1, 1, 0, 1, 1, 0, 1)$&$(94, 61)$&$Spin(7)_{24} \times SU(2)_{13} \times SU(2)_9$\\
\hline
25&$\begin{matrix}E_6(a_3)\\A_3+A_1\end{matrix}\quad 2A_1$&$(0, 1, 1, 0, 1, 2, 0, 1)$&$(113, 76)$&$Spin(7)_{16} \times SU(2)_{48}\times SU(2)_{24}\times SU(2)_9$\\
\hline
26&$\begin{matrix}E_6(a_3)\\2A_2+A_1\end{matrix}\quad A_2$&$(0, 1, 0, 0, 2, 1, 0, 1)$&$(98, 65)$&${(G_2)}_{12}^2 \times SU(2)_{26}$\\
\hline
27&$\begin{matrix}E_6(a_3)\\2A_2+A_1\end{matrix}\quad 3A_1$&$(0, 1, 0, 0, 1, 1, 0, 2)$&$(111, 77)$&$(G_2)_{24} \times SU(2)_{26} \times SU(2)_{13}$\\
\hline
28&$\begin{matrix}E_6(a_3)\\2A_2+A_1\end{matrix}\quad 2A_1$&$(0, 1, 0, 0, 1, 2, 0, 2)$&$(130, 92)$&$Spin(7)_{16} \times SU(2)_{26} \times SU(2)_{72}$\\
\hline
29&$\begin{matrix}E_6(a_3)\\A_3\end{matrix}\quad A_2$&$(0, 1, 1, 1, 2, 1, 0, 0)$&$(92, 58)$&${SU(4)}_{12}^2 \times Sp(2)_{10} \times U(1)$\\
\hline
30&$\begin{matrix}E_6(a_3)\\A_3\end{matrix}\quad 3A_1$&$(0, 1, 1, 1, 1, 1, 0, 1)$&$(105, 70)$&$SU(4)_{24} \times Sp(2)_{10} \times SU(2)_{13} $\\
\hline
31&$\begin{matrix}E_6(a_3)\\A_3\end{matrix}\quad 2A_1$&$(0, 1, 1, 1, 1, 2, 0, 1)$&$(124, 85)$&$Spin(7)_{16} \times Sp(2)_{10} \times SU(2)_{24} \times U(1)$\\
\hline
32&$\begin{matrix}E_6(a_3)\\A_2+2A_1\end{matrix}\quad A_2+2A_1$&$(0, 1, 0, 1, 0, 1, 1, 1)$&$(100, 69)$&$SU(4)_{54} \times U(1)$\\
\hline
33&$\begin{matrix}E_6(a_3)\\A_2+2A_1\end{matrix}\quad A_2+A_1$&$(0, 1, 0, 1, 1, 1, 1, 1)$&$(113, 80)$&$SU(3)_{54} \times SU(3)_{12} \times U(1)$\\
\hline
34&$\begin{matrix}E_6(a_3)\\2A_2\end{matrix}\quad A_2$&$(0, 1, 0, 0, 3, 1, 0, 1)$&$(112, 76)$&${(G_2)}_{12}^3$\\
\hline
35&$\begin{matrix}E_6(a_3)\\2A_2\end{matrix}\quad 3A_1$&$(0, 1, 0, 0, 2, 1, 0, 2)$&$(125, 88)$&$(G_2)_{24} \times (G_2)_{12} \times SU(2)_{13}$\\
\hline
36&$\begin{matrix}E_6(a_3)\\2A_2\end{matrix}\quad 2A_1$&$(0, 1, 0, 0, 2, 2, 0, 2)$&$(144, 103)$&$Spin(7)_{16} \times (G_2)_{12} \times SU(2)_{72}$\\
\hline
37&$\begin{matrix}E_6(a_3)\\A_2+A_1\end{matrix}\quad A_2+A_1$&$(0, 1, 0, 1, 2, 1, 1, 1)$&$(126, 91)$&${SU(3)}_{12}^2 \times SU(2)_{24} \times U(1)$\\
\hline
38&$\begin{matrix}A_5\\A_4+A_1\end{matrix}\quad 0$&$(0, 0, 0, 0, 2, 1, 0, 1)$&$(111, 60)$&$(E_7)_{24} \times SU(2)_7$\\
\hline\htarget{IntFixture39}
39&$\begin{matrix}A_5\\D_4\end{matrix}\quad 0$&$(0, 1, 0, 0, 2, 0, 0, 0)$&$(79, 27)$&$[(E_8)_{12} \text{ SCFT}] \times [(E_6)_{12} \times SU(2)_7 \text{ SCFT}]$\\
\hline
40&$\begin{matrix}A_5\\A_4\end{matrix}\quad 0$&$(0, 0, 1, 0, 2, 1, 0, 1)$&$(119, 67)$&$(E_7)_{24} \times SU(2)_8 \times SU(2)_7$\\
\hline
41&$\begin{matrix}A_5\\D_4(a_1)\end{matrix}\quad A_2$&$(0, 0, 2, 0, 3, 0, 0, 0)$&$(79, 47)$&${Spin(8)}_{12}^2 \times SU(2)_7$\\
\hline
42&$\begin{matrix}A_5\\D_4(a_1)\end{matrix}\quad 3A_1$&$(0, 0, 2, 0, 2, 0, 0, 1)$&$(92, 59)$&$Spin(8)_{24} \times SU(2)_{13} \times SU(2)_7$\\
\hline
43&$\begin{matrix}A_5\\D_4(a_1)\end{matrix}\quad 2A_1$&$(0, 0, 2, 0, 2, 1, 0, 1)$&$(111, 74)$&$Spin(7)_{16}\times {SU(2)}_{24}^3 \times SU(2)_7$\\
\hline
44&$\begin{matrix}A_5\\A_3+A_1\end{matrix}\quad A_2$&$(0, 0, 1, 0, 3, 1, 0, 0)$&$(88, 55)$&${Spin(7)}_{12}^2 \times SU(2)_9 \times SU(2)_7$\\
\hline
45&$\begin{matrix}A_5\\A_3+A_1\end{matrix}\quad 3A_1$&$(0, 0, 1, 0, 2, 1, 0, 1)$&$(101, 67)$&$Spin(7)_{24} \times SU(2)_{13} \times SU(2)_9 \times SU(2)_7$\\
\hline
46&$\begin{matrix}A_5\\A_3+A_1\end{matrix}\quad 2A_1$&$(0, 0, 1, 0, 2, 2, 0, 1)$&$(120, 82)$&${\begin{aligned}Spin(7)_{16}&\times SU(2)_{48} \times SU(2)_{24}\\&\times SU(2)_9 \times SU(2)_7\end{aligned}}$\\
\hline
47&$\begin{matrix}A_5\\2A_2+A_1\end{matrix}\quad A_2$&$(0, 0, 0, 0, 3, 1, 0, 1)$&$(105, 71)$&${(G_2)}_{12}^2 \times SU(2)_{26} \times SU(2)_7$\\
\hline
48&$\begin{matrix}A_5\\2A_2+A_1\end{matrix}\quad 3A_1$&$(0, 0, 0, 0, 2, 1, 0, 2)$&$(118, 83)$&$(G_2)_{24} \times SU(2)_{26} \times SU(2)_{13} \times SU(2)_7$\\
\hline
49&$\begin{matrix}A_5\\2A_2+A_1\end{matrix}\quad 2A_1$&$(0, 0, 0, 0, 2, 2, 0, 2)$&$(137, 98)$&$Spin(7)_{16}\times SU(2)_{72} \times SU(2)_{26} \times SU(2)_7 $\\
\hline
50&$\begin{matrix}A_5\\A_3\end{matrix}\quad A_2$&$(0, 0, 1, 1, 3, 1, 0, 0)$&$(99, 64)$&${SU(4)}_{12}^2 \times Sp(2)_{10} \times SU(2)_7$\\
\hline
51&$\begin{matrix}A_5\\A_3\end{matrix}\quad 3A_1$&$(0, 0, 1, 1, 2, 1, 0, 1)$&$(112, 76)$&$SU(4)_{24} \times Sp(2)_{10} \times SU(2)_{13} \times SU(2)_7$\\
\hline
52&$\begin{matrix}A_5\\A_3\end{matrix}\quad 2A_1$&$(0, 0, 1, 1, 2, 2, 0, 1)$&$(131, 91)$&${\begin{aligned}Spin(7)_{16}&\times Sp(2)_{10} \times SU(2)_{24}\\& \times SU(2)_7 \times U(1)\end{aligned}}$\\
\hline
53&$\begin{matrix}A_5\\A_2+2A_1\end{matrix}\quad A_2+2A_1$&$(0, 0, 0, 1, 1, 1, 1, 1)$&$(107, 75)$&$SU(4)_{54} \times SU(2)_7 \times U(1)$\\
\hline
54&$\begin{matrix}A_5\\A_2+2A_1\end{matrix}\quad A_2+A_1$&$(0, 0, 0, 1, 2, 1, 1, 1)$&$(120, 86)$&$SU(3)_{54} \times SU(3)_{12} \times SU(2)_7 \times U(1)$\\
\hline
55&$\begin{matrix}A_5\\2A_2\end{matrix}\quad A_2$&$(0, 0, 0, 0, 4, 1, 0, 1)$&$(119, 82)$&${(G_2)}_{12}^3 \times SU(2)_7$\\
\hline
56&$\begin{matrix}A_5\\2A_2\end{matrix}\quad 3A_1$&$(0, 0, 0, 0, 3, 1, 0, 2)$&$(132, 94)$&$(G_2)_{24} \times (G_2)_{12} \times SU(2)_{13} \times SU(2)_7$\\
\hline
57&$\begin{matrix}A_5\\2A_2\end{matrix}\quad 2A_1$&$(0, 0, 0, 0, 3, 2, 0, 2)$&$(151, 109)$&$Spin(7)_{16} \times (G_2)_{12} \times SU(2)_{72} \times SU(2)_7$\\
\hline
58&$\begin{matrix}A_5\\A_2+A_1\end{matrix}\quad A_2+A_1$&$(0, 0, 0, 1, 3, 1, 1, 1)$&$(133, 97)$&${SU(3)}_{12}^2\times SU(2)_{24} \times SU(2)_7  \times U(1)$\\
\hline
59&$\begin{matrix}D_5(a_1)\\D_5(a_1)\end{matrix}\quad 0$&$(0, 2, 0, 1, 0, 0, 1, 0)$&$(86, 36)$&${(E_7)}_{18}\times {(E_6)}_6 \times U(1)$\\
\hline
60&$\begin{matrix}D_5(a_1)\\A_4+A_1\end{matrix}\quad A_1$&$(0, 1, 0, 1, 1, 1, 1, 0)$&$(97, 57)$&$SU(7)_{18} \times U(1)^2$\\
\hline
61&$\begin{matrix}D_5(a_1)\\D_4\end{matrix}\quad 0$&$(0, 2, 0, 1, 1, 0, 1, 0)$&$(99, 47)$&${(E_6)}_{18} \times {(E_6)}_{6} \times SU(3)_{12} \times U(1)$\\
\hline
62&$\begin{matrix}D_5(a_1)\\A_4\end{matrix}\quad A_1$&$(0, 1, 1, 1, 1, 1, 1, 0)$&$(105, 64)$&$SU(7)_{18} \times SU(2)_8 \times U(1)^2$\\
\hline\htarget{IntFixture63}
63&$\begin{matrix}D_5(a_1)\\D_4(a_1)\end{matrix}\quad A_2+2A_1$&$(0, 1, 2, 1, 0, 0, 1, 0)$&$(73, 45)$&$SU(3)_{54-k-k'} \times SU(3)_{k} \times SU(3)_{k'} \times U(1)$\\
\hline
64&$\begin{matrix}D_5(a_1)\\D_4(a_1)\end{matrix}\quad A_2+A_1$&$(0, 1, 2, 1, 1, 0, 1, 0)$&$(86, 56)$&$SU(3)_{12} \times {SU(2)}_{18}^3 \times U(1)^3$\\
\hline
65&$\begin{matrix}D_5(a_1)\\D_4(a_1)\end{matrix}\quad A_2$&$(0, 1, 2, 1, 2, 0, 1, 0)$&$(99, 67)$&${SU(3)}_{12}^2 \times U(1)^5$\\
\hline\htarget{IntFixture66}
66&$\begin{matrix}D_5(a_1)\\A_3+A_1\end{matrix}\quad A_2+2A_1$&$(0, 1, 1, 1, 0, 1, 1, 0)$&$(82, 53)$&$SU(3)_{54-k} \times SU(3)_k \times SU(2)_9 \times U(1)$\\
\hline
67&$\begin{matrix}D_5(a_1)\\A_3+A_1\end{matrix}\quad A_2+A_1$&$(0, 1, 1, 1, 1, 1, 1, 0)$&$(95, 64)$&${\begin{aligned}SU(3)_{12} &\times SU(2)_{36} \times SU(2)_{18}\\& \times SU(2)_9 \times U(1)^2\end{aligned}}$\\
\hline
68&$\begin{matrix}D_5(a_1)\\A_3+A_1\end{matrix}\quad A_2$&$(0, 1, 1, 1, 2, 1, 1, 0)$&$(108, 75)$&${SU(3)}_{12}^2 \times SU(2)_9 \times U(1)^3$\\
\hline
69&$\begin{matrix}D_5(a_1)\\2A_2+A_1\end{matrix}\quad A_2+2A_1$&$(0, 1, 0, 1, 0, 1, 1, 1)$&$(99, 69)$&$SU(3)_{54} \times SU(2)_{26} \times U(1)$\\
\hline
70&$\begin{matrix}D_5(a_1)\\2A_2+A_1\end{matrix}\quad A_2+A_1$&$(0, 1, 0, 1, 1, 1, 1, 1)$&$(112, 80)$&$SU(3)_{12} \times SU(2)_{54}\times SU(2)_{26}  \times U(1)$\\
\hline
71&$\begin{matrix}D_5(a_1)\\A_3\end{matrix}\quad A_2+2A_1$&$(0, 1, 1, 2, 0, 1, 1, 0)$&$(93, 62)$&$SU(3)_{18} \times SU(2)_{36} \times Sp(2)_{10} \times U(1)^2$\\
\hline
72&$\begin{matrix}D_5(a_1)\\A_3\end{matrix}\quad A_2+A_1$&$(0, 1, 1, 2, 1, 1, 1, 0)$&$(106, 73)$&$SU(3)_{12} \times Sp(2)_{10} \times SU(2)_{18} \times U(1)^3$\\
\hline
73&$\begin{matrix}D_5(a_1)\\A_3\end{matrix}\quad A_2$&$(0, 1, 1, 2, 2, 1, 1, 0)$&$(119, 84)$&${SU(3)}_{12}^2 \times Sp(2)_{10} \times U(1)^3$\\
\hline
74&$\begin{matrix}D_5(a_1)\\A_2+2A_1\end{matrix}\quad 2A_2$&$(0, 1, 0, 1, 1, 1, 1, 1)$&$(113, 80)$&$(G_2)_{12} \times SU(3)_{54} \times U(1)$\\
\hline
75&$\begin{matrix}D_5(a_1)\\2A_2\end{matrix}\quad A_2+A_1$&$(0, 1, 0, 1, 2, 1, 1, 1)$&$(126, 91)$&$(G_2)_{12} \times SU(3)_{12} \times SU(2)_{54} \times U(1)$\\
\hline
76&$\begin{matrix}A_4+A_1\\A_4+A_1\end{matrix}\quad A_2$&$(0, 0, 0, 1, 3, 1, 0, 0)$&$(88, 57)$&${SU(4)}_{12}^2 \times U(1)$\\
\hline
77&$\begin{matrix}A_4+A_1\\A_4+A_1\end{matrix}\quad 3A_1$&$(0, 0, 0, 1, 2, 1, 0, 1)$&$(101, 69)$&$SU(4)_{24} \times SU(2)_{13} \times U(1)$\\
\hline
78&$\begin{matrix}A_4+A_1\\A_4+A_1\end{matrix}\quad 2A_1$&$(0, 0, 0, 1, 2, 2, 0, 1)$&$(120, 84)$&$Spin(7)_{16} \times SU(2)_{24} \times U(1)^2$\\
\hline
79&$\begin{matrix}A_4+A_1\\D_4\end{matrix}\quad 2A_1$&$(0, 1, 0, 1, 2, 1, 0, 0)$&$(88, 51)$&$Spin(8)_{16} \times SU(4)_{12} \times U(1)^2$\\
\hline
80&$\begin{matrix}A_4+A_1\\D_4\end{matrix}\quad A_1$&$(0, 1, 0, 1, 2, 1, 1, 0)$&$(110, 68)$&$SU(6)_{18} \times SU(3)_{12} \times U(1)^2$\\
\hline
81&$\begin{matrix}A_4+A_1\\A_4\end{matrix}\quad A_2$&$(0, 0, 1, 1, 3, 1, 0, 0)$&$(96, 64)$&${SU(4)}_{12}^2 \times SU(2)_8 \times U(1)$\\
\hline
82&$\begin{matrix}A_4+A_1\\A_4\end{matrix}\quad 3A_1$&$(0, 0, 1, 1, 2, 1, 0, 1)$&$(109, 76)$&$SU(4)_{24} \times SU(2)_{13} \times SU(2)_8 \times U(1)$\\
\hline
83&$\begin{matrix}A_4+A_1\\A_4\end{matrix}\quad 2A_1$&$(0, 0, 1, 1, 2, 2, 0, 1)$&$(128, 91)$&$Spin(7)_{16} \times SU(2)_8 \times SU(2)_{24} \times U(1)^2$\\
\hline
84&$\begin{matrix}A_4+A_1\\D_4(a_1)\end{matrix}\quad D_4(a_1)$&$(0, 0, 4, 0, 1, 0, 0, 0)$&$(64, 39)$&${SU(2)}_8^9$\\
\hline
85&$\begin{matrix}A_4+A_1\\D_4(a_1)\end{matrix}\quad A_3+A_1$&$(0, 0, 3, 0, 1, 1, 0, 0)$&$(73, 47)$&${SU(2)}_{16}^3 \times {SU(2)}_9 \times {SU(2)}_8^3$\\
\hline
86&$\begin{matrix}A_4+A_1\\D_4(a_1)\end{matrix}\quad 2A_2+A_1$&$(0, 0, 2, 0, 1, 1, 0, 1)$&$(90, 63)$&${SU(2)}_{26}\times {SU(2)}_{24}^3$\\
\hline
87&$\begin{matrix}A_4+A_1\\D_4(a_1)\end{matrix}\quad A_3$&$(0, 0, 3, 1, 1, 1, 0, 0)$&$(84, 56)$&$Sp(2)_{10} \times {SU(2)}_8^3 \times U(1)^3$\\
\hline
88&$\begin{matrix}A_4+A_1\\D_4(a_1)\end{matrix}\quad 2A_2$&$(0, 0, 2, 0, 2, 1, 0, 1)$&$(104, 74)$&$(G_2)_{12} \times SU(2)_{24}^3$\\
\hline
89&$\begin{matrix}A_4+A_1\\A_3+A_1\end{matrix}\quad A_3+A_1$&$(0, 0, 2, 0, 1, 2, 0, 0)$&$(82, 55)$&${SU(2)}_{32}\times {SU(2)}_{16}^2 \times {SU(2)}_9^2 \times {SU(2)}_8^2$\\
\hline
90&$\begin{matrix}A_4+A_1\\A_3+A_1\end{matrix}\quad 2A_2+A_1$&$(0, 0, 1, 0, 1, 2, 0, 1)$&$(99, 71)$&${SU(2)}_{48} \times {SU(2)}_{26}\times {SU(2)}_{24} \times {SU(2)}_9$\\
\hline
91&$\begin{matrix}A_4+A_1\\A_3+A_1\end{matrix}\quad A_3$&$(0, 0, 2, 1, 1, 2, 0, 0)$&$(93, 64)$&${\begin{aligned}{Sp(2)}_{10}&\times {SU(2)}_{16} \times SU(2)_9\\&  \times {SU(2)}_8\times U(1)^2\end{aligned}}$\\
\hline
92&$\begin{matrix}A_4+A_1\\A_3+A_1\end{matrix}\quad 2A_2$&$(0, 0, 1, 0, 2, 2, 0, 1)$&$(113, 82)$&$(G_2)_{12}\times SU(2)_{48} \times SU(2)_{24}\times SU(2)_9  $\\
\hline
93&$\begin{matrix}A_4+A_1\\2A_2+A_1\end{matrix}\quad 2A_2+A_1$&$(0, 0, 0, 0, 1, 2, 0, 2)$&$(116, 87)$&${SU(2)}_{72}\times {SU(2)}_{26}^2$\\
\hline
94&$\begin{matrix}A_4+A_1\\2A_2+A_1\end{matrix}\quad A_3$&$(0, 0, 1, 1, 1, 2, 0, 1)$&$(110, 80)$&${Sp(2)}_{10} \times {SU(2)}_{26} \times  {SU(2)}_{24} \times U(1)$\\
\hline
95&$\begin{matrix}A_4+A_1\\2A_2+A_1\end{matrix}\quad 2A_2$&$(0, 0, 0, 0, 2, 2, 0, 2)$&$(130, 98)$&$(G_2)_{12}\times SU(2)_{72} \times SU(2)_{26} $\\
\hline
96&$\begin{matrix}A_4+A_1\\A_3\end{matrix}\quad A_3$&$(0, 0, 2, 2, 1, 2, 0, 0)$&$(104, 73)$&${Sp(2)}_{10}^2 \times SU(2)_8 \times U(1)^3$\\
\hline
97&$\begin{matrix}A_4+A_1\\A_3\end{matrix}\quad 2A_2$&$(0, 0, 1, 1, 2, 2, 0, 1)$&$(124, 91)$&$(G_2)_{12} \times Sp(2)_{10} \times SU(2)_{24} \times U(1)$\\
\hline
98&$\begin{matrix}A_4+A_1\\2A_2\end{matrix}\quad 2A_2$&$(0, 0, 0, 0, 3, 2, 0, 2)$&$(144, 109)$&${(G_2)}_{12}^2 \times SU(2)_{72}$\\
\hline
99&$\begin{matrix}D_4\\D_4\end{matrix}\quad 0$&$(0, 2, 0, 1, 2, 0, 1, 0)$&$(112, 58)$&${(E_6)}_{18} \times {(E_6)}_{6} \times {SU(3)}_{12}^2$\\
\hline
100&$\begin{matrix}D_4\\A_4\end{matrix}\quad 2A_1$&$(0, 1, 1, 1, 2, 1, 0, 0)$&$(96, 58)$&$Spin(8)_{16} \times SU(4)_{12} \times SU(2)_8 \times U(1)^2$\\
\hline
101&$\begin{matrix}D_4\\A_4\end{matrix}\quad A_1$&$(0, 1, 1, 1, 2, 1, 1, 0)$&$(118, 75)$&$SU(6)_{18} \times SU(3)_{12} \times SU(2)_8 \times U(1)^2$\\
\hline
102&$\begin{matrix}D_4\\D_4(a_1)\end{matrix}\quad A_2+2A_1$&$(0, 1, 2, 1, 1, 0, 1, 0)$&$(86, 56)$&$SU(3)_{12} \times {SU(2)}_{18}^3 \times  U(1)^3$\\
\hline
103&$\begin{matrix}D_4\\D_4(a_1)\end{matrix}\quad 2A_2$&$(0, 1, 2, 0, 2, 0, 0, 0)$&$(72, 41)$&${Spin(8)}_{12}^2 \times U(1)^2$\\
\hline
104&$\begin{matrix}D_4\\D_4(a_1)\end{matrix}\quad A_2+A_1$&$(0, 1, 2, 1, 2, 0, 1, 0)$&$(99, 67)$&${SU(3)}_{12}^2 \times U(1)^5$\\
\hline
105&$\begin{matrix}D_4\\D_4(a_1)\end{matrix}\quad A_2$&$(0, 1, 2, 1, 3, 0, 1, 0)$&$(112, 78)$&${SU(3)}_{12}^3 \times U(1)^4$\\
\hline
106&$\begin{matrix}D_4\\A_3+A_1\end{matrix}\quad A_2+2A_1$&$(0, 1, 1, 1, 1, 1, 1, 0)$&$(95, 64)$&${\begin{aligned}SU(3)_{12}&\times SU(2)_{36} \times SU(2)_{18}\\& \times SU(2)_9 \times U(1)^2\end{aligned}}$\\
\hline
107&$\begin{matrix}D_4\\A_3+A_1\end{matrix}\quad 2A_2$&$(0, 1, 1, 0, 2, 1, 0, 0)$&$(81, 49)$&${Spin(7)}_{12}^2 \times SU(2)_9 \times U(1)$\\
\hline
108&$\begin{matrix}D_4\\A_3+A_1\end{matrix}\quad A_2+A_1$&$(0, 1, 1, 1, 2, 1, 1, 0)$&$(108, 75)$&${SU(3)}_{12}^2 \times SU(2)_9 \times U(1)^3$\\
\hline
109&$\begin{matrix}D_4\\A_3+A_1\end{matrix}\quad A_2$&$(0, 1, 1, 1, 3, 1, 1, 0)$&$(121, 86)$&${SU(3)}_{12}^3 \times SU(2)_9 \times U(1)^2$\\
\hline
110&$\begin{matrix}D_4\\2A_2+A_1\end{matrix}\quad 2A_2+A_1$&$(0, 1, 0, 0, 1, 1, 0, 1)$&$(84, 54)$&${(G_2)}_{12}\times {Sp(2)}_{26}$\\
\hline
111&$\begin{matrix}D_4\\2A_2+A_1\end{matrix}\quad 2A_2$&$(0, 1, 0, 0, 2, 1, 0, 1)$&$(98, 65)$&${(G_2)}_{12}^2 \times SU(2)_{26}$\\
\hline
112&$\begin{matrix}D_4\\A_3\end{matrix}\quad A_2+2A_1$&$(0, 1, 1, 2, 1, 1, 1, 0)$&$(106, 73)$&${\begin{aligned} Sp(2)_{10}&\times SU(3)_{12}\times SU(2)_{36} \\& \times SU(2)_{18} \times U(1)^2\end{aligned}}$\\
\hline
113&$\begin{matrix}D_4\\A_3\end{matrix}\quad 2A_2$&$(0, 1, 1, 1, 2, 1, 0, 0)$&$(92, 58)$&$Spin(7)_{12} \times SU(4)_{12} \times Sp(2)_{10} \times U(1)$\\
\hline
114&$\begin{matrix}D_4\\A_3\end{matrix}\quad A_2+A_1$&$(0, 1, 1, 2, 2, 1, 1, 0)$&$(119, 84)$&${SU(3)}_{12}^2 \times Sp(2)_{10} \times U(1)^3$\\
\hline
115&$\begin{matrix}D_4\\A_3\end{matrix}\quad A_2$&$(0, 1, 1, 2, 3, 1, 1, 0)$&$(132, 95)$&${SU(3)}_{12}^3 \times Sp(2)_{10} \times U(1)^2$\\
\hline
116&$\begin{matrix}D_4\\2A_2\end{matrix}\quad 2A_2$&$(0, 1, 0, 0, 3, 1, 0, 1)$&$(112, 76)$&${(G_2)}_{12}^3$\\
\hline
117&$\begin{matrix}A_4\\A_4\end{matrix}\quad A_2$&$(0, 0, 2, 1, 3, 1, 0, 0)$&$(104, 71)$&${SU(4)}_{12}^2 \times {SU(2)}_8^2 \times U(1)$\\
\hline
118&$\begin{matrix}A_4\\A_4\end{matrix}\quad 3A_1$&$(0, 0, 2, 1, 2, 1, 0, 1)$&$(117, 83)$&$SU(4)_{24} \times SU(2)_{13} \times {SU(2)}_8^2 \times U(1)$\\
\hline
119&$\begin{matrix}A_4\\A_4\end{matrix}\quad 2A_1$&$(0, 0, 2, 1, 2, 2, 0, 1)$&$(136, 98)$&$Spin(7)_{16} \times {SU(2)}_8^2 \times SU(2)_{24} \times U(1)^2$\\
\hline
120&$\begin{matrix}A_4\\D_4(a_1)\end{matrix}\quad D_4(a_1)$&$(0, 0, 5, 0, 1, 0, 0, 0)$&$(72, 46)$&${SU(2)}_8^{10}$\\
\hline
121&$\begin{matrix}A_4\\D_4(a_1)\end{matrix}\quad A_3+A_1$&$(0, 0, 4, 0, 1, 1, 0, 0)$&$(81, 54)$&${SU(2)}_{16}^3 \times {SU(2)}_9 \times {SU(2)}_8^4$\\
\hline
122&$\begin{matrix}A_4\\D_4(a_1)\end{matrix}\quad 2A_2+A_1$&$(0, 0, 3, 0, 1, 1, 0, 1)$&$(98, 70)$&${SU(2)}_{26}\times {SU(2)}_{24}^3 \times {SU(2)}_8$\\
\hline
123&$\begin{matrix}A_4\\D_4(a_1)\end{matrix}\quad A_3$&$(0, 0, 4, 1, 1, 1, 0, 0)$&$(92, 63)$&$Sp(2)_{10} \times {SU(2)}_8^4 \times U(1)^3$\\
\hline
124&$\begin{matrix}A_4\\D_4(a_1)\end{matrix}\quad 2A_2$&$(0, 0, 3, 0, 2, 1, 0, 1)$&$(112, 81)$&$(G_2)_{12} \times SU(2)_8 \times {SU(2)}_{24}^3$\\
\hline
125&$\begin{matrix}A_4\\A_3+A_1\end{matrix}\quad A_3+A_1$&$(0, 0, 3, 0, 1, 2, 0, 0)$&$(90, 62)$&${SU(2)}_{32}\times {SU(2)}_{16}^2 \times {SU(2)}_9^2 \times {SU(2)}_8^2$\\
\hline
126&$\begin{matrix}A_4\\A_3+A_1\end{matrix}\quad 2A_2+A_1$&$(0, 0, 2, 0, 1, 2, 0, 1)$&$(107, 78)$&${\begin{aligned}{SU(2)}_{48} &\times {SU(2)}_{26}\times {SU(2)}_{24}\\& \times {SU(2)}_9 \times SU(2)_8\end{aligned}}$\\
\hline
127&$\begin{matrix}A_4\\A_3+A_1\end{matrix}\quad A_3$&$(0, 0, 3, 1, 1, 2, 0, 0)$&$(101, 71)$&${\begin{aligned}{Sp(2)}_{10}&\times {SU(2)}_{16} \times SU(2)_9\\&  \times {SU(2)}_8^2\times U(1)^2\end{aligned}}$\\
\hline
128&$\begin{matrix}A_4\\A_3+A_1\end{matrix}\quad 2A_2$&$(0, 0, 2, 0, 2, 2, 0, 1)$&$(121, 89)$&${\begin{aligned}(G_2)_{12}&\times SU(2)_{48} \times SU(2)_{24}\\& \times SU(2)_9 \times SU(2)_8\end{aligned}} $\\
\hline
129&$\begin{matrix}A_4\\2A_2+A_1\end{matrix}\quad 2A_2+A_1$&$(0, 0, 1, 0, 1, 2, 0, 2)$&$(124, 94)$&${SU(2)}_{72}\times {SU(2)}_{26}^2 \times SU(2)_8$\\
\hline
130&$\begin{matrix}A_4\\2A_2+A_1\end{matrix}\quad A_3$&$(0, 0, 2, 1, 1, 2, 0, 1)$&$(118, 87)$&${\begin{aligned}{Sp(2)}_{10}&\times {SU(2)}_{26} \times {SU(2)}_{24}\\& \times {SU(2)}_8 \times U(1)\end{aligned}}$\\
\hline
131&$\begin{matrix}A_4\\2A_2+A_1\end{matrix}\quad 2A_2$&$(0, 0, 1, 0, 2, 2, 0, 2)$&$(138, 105)$&$(G_2)_{12}\times SU(2)_{72} \times SU(2)_{26} \times SU(2)_8 $\\
\hline
132&$\begin{matrix}A_4\\A_3\end{matrix}\quad A_3$&$(0, 0, 3, 2, 1, 2, 0, 0)$&$(112, 80)$&${Sp(2)}_{10}^2 \times {SU(2)}_8^2 \times U(1)^3$\\
\hline
133&$\begin{matrix}A_4\\A_3\end{matrix}\quad 2A_2$&$(0, 0, 2, 1, 2, 2, 0, 1)$&$(132, 98)$&${\begin{aligned}(G_2)_{12}&\times Sp(2)_{10} \times SU(2)_{24}\\& \times SU(2)_8 \times U(1)\end{aligned}}$\\
\hline
134&$\begin{matrix}A_4\\2A_2\end{matrix}\quad 2A_2$&$(0, 0, 1, 0, 3, 2, 0, 2)$&$(152, 116)$&${(G_2)}_{12}^2\times SU(2)_{72} \times SU(2)_8 $\\
\hline
\end{longtable}

}

We were unable to determine the $SU(3)$ levels in fixtures \hyperlink{IntFixture63}{63} and \hyperlink{IntFixture66}{66}.

\subsection{Mixed fixtures}\label{mixed_fixtures}

We find many ``new'' SCFTs in our list of mixed fixtures. For each fixture in the table below, we list the global symmetry group, the graded Coulomb branch dimensions, and the effective number of vector and hypermultiplets of the SCFT. The effective number of hypermultiplets, for the fixture as a whole, is the sum of the $n_h$ listed in the table and the number of free hypermultiplets in the last column. When the hypermultiplets transform under the nonabelian part of the ``manifest'' global symmetry of the fixture, we list that representation. Otherwise, we just give their number.

All SCFTs in the list below are ``new'', except for the $(E_6)_{6}$ SCFT, the $(E_6)_{12} \times SU(2)_7$ SCFT, the $SU(4)_8^3$ SCFT, and the $(E_8)_{12}$ SCFT, which have previously appeared in the classification of the A- and D-series fixtures, and the ${(E_7)}_{16}\times {SU(2)}_9$ and ${(G_2)}_{12}\times {Sp(2)}_{26}$ SCFTs, which appeared above.

{\footnotesize
\renewcommand{\arraystretch}{1.6}

\begin{longtable}{|c|l|c|c|c|}
\hline
\#&Fixture&$\scriptsize(n_2,n_3,n_4,n_5,n_6,n_8,n_9,n_{12})$&$(n_h,n_v)$&Theory\\
\hline 
\endhead
1&$\begin{matrix}E_6(a_1)\\A_2+A_1\end{matrix}\quad 0$&$(0,0,0,0,1,0,0,0)$&$(40,11)$&$(E_8)_{12}\,\text{SCFT} + 1(1,27)$\\
\hline
2&$\begin{matrix}E_6(a_1)\\ 3A_1\end{matrix}\quad A_1$&$(0,0,0,0,1,0,0,0)$&$(40,11)$&$(E_8)_{12}\,\text{SCFT} + \frac{1}{2}(1,2,1)+1(3,1,6)$\\
\hline
3&$\begin{matrix}E_6(a_1)\\ 2A_1\end{matrix}\quad A_1$&$(0,0,0,0,1,1,0,0)$&$(72,26)$&$Spin(20)_{16}\,\text{SCFT} + 1(6,1)$\\
\hline\htarget{MixedFixture4}
4&$\begin{matrix}D_5\\ 2A_2+A_1\end{matrix}\quad A_1$&$(0,0,1,0,0,1,0,0)$&$(57,22)$&$(E_7)_{16} \times SU(2)_9 + \frac{1}{2}(2,1)+1(1,6)$\\
\hline
5&$\begin{matrix}D_5\\A_2+2A_1\end{matrix}\quad 3A_1$&$(0,0,1,1,0,0,0,0)$&$(42,16)$&$SU(8)_{10}\times SU(3)_{12}+1(2;3,1)+\frac{1}{2}(3;1,2)$\\
\hline
6&$\begin{matrix}D_5\\A_2+2A_1\end{matrix}\quad 2A_1$&$(0,0,1,1,0,1,0,0)$&$(68,31)$&$(E_6)_{16} \times Sp(2)_{10} \times U(1) + 1(2,1)$\\
\hline
7&$\begin{matrix}D_5\\ 2A_2\end{matrix}\quad A_1$&$(0,0,1,0,1,1,0,0)$&$(72,33)$&$Spin(7)_{12} \times Spin(12)_{16}\,\text{SCFT} + 1(1,6)$\\
\hline
8&$\begin{matrix}D_5\\A_2+A_1\end{matrix}\quad 3A_1$&$(0,0,1,1,1,0,0,0)$&$(60,27)$&$SU(8)_{12} \times SU(4)_{10} + \frac{1}{2}(1;1,2)+1(1;3,1)$\\
\hline
9&$\begin{matrix}D_5\\A_2+A_1\end{matrix}\quad 2A_1$&$(0,0,1,1,1,1,0,0)$&$(82,42)$&${\begin{gathered}Spin(10)_{16} \times SU(4)_{12} \times SU(2)_{10} \times U(1)\\ + \text{1 free hyper}\end{gathered}}$\\
\hline
10&$\begin{matrix}D_5\\A_2\end{matrix}\quad 3A_1$&$(0,0,1,1,2,0,0,0)$&$(76,38)$&${SU(6)}_{12}^2 \times SU(2)_{12} + \frac{1}{2}(1,1;1,2)$\\
\hline
11&$\begin{matrix}E_6(a_3)\\D_5(a1)\end{matrix}\quad 0$&$(0,2,0,0,0,0,0,0)$&$(32,10)$&${[(E_6)_6)\,\text{SCFT}]}^2 + 1(27)$\\
\hline
12&$\begin{matrix}E_6(a3)\\A_4+A_1\end{matrix}\quad A_1$&$(0,1,0,0,1,1,0,0)$&$(64,31)$&$Spin(13)_{16}\times U(1) + 1(6)$\\
\hline
13&$\begin{matrix}E_6(a3)\\A_4\end{matrix}\quad A_1$&$(0,1,1,0,1,1,0,0)$&$(72,38)$&$Spin(12)_{16} \times SU(2)_8 \times U(1) \,\text{SCFT} + 1(1,6)$\\
\hline
14&$\begin{matrix}E_6(a3)\\D_4(a1)\end{matrix}\quad A_2+2A_1$&$(0,1,2,0,0,0,0,0)$&$(40,19)$&${SU(4)}_{8}^3\, \text{SCFT} + 3(2)$\\
\hline
15&$\begin{matrix}E_6(a3)\\D_4(a1)\end{matrix}\quad A_2+A_1$&$(0,1,2,0,1,0,0,0)$&$(56,30)$&$Spin(8)_{12} \times {SU(2)}_8^3 \times U(1)^2 + \text{3 free hypers}$\\
\hline
16&$\begin{matrix}E_6(a3)\\A_3+A_1\end{matrix}\quad A_2+2A_1$&$(0,1,1,0,0,1,0,0)$&$(51,27)$&$Sp(3)_9 \times SU(4)_{16} + 2(1,2)$\\
\hline
17&$\begin{matrix}E_6(a3)\\A_3+A_1\end{matrix}\quad A_2+A_1$&$(0,1,1,0,1,1,0,0)$&$(66,38)$&${\begin{gathered}Spin(7)_{12} \times Sp(2)_9 \times SU(2)_{32} \times U(1)\\ + \text{2 free hypers}\end{gathered}}$\\
\hline
18&$\begin{matrix}E_6(a3)\\2A_2+A_1\end{matrix}\quad A_2+2A_1$&$(0,1,0,0,0,1,0,1)$&$(70,43)$&$Sp(3)_{26} + 1(1,2)$\\
\hline
19&$\begin{matrix}E_6(a3)\\2A_2+A_1\end{matrix}\quad A_2+A_1$&$(0,1,0,0,1,1,0,1)$&$(84,54)$&$(G_2)_{12} \times Sp(2)_{26} + \text{1 free hyper}$\\
\hline
20&$\begin{matrix}E_6(a3)\\A_3\end{matrix}\quad A_2+2A_1$&$(0,1,1,1,0,1,0,0)$&$(64,36)$&$Sp(4)_{10} \times {SU(2)}_{16}^2 \times U(1)^2 + 1(1,2)$\\
\hline
21&$\begin{matrix}E_6(a3)\\A_3\end{matrix}\quad A_2+A_1$&$(0,1,1,1,1,1,0,0)$&$(78,47)$&$SU(4)_{12} \times Sp(3)_{10} \times {U(1)}^2 + \text{1 free hyper}$\\
\hline
22&$\begin{matrix}E_6(a3)\\A_2+2A_1\end{matrix}\quad 2A_2$&$(0,1,0,0,1,1,0,1)$&$(84,54)$&$(G_2)_{12} \times Sp(2)_{26} + 1(2,1)$\\
\hline
23&$\begin{matrix}E_6(a3)\\2A_2\end{matrix}\quad A_2+A_1$&$(0,1,0,0,2,1,0,1)$&$(98,65)$&${(G_2)}_{12}^2 \times SU(2)_{26} + \text{1 free hyper}$\\
\hline
24&$\begin{matrix}A_5\\ D_5(a_1)\end{matrix}\quad 0$&$(0,1,0,0,1,0,0,0)$&$(39,16)$&$(E_6)_{12} \times SU(2)_7\,\text{SCFT} + 1(1,27)$\\
\hline
25&$\begin{matrix}A_5\\A_4+A_1\end{matrix}\quad A_1$&$(0,0,0,0,2,1,0,0)$&$(71,37)$&$Spin(13)_{16} \times SU(2)_7 + 1(1,6)$\\
\hline
26&$\begin{matrix}A_5\\A_4\end{matrix}\quad A_1$&$(0,0,1,0,2,1,0,0)$&$(79,44)$&$\begin{gathered}Spin(12)_{16} \times SU(2)_8 \times SU(2)_7 \,\text{SCFT}\\ + 1(1,1,6)\end{gathered}$\\
\hline
27&$\begin{matrix}A_5\\D_4(a1)\end{matrix}\quad A_2+2A_1$&$(0,0,2,0,1,0,0,0)$&$(47,25)$&${Sp(2)}_8^3 \times SU(2)_7 + 3(1,2)$\\
\hline
28&$\begin{matrix}A_5\\ D_4(a1)\end{matrix}\quad A_2+A_1$&$(0,0,2,0,2,0,0,0)$&$(63,36)$&$\begin{gathered}Spin(8)_{12} \times {SU(2)}_8^3 \times SU(2)_7\\ + \text{3 free hypers}\end{gathered}$\\
\hline
29&$\begin{matrix}A_5\\A_3+A_1\end{matrix}\quad A_2+2A_1$&$(0,0,1,0,1,1,0,0)$&$(58,33)$&$Sp(3)_{9} \times Sp(2)_{16} \times SU(2)_7 + 2(1,1,2)$\\
\hline
30&$\begin{matrix}A_5\\A_3+A_1\end{matrix}\quad A_2+A_1$&$(0,0,1,0,2,1,0,0)$&$(73,44)$&${\begin{gathered}Spin(7)_{12} \times Sp(2)_9 \times SU(2)_{32} \times SU(2)_7\\ + \text{2 free hypers}\end{gathered}}$\\
\hline
31&$\begin{matrix}A_5\\2A_2+A_1\end{matrix}\quad A_2+2A_1$&$(0,0,0,0,1,1,0,1)$&$(77,49)$&$Sp(3)_{26} \times SU(2)_7 + 1(1,1,2)$\\
\hline
32&$\begin{matrix}A_5\\ 2A_2+A_1\end{matrix}\quad A_2+A_1$&$(0,0,0,0,2,1,0,1)$&$(91,60)$&$(G_2)_{12} \times Sp(2)_{26} \times SU(2)_7 + \text{1 free hyper}$\\
\hline
33&$\begin{matrix}A_5\\A_3\end{matrix}\quad A_2+2A_1$&$(0,0,1,1,1,1,0,0)$&$(71,42)$&$\begin{gathered}Sp(4)_{10} \times SU(2)_{32} \times SU(2)_7 \times U(1)\\ + 1(1,1,2)\end{gathered}$\\
\hline
34&$\begin{matrix}A_5\\A_3\end{matrix}\quad A_2+A_1$&$(0,0,1,1,2,1,0,0)$&$(85,53)$&$\begin{gathered}SU(4)_{12} \times Sp(3)_{10} \times SU(2)_7 \times U(1)\\ + \text{1 free hyper}\end{gathered}$\\
\hline
35&$\begin{matrix}A_5\\A_2+2A_1\end{matrix}\quad 2A_2$&$(0,0,0,0,2,1,0,1)$&$(91,60)$&$(G_2)_{12} \times Sp(2)_{26} \times SU(2)_7 + 1(1,2,1)$\\
\hline
36&$\begin{matrix}A_5\\ 2A_2\end{matrix}\quad A_2+A_1$&$(0,0,0,0,3,1,0,1)$&$(105,71)$&${(G_2)}_{12}^2 \times SU(2)_{26} \times SU(2)_7 + \text{1 free hyper}$\\
\hline
37&$\begin{matrix}D_5(a1)\\A_4+A_1\end{matrix}\quad 3A_1$&$(0,1,0,1,1,0,0,0)$&$(52,25)$&$SU(6)_{12} \times Spin(7)_{10} + \frac{1}{2}(1,2)+1(3,1)$\\
\hline
38&$\begin{matrix}D_5(a1)\\A_4+A_1\end{matrix}\quad 2A_1$&$(0,1,0,1,1,1,0,0)$&$(74,40)$&${\begin{gathered}Spin(10)_{16} \times SU(2)_{10} \times SU(2)_{32} \times U(1)\\ + \text{1 free hyper}\end{gathered}}$\\
\hline
39&$\begin{matrix}D_5(a1)\\A_4\end{matrix}\quad 3A_1$&$(0,1,1,1,1,0,0,0)$&$(60,32)$&${\begin{gathered}SU(5)_{12} \times SU(4)_{10} \times SU(2)_8 \times U(1)\\ + \frac{1}{2}(1;1,2)+1(1;3,1)\end{gathered}}$\\
\hline
40&$\begin{matrix}D_5(a1)\\A_4\end{matrix}\quad 2A_1$&$(0,1,1,1,1,1,0,0)$&$(82,47)$&${\begin{gathered}Spin(10)_{16} \times SU(2)_8 \times SU(2)_{10} \times U(1)^2\\ + \text{1 free hyper}\end{gathered}}$\\
\hline
41&$\begin{matrix}D_5(a1)\\D_4(a1)\end{matrix}\quad 2A_2+A_1$&$(0,1,2,0,0,0,0,0)$&$(40,19)$&${SU(4)}_{8}^3\,\text{SCFT} + 1(2) + \text{3 free hypers}$\\
\hline
42&$\begin{matrix}D_5(a1)\\D_4(a1)\end{matrix}\quad 2A_2$&$(0,1,2,0,1,0,0,0)$&$(56,30)$&$\begin{gathered}Spin(8)_{12} \times {SU(2)}_8^3 \times U(1)^2\\ + \text{3 free hypers}\end{gathered}$\\
\hline
43&$\begin{matrix}D_5(a1)\\A_3+A_1\end{matrix}\quad 2A_2+A_1$&$(0,1,1,0,0,1,0,0)$&$(51,27)$&$SU(4)_{16} \times Sp(3)_{9} + \frac{1}{2}(2,1)+\text{2 free hypers}$\\
\hline
44&$\begin{matrix}D_5(a1)\\A_3+A_1\end{matrix}\quad 2A_2$&$(0,1,1,0,1,1,0,0)$&$(66,38)$&${\begin{gathered}Spin(7)_{12} \times Sp(2)_9 \times SU(2)_{32} \times U(1)\\ + \text{2 free hypers}\end{gathered}}$\\
\hline
45&$\begin{matrix}D_5(a1)\\2A_2+A_1\end{matrix}\quad 2A_2+A_1$&$(0,1,0,0,0,1,0,1)$&$(70,43)$&$Sp(3)_{26} + \text{1 free hyper}$\\
\hline
46&$\begin{matrix}D_5(a1)\\2A_2+A_1\end{matrix}\quad A_3$&$(0,1,1,1,0,1,0,0)$&$(63,36)$&${\begin{gathered}Sp(3)_{10} \times SU(3)_{16} \times SU(2)_{9} \times U(1)\\ + \frac{1}{2}(2,1) + \text{1 free hyper}\end{gathered}}$\\
\hline
47&$\begin{matrix}D_5(a1)\\2A_2+A_1\end{matrix}\quad 2A_2$&$(0,1,0,0,1,1,0,1)$&$(84,54)$&$(G_2)_{12} \times Sp(2)_{26} + \text{1 free hyper}$\\
\hline
48&$\begin{matrix}D_5(a1)\\A_3\end{matrix}\quad 2A_2$&$(0,1,1,1,1,1,0,0)$&$(78,47)$&$Spin(7)_{12} \times Sp(3)_{10} \times {U(1)}^2 + \text{1 free hyper}$\\
\hline
49&$\begin{matrix}D_5(a1)\\2A_2\end{matrix}\quad 2A_2$&$(0,1,0,0,2,1,0,1)$&$(98,65)$&${(G_2)}_{12}^2 \times SU(2)_{26} + \text{1 free hyper}$\\
\hline
50&$\begin{matrix}A_4+A_1\\A_4+A_1\end{matrix}\quad A_2+2A_1$&$(0,0,0,1,1,1,0,0)$&$(60,35)$&$SU(3)_{32} \times Sp(3)_{10} + 1(2)$\\
\hline
51&$\begin{matrix}A_4+A_1\\A_4+A_1\end{matrix}\quad A_2+A_1$&$(0,0,0,1,2,1,0,0)$&$(74,46)$&$SU(4)_{12} \times SU(2)_{10} \times {SU(2)_{32}}^2 + \text{1 free hyper}$\\
\hline
52&$\begin{matrix}A_4+A_1\\A_4\end{matrix}\quad A_2+2A_1$&$(0,0,1,1,1,1,0,0)$&$(68,42)$&$SU(3)_{32} \times Sp(2)_{10} \times SU(2)_8 \times U(1) + 1(1,2)$\\
\hline
53&$\begin{matrix}A_4+A_1\\A_4\end{matrix}\quad A_2+A_1$&$(0,0,1,1,2,1,0,0)$&$(82,53)$&${\begin{gathered}SU(4)_{12}\times SU(2)_{32} \times SU(2)_{10} \times SU(2)_8 \times U(1)\\ + \text{1 free hyper}\end{gathered}}$\\
\hline
54&$\begin{matrix}D_4\\A_4+A_1\end{matrix}\quad 3A_1$&$(0,1,0,1,2,0,0,0)$&$(68,36)$&$SU(6)_{12} \times {SU(3)}_{12}^2 + \frac{1}{2}(1;1,2)$\\
\hline
55&$\begin{matrix}D_4\\A_4\end{matrix}\quad 3A_1$&$(0,1,1,1,2,0,0,0)$&$(76,43)$&${\begin{gathered}SU(6)_{12} \times SU(3)_{12} \times SU(2)_8 \times U(1)\\ + \frac{1}{2}(1,1;1,2)\end{gathered}}$\\
\hline
56&$\begin{matrix}D_4\\ D_4(a1)\end{matrix}\quad 2A_2+A_1$&$(0,1,2,0,1,0,0,0)$&$(56,30)$&$Spin(8)_{12}\times SU(2)_{8}^3  + 1(1,2)$\\
\hline
57&$\begin{matrix}D_4\\A_3+A_1\end{matrix}\quad 2A_2+A_1$&$(0,1,1,0,1,1,0,0)$&$(66,38)$&$\begin{gathered}Spin(7)_{12} \times Sp(2)_9 \times SU(2)_{16} \times U(1)\\ + \frac{1}{2}(1,1,2)\end{gathered}$\\
\hline
58&$\begin{matrix}D_4\\2A_2+A_1\end{matrix}\quad A_3$&$(0,1,1,1,1,1,0,0)$&$(77,47)$&${\begin{gathered}SU(4)_{12}\times Sp(2)_{10} \times SU(2)_{16} \times SU(2)_9 \times U(1)\\ + \frac{1}{2}(1,2,1)\end{gathered}}$\\
\hline
59&$\begin{matrix}A_4\\A_4\end{matrix}\quad A_2+2A_1$&$(0,0,2,1,1,1,0,0)$&$(76,49)$&$\begin{gathered}Sp(2)_{10} \times SU(2)_{32} \times {SU(2)}_8^2 \times U(1)^2\\ + 1(1,1,2)\end{gathered}$\\
\hline
60&$\begin{matrix}A_4\\A_4\end{matrix}\quad A_2+A_1$&$(0,0,2,1,2,1,0,0)$&$(90,60)$&${\begin{gathered}SU(4)_{12} \times SU(2)_{10} \times {SU(2)}_8^2 \times U(1)^2\\ + \text{1 free hyper}\end{gathered}}$\\
\hline
\end{longtable}
}

\section{A Detour Through the Twisted Sector}\label{a_detour_through_the_twisted_sector}

There are several fixtures on our list, where the levels of the enhanced flavour symmetry group cannot be determined by considerations from the untwisted sector alone. For instance, consider the pair of fixtures,

\begin{displaymath}
\begin{matrix} \includegraphics[width=153pt]{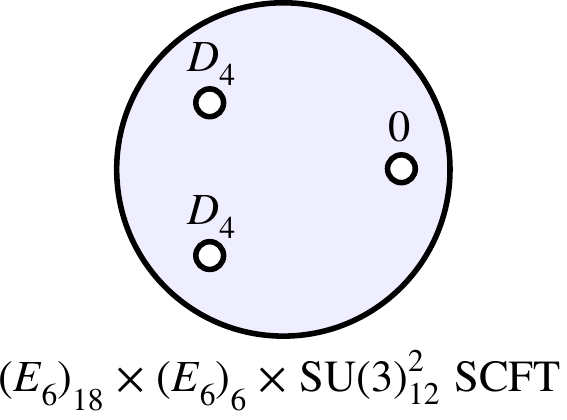}\end{matrix}\qquad\text{and}\qquad
\begin{matrix} \includegraphics[width=188pt]{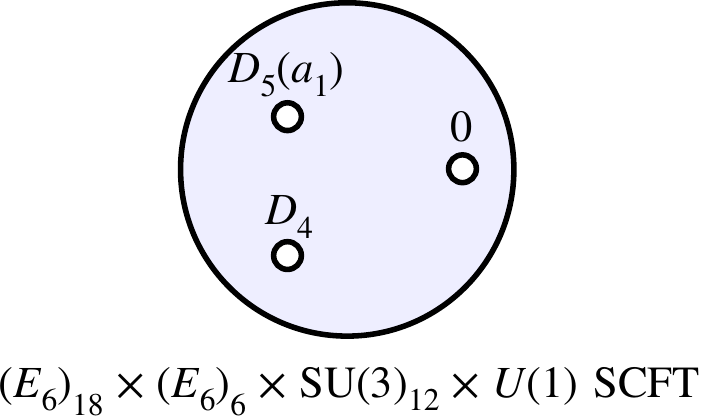}\end{matrix}
\end{displaymath}
In each case, only the diagonal ${(E_6)}_{24}\subset{(E_6)}_{24-k}\times{(E_6)}_{k}$ is manifest. Moreover, the only gaugings, available in the untwisted sector, have Abelian centralizers in ${(E_6)}_{24-k}\times{(E_6)}_{k}$, which makes determining the individual levels (as opposed to their sum) difficult.

To fix the ambiguity, we need to make recourse to the $\mathbb{Z}_2$-twisted sector. While a full discussion of the $\mathbb{Z}_2$-twisted sector is beyond the scope of this paper, we will borrow a few results of that analysis, deferring a full discussion to a future paper.

The twisted punctures are labeled by nilpotent orbits in $F_4$. We will denote them by their Bala-Carter labels, and colour them grey. The empty fixture

\begin{displaymath}
 \includegraphics[width=92pt]{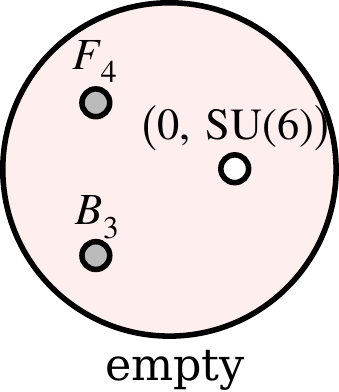}
\end{displaymath}
will allow us to gauge an ${SU(6)}_{24}\subset {(E_6)}_{24-k}\times{(E_6)}_{k}$. The centralizer is ${SU(2)}_{24-k}\times {SU(2)}_k$, from which we can read off the ``missing'' information about the levels.

We will also need the free-field fixture

\begin{displaymath}
 \includegraphics[width=92pt]{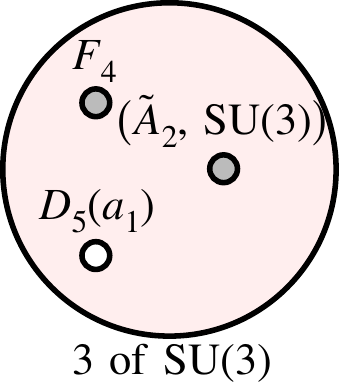}
\end{displaymath}
and the interacting fixture

\begin{displaymath}
 \includegraphics[width=92pt]{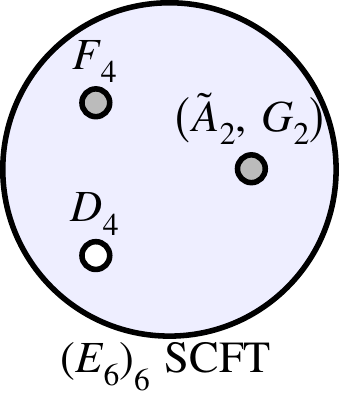}
\end{displaymath}
which is a realization of the ${(E_6)}_6$ SCFT. Finally, we will also need two ``new'' interacting fixtures

\bigskip
\noindent
\begin{tabular}{|c|c|c|l|}
\hline
Fixture&$(n_2,n_3,n_4,n_5,n_6,n_8,n_9,n_{12})$&$(n_h,n_v)$&Global Symmetry\\
\hline 
$\begin{matrix} \includegraphics[width=75pt]{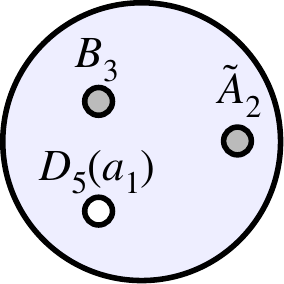}\end{matrix}$&$(0,2,1,2,1,0,1,0)$&$(83,63)$&${(G_2)}_{10}\times{SU(2)}_{18}\times{SU(2)}_{6}\times U(1)$\\
\hline
$\begin{matrix} \includegraphics[width=75pt]{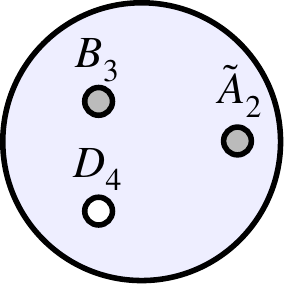}\end{matrix}$&$(0,2,1,2,2,0,1,0)$&$(96,74)$&${(G_2)}_{10}\times{SU(3)}_{12}\times{SU(2)}_{18}\times{SU(2)}_{6}$\\
\hline
\end{tabular}

\bigskip
In both cases, all of the global symmetry \emph{except} the ${SU(2)}_{18}$ is manifest (in particular, the ${SU(2)}_{6}$ is manifest). The 4-punctured sphere

\begin{displaymath}
 \includegraphics[width=241pt]{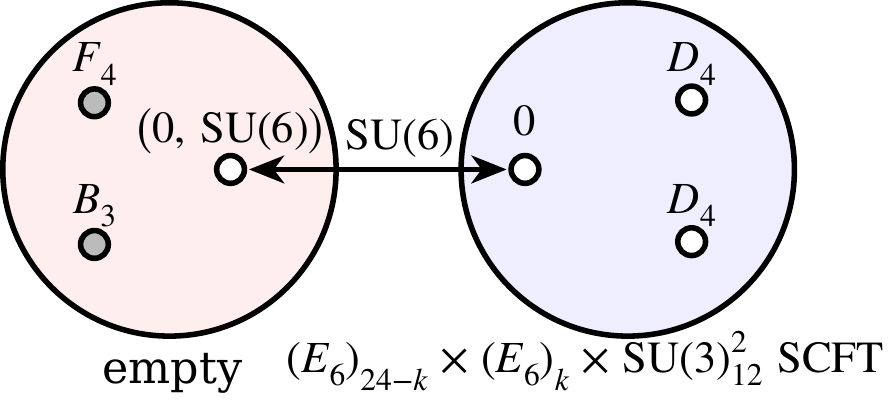}
\end{displaymath}
has global symmetry

\begin{displaymath}
F = {SU(3)}_{12}^2\times {SU(2)}_{24-k}\times{SU(2)}_{k}
\end{displaymath}
The S-dual

\begin{displaymath}
 \includegraphics[width=296pt]{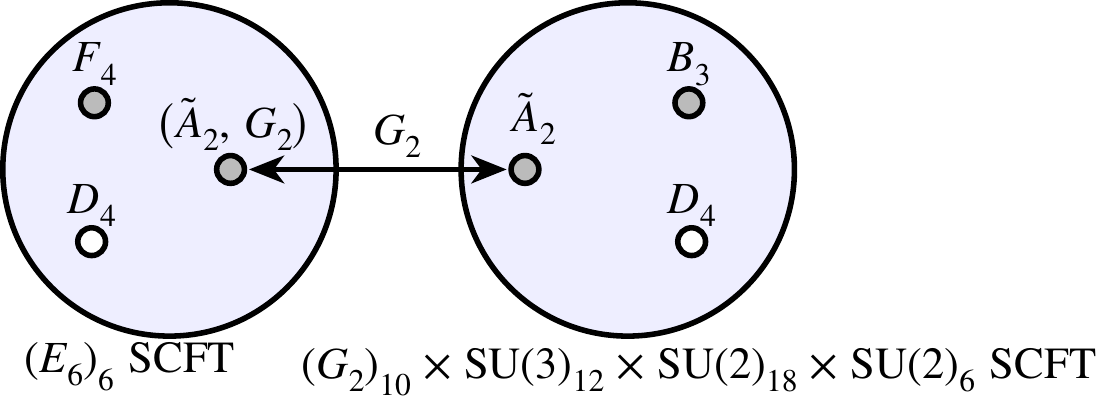}
\end{displaymath}
manifestly has one of the $SU(2)$ levels as $k=6$, which determines the other level to be $18$.

Similarly, for

\begin{displaymath}
 \includegraphics[width=265pt]{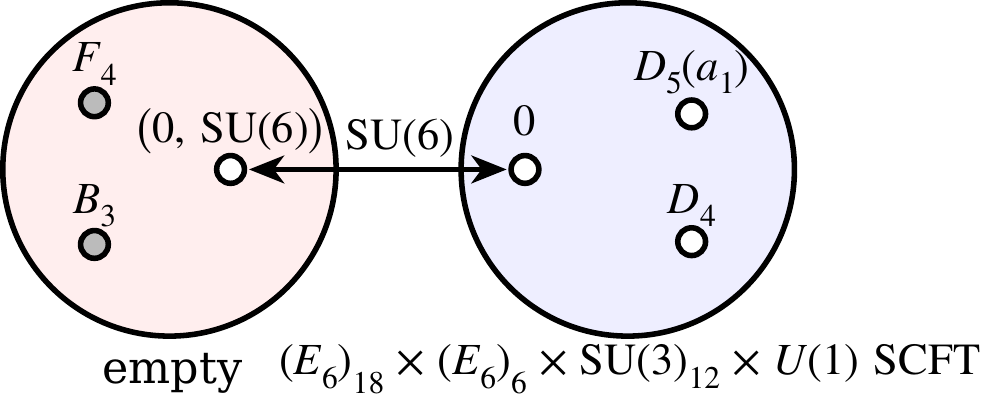}
\end{displaymath}
the global symmetry group is

\begin{displaymath}
F = {SU(3)}_{12}\times {SU(2)}_{24-k}\times{SU(2)}_{k}\times U(1)
\end{displaymath}
Now there are two S-dual presentations of the theory:

\begin{displaymath}
 \includegraphics[width=264pt]{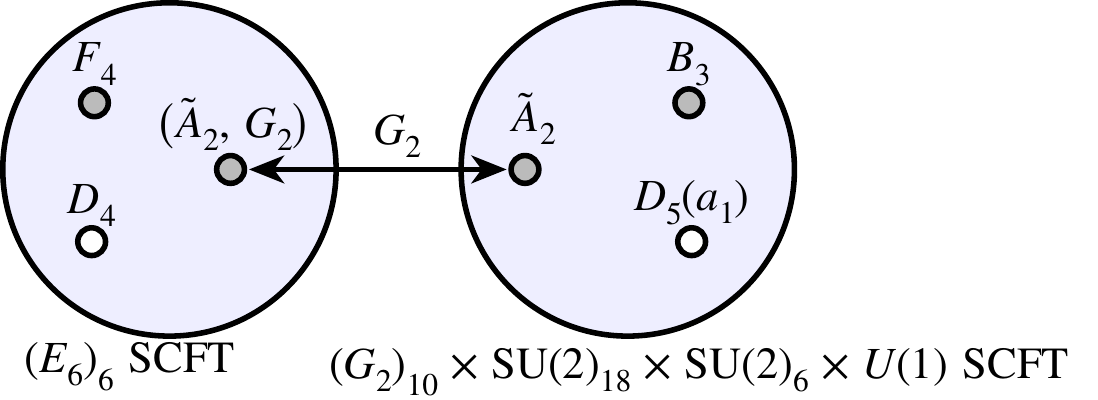}
\end{displaymath}
and
\begin{displaymath}
 \includegraphics[width=264pt]{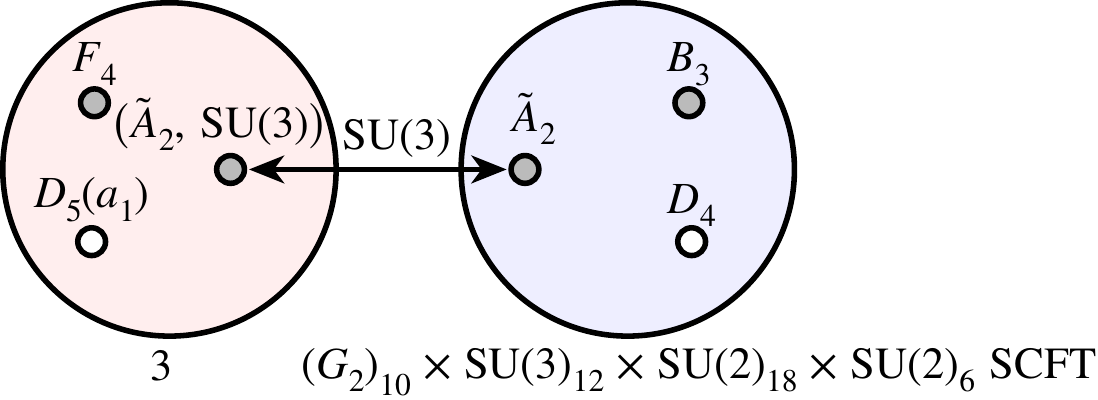}
\end{displaymath}
Again, the fact that one of the $SU(2)$ levels is manifest suffices to determine the other.

As another example, consider the pair of fixtures

\begin{equation}
\begin{matrix} \includegraphics[width=288pt]{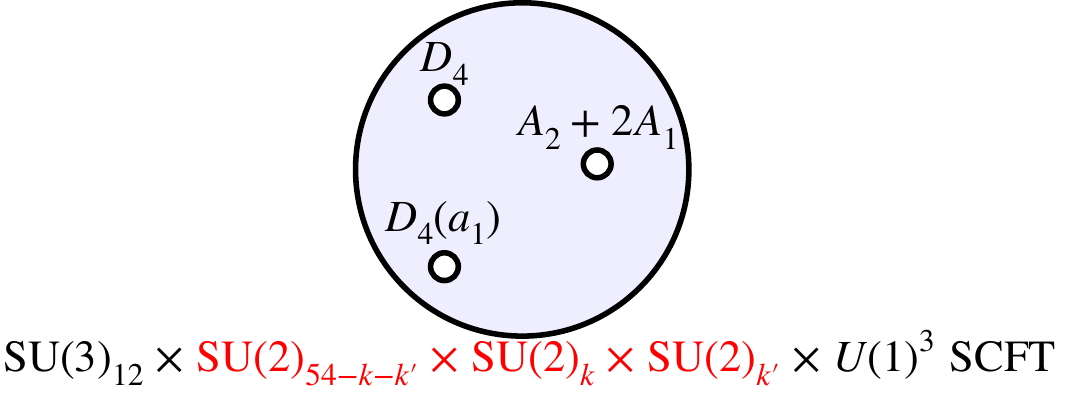}\end{matrix}
\label{D4_D4a1_A22A1}\end{equation}
and

\begin{equation}
 \includegraphics[width=273pt]{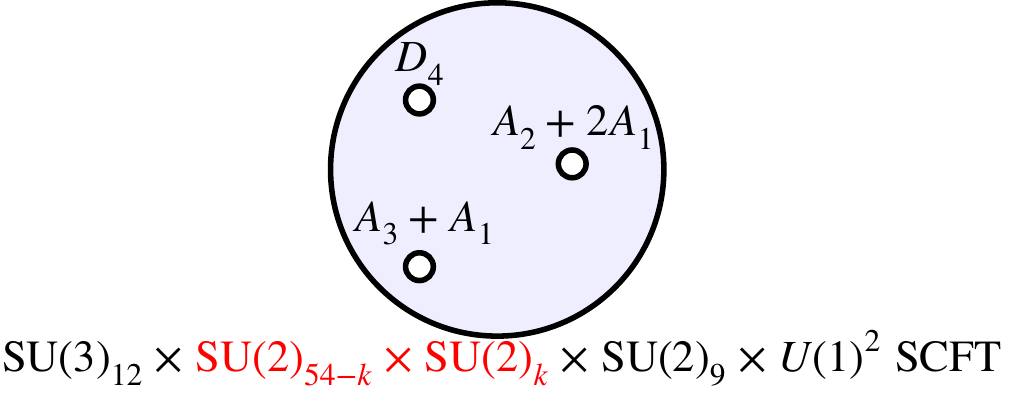}
\label{D4_A3A1_A22A1}\end{equation}
In each case, only the diagonal ${SU(2)}_{54}$ subgroup, of the indicated $\color{red}SU(2)$s, is manifest. Moreover, these fixtures are not gaugeable within the untwisted theory. So there is no obvious way to determine the individual $\color{red}SU(2)$ levels. Fortunately, the twisted sector provides the empty fixture

\begin{displaymath}
 \includegraphics[width=92pt]{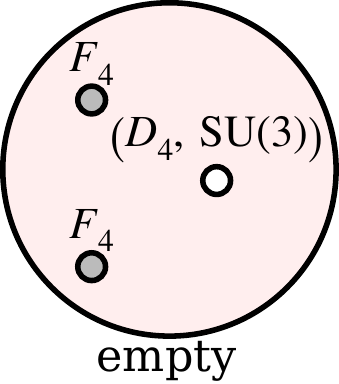}
\end{displaymath}
which allows us to gauge the ${SU(3)}_{12}$ symmetry of each of these fixtures:

\begin{displaymath}
 \includegraphics[width=215pt]{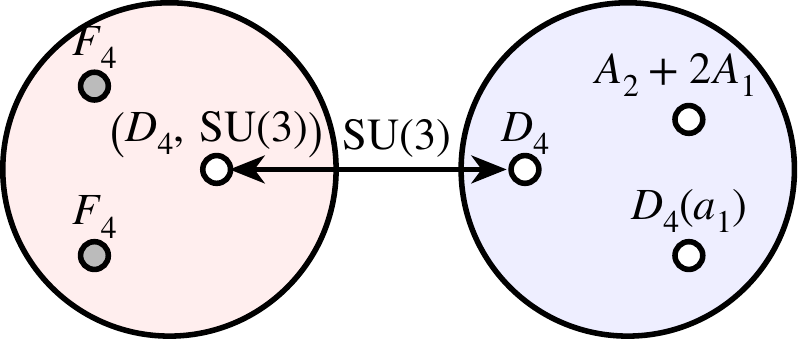}
\end{displaymath}
and

\begin{displaymath}
 \includegraphics[width=215pt]{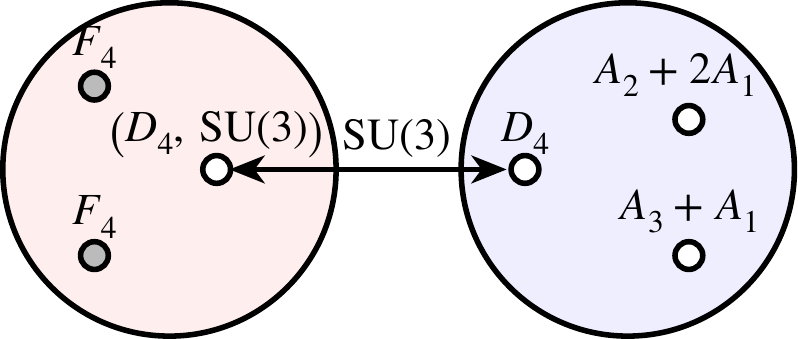}
\end{displaymath}
From the S-duals

\begin{displaymath}
 \includegraphics[width=219pt]{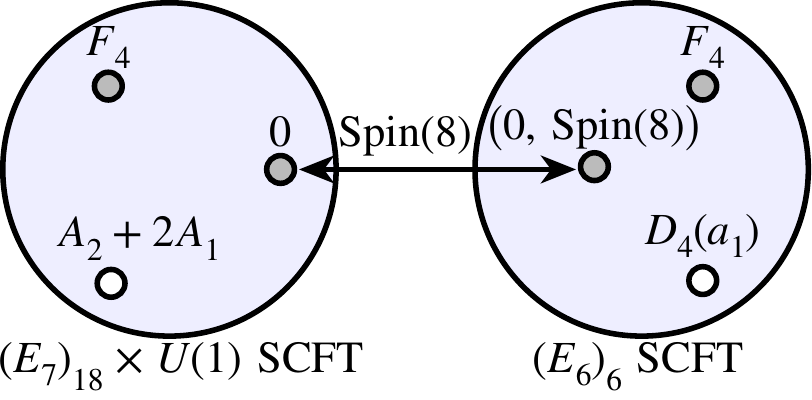}
\end{displaymath}
and

\begin{displaymath}
 \includegraphics[width=219pt]{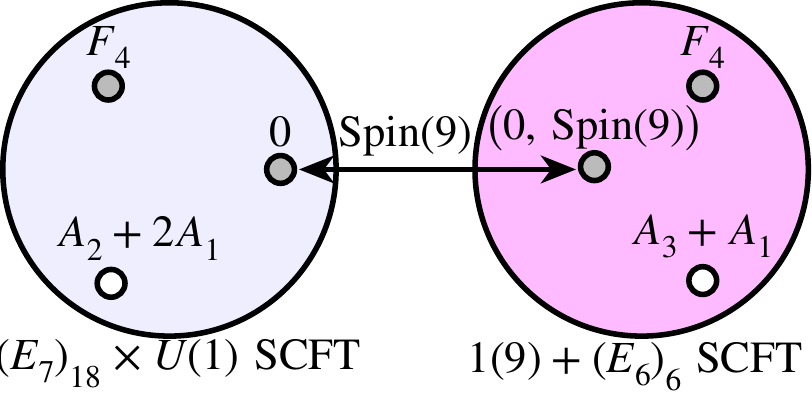}
\end{displaymath}
and the Lie-algebra embeddings

\begin{displaymath}
\begin{split}
{(\mathfrak{e}_7)}_k&\supset {(\mathfrak{f}_4)}_k \oplus {\mathfrak{su}(2)}_{3k}\\
{(\mathfrak{e}_7)}_k&\supset {\mathfrak{so}(9)}_k \oplus {\mathfrak{su}(2)}_{2k}\oplus {\mathfrak{su}(2)}_{k}\\
{(\mathfrak{e}_7)}_k&\supset {\mathfrak{so}(8)}_k \oplus {\mathfrak{su}(2)}_{k}\oplus {\mathfrak{su}(2)}_{k}\oplus {\mathfrak{su}(2)}_{k}
\end{split}
\end{displaymath}
we determine the levels in \eqref{D4_D4a1_A22A1} and \eqref{D4_A3A1_A22A1} to be $k=k'=18$.

Finally, let us turn to the mixed fixture
\begin{displaymath}
 \includegraphics[width=266pt]{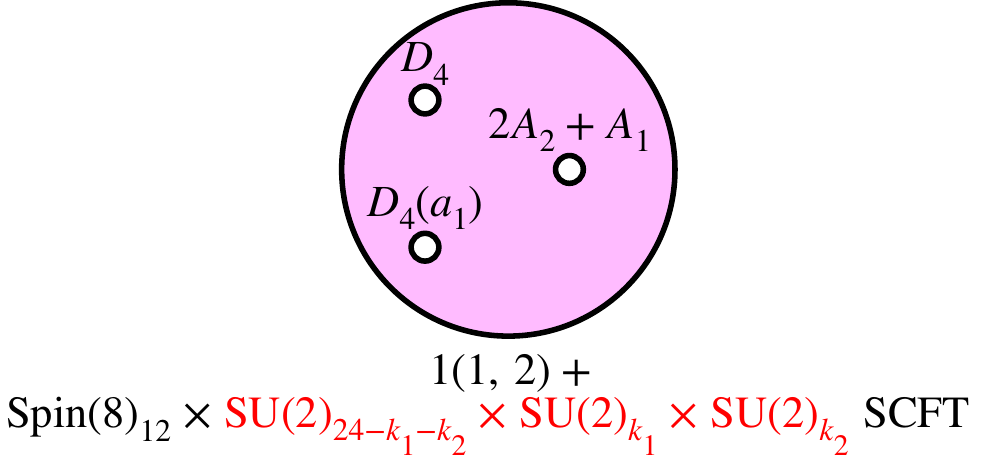}
\end{displaymath}
Only the diagonal ${SU(2)}_{24}\subset {\color{red}{SU(2)}_{24-k_1-k_2}\times {SU(2)}_{k_2}\times {SU(2)}_{k_2}}$ is manifest. Gauging the ${SU(3)}_{12}$ symmetry of the $D_4$ puncture, as before, we find that the S-dual is a $Spin(8)$ gauge theory, with matter in the $1(8_v)+1(8_s)+1(8_c)+2(1)$, coupled to two copies of the ${(E_6)}_6$ SCFT.

\begin{displaymath}
 \includegraphics[width=230pt]{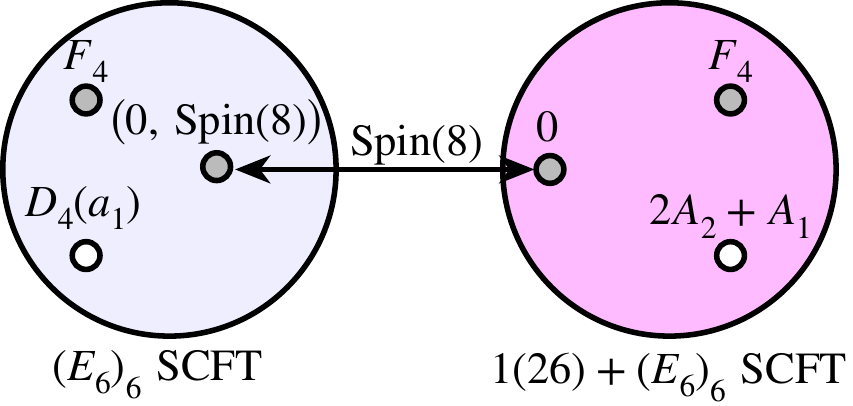}
\end{displaymath}
From this, we read off the levels of the three $SU(2)$s: $k_1=k_2=24-k_1-k_2=8$.

\section{Applications}\label{applications}

\subsection{$E_6$ and $F_4$ gauge theory}\label{_and__gauge_theory}

\subsubsection{$E_6 + 4(27)$}\label{}

$E_6$ gauge theory, with four fundamental hypermultiplets, is superconformal. It is realized as the 4-punctured sphere

\begin{displaymath}
 \includegraphics[width=240pt]{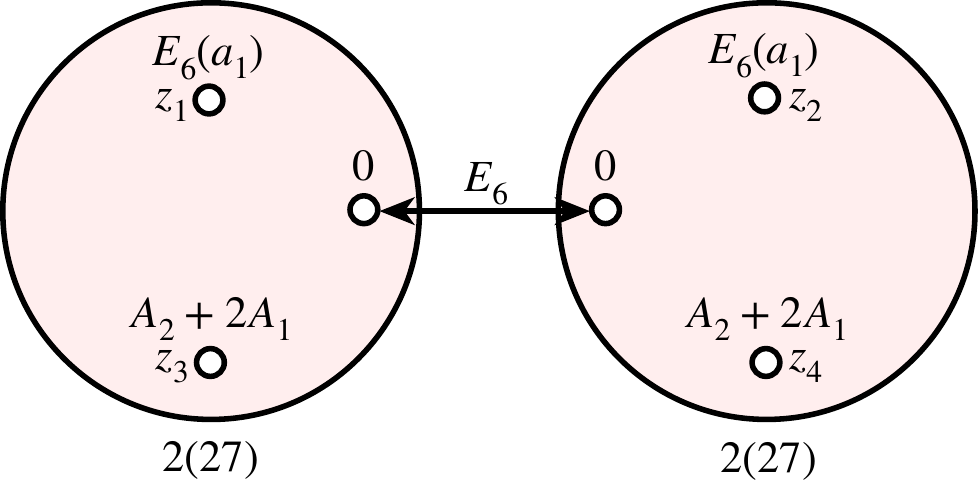}
\end{displaymath}
The S-dual theory is an $SU(2)$ gauging of the $SU(4)_{54}\times SU(2)_7\times U(1)$ SCFT, with an additional half-hypermultiplet in the fundamental.

\begin{displaymath}
 \includegraphics[width=240pt]{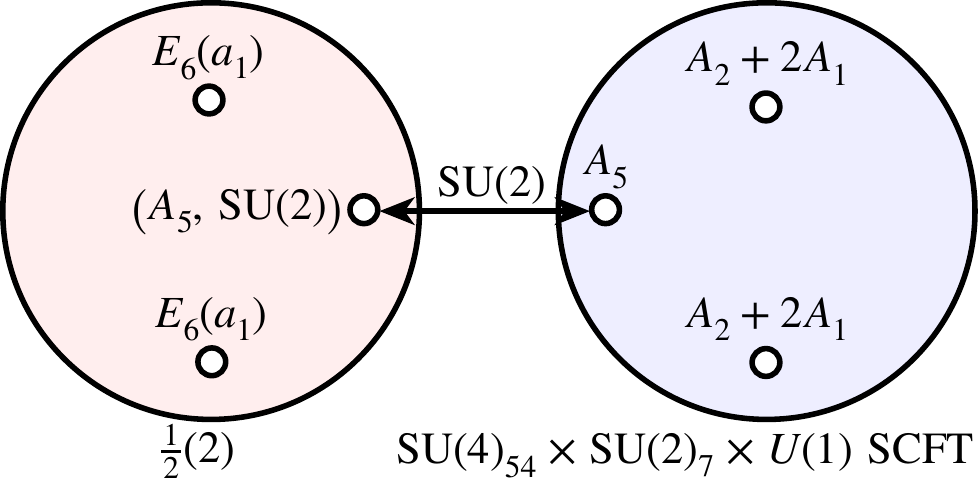}
\end{displaymath}
The $k$-differentials, which determine the Seiberg-Witten solution, are

\begin{equation}
\begin{aligned}
\phi_2(z) &= \frac{u_2\, z_{1 2} z_{3 4}\, {(d z)}^2}{ {(z-z_1)}{(z-z_2)}{(z-z_3)}{(z-z_4)} }\\
\phi_5(z) &= \frac{u_5\, z_{1 2} z_{3 4}^4\, {(d z)}^5}{ {(z-z_1)}{(z-z_2)}{(z-z_3)}^4{(z-z_4)}^4 }\\
\phi_6(z) &= \frac{u_6\, z_{1 2}^2 z_{3 4}^4\, {(d z)}^6}{ {(z-z_1)}^2{(z-z_2)}^2{(z-z_3)}^4{(z-z_4)}^4 }\\
\phi_8(z) &= \frac{u_8\, z_{1 2}^2 z_{3 4}^6\, {(d z)}^8}{ {(z-z_1)}^2{(z-z_2)}^2{(z-z_3)}^6{(z-z_4)}^6 }\\
\phi_9(z) &= \frac{u_9\, z_{1 2}^2 z_{3 4}^7\, {(d z)}^9}{ {(z-z_1)}^2{(z-z_2)}^2{(z-z_3)}^7{(z-z_4)}^7 }\\
\phi_{12}(z) &= \frac{u_{12}\, z_{1 2}^3 z_{3 4}^9\, {(d z)}^{12}}{ {(z-z_1)}^3{(z-z_2)}^3{(z-z_3)}^9{(z-z_4)}^9 }
\end{aligned}
\label{E6sol}\end{equation}
The gauge coupling, $\tau = \frac{\theta}{\pi}+\frac{8\pi i}{g^2}$, is determined by the $SL(2,\mathbb{C})$-invariant cross-ratio
\begin{equation}\label{gaugecoupling}
f(\tau)\equiv - \frac{\theta_2^4(0,\tau)}{\theta_4^4(0,\tau)} = \frac{z_{1 3}z_{2 4}}{z_{1 4} z_{2 3}}
\end{equation}
and, for calculational purposes, it is usually convenient to use $SL(2,\mathbb{C})$ to fix $(z_1,z_2,z_3,z_4)=(0,\infty,f(\tau),1)$ in \eqref{E6sol}.

The solution to $E_6$ gauge theory with $N_f \leq 3$ fundamental hypermultiplets was first found in \cite{Terashima:1998iz}.

\subsubsection{$F_4+3(26)$}\label{F4_2}
$F_4$ gauge theory, with three fundamentals, is also superconformal. It is realized as
\begin{displaymath}
 \includegraphics[width=240pt]{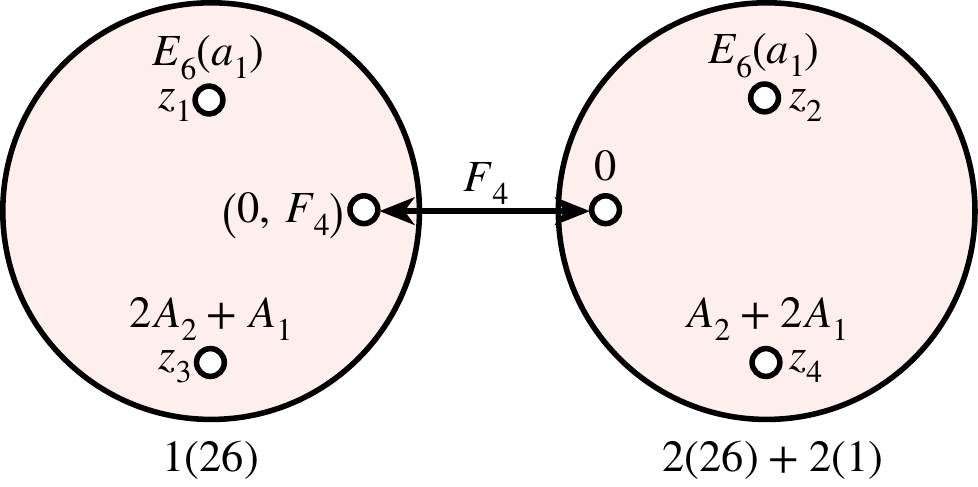}
\end{displaymath}
The S-dual theory is an $SU(2)$ gauging of the $Sp(3)_{26}\times SU(2)_7$ SCFT, with additional matter in the $\tfrac{1}{2}(2)+2(1)$.

\begin{displaymath}
 \includegraphics[width=240pt]{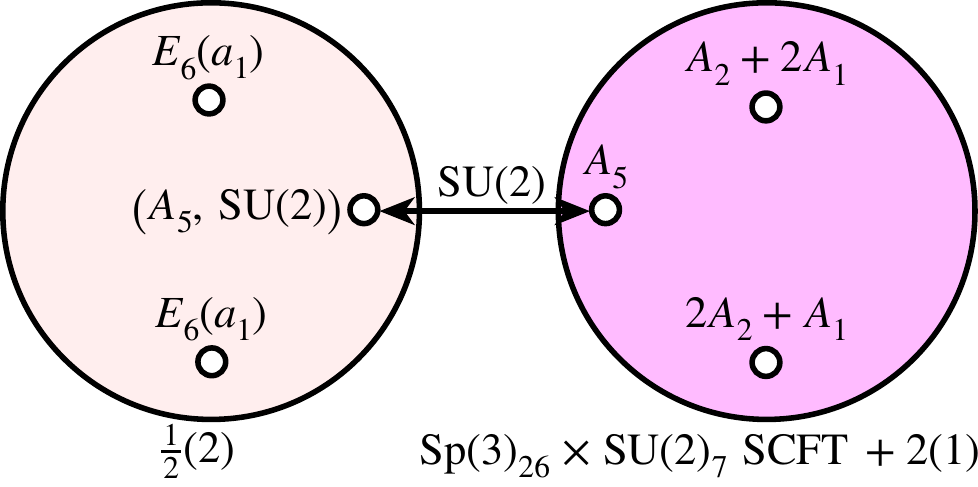}
\end{displaymath}
The nonzero $k$-differentials, which determine the Seiberg-Witten solution, are the same as in \eqref{E6sol} but with $\phi_5(z)\equiv 0\equiv\phi_9(z)$. The gauge coupling is again given by \eqref{gaugecoupling}. Physically, this theory is obtained by Higgsing $E_6\to F_4$, using one of the hypermultiplets in the $27$.

In practice, given the solution to $E_6+4(27)$, the solution to $F_4+3(26)+2(1)$ is obtained by noting that
\begin{itemize}%
\item There is a $\mathbb{Z}_2$ symmetry, $\sigma: (u_5,u_9)\mapsto (-u_5,-u_9)$, acting on the Coulomb branch of the $E_6+4(27)$.
\item The Coulomb branch geometry of $F_4+3(26)+2(1)$ is the geometry of the fixed-locus of $\sigma$.

\end{itemize}
\subsection{Adding ${(E_8)}_{12}$ SCFTs}\label{adding__scfts}
Starting with the $E_6 + 4(27)$ Lagrangian field theory, we can start replacing hypermultiplets in the $27$ with copies of the ${(E_8)}_{12}$ SCFT. For $n$ $27$s and $4-n$ copies of the ${(E_8)}_{12}$ SCFT, the flavour symmetry group of the theory is
\begin{displaymath}
F = {SU(3)}_{12}^{4-n}\times U(n)_{54}
\end{displaymath}
In each of these cases, the S-dual theory is an $SU(2)$ gauging of the ${SU(3)}_{12}^{4-n}\times SU(n)_{54}\times SU(2)_7\times U(1)$ SCFT, with an additional half-hypermultiplet in the fundamental (the $U(1)$ is absent for $n=0$).

\subsubsection{$n=3$}\label{_3}
With one copy of the ${(E_8)}_{12}$ SCFT,
\begin{displaymath}
 \includegraphics[width=240pt]{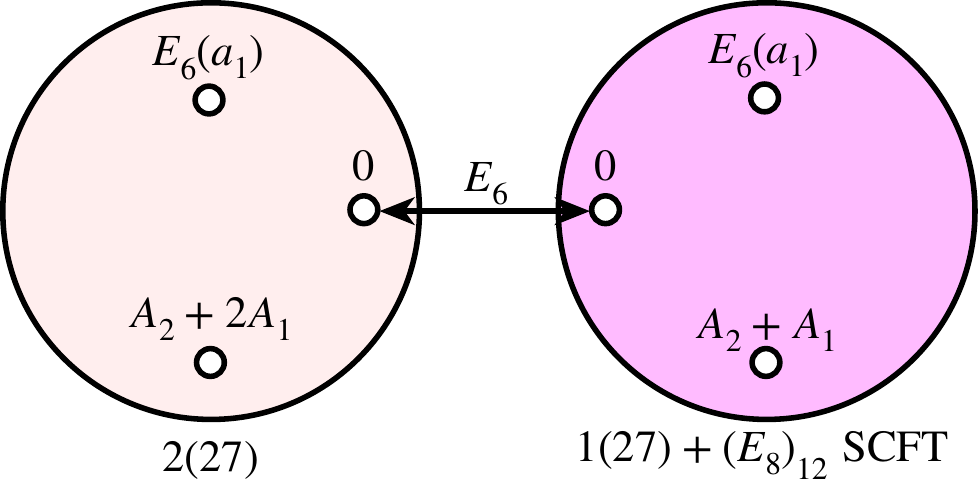}
\end{displaymath}
is dual to
\begin{displaymath}
 \includegraphics[width=283pt]{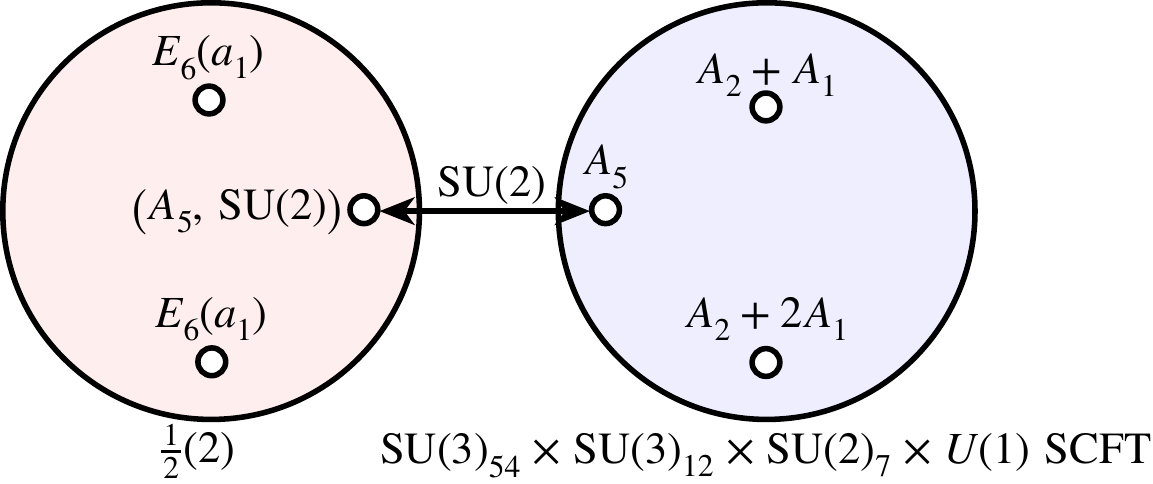}
\end{displaymath}

\subsubsection{$n=2$}\label{_4}
With two copies of the ${(E_8)}_{12}$ SCFT, there are two possible realizations. Either
\begin{displaymath}
 \includegraphics[width=240pt]{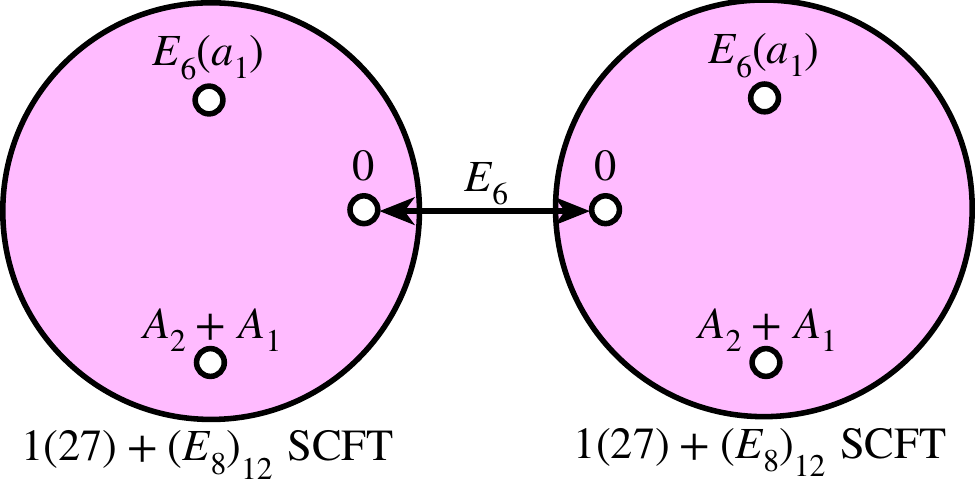}
\end{displaymath}
dual to
\begin{displaymath}
 \includegraphics[width=282pt]{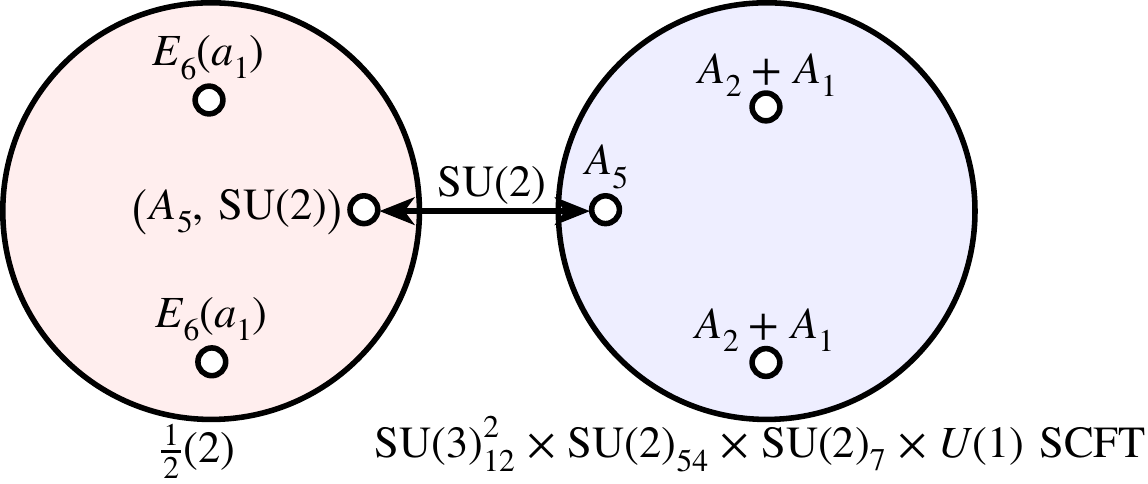}
\end{displaymath}
or
\begin{displaymath}
 \includegraphics[width=240pt]{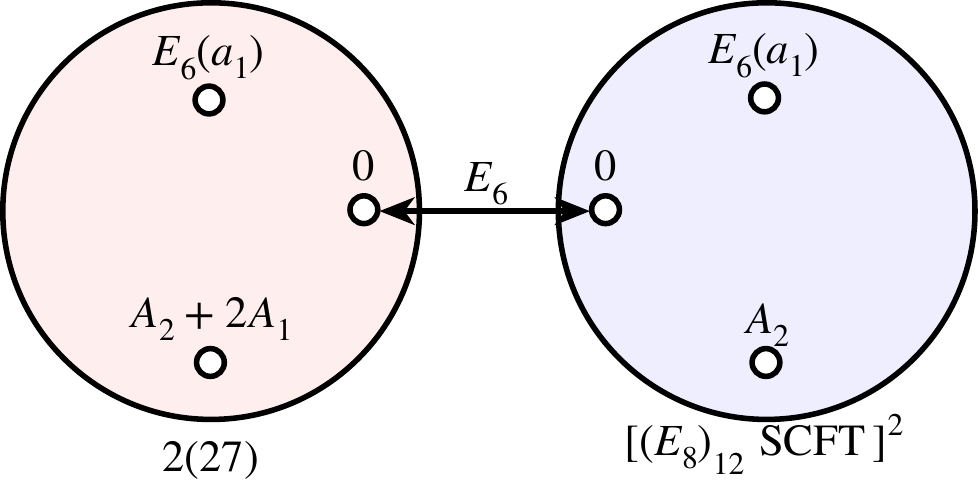}
\end{displaymath}
dual to
\begin{displaymath}
 \includegraphics[width=282pt]{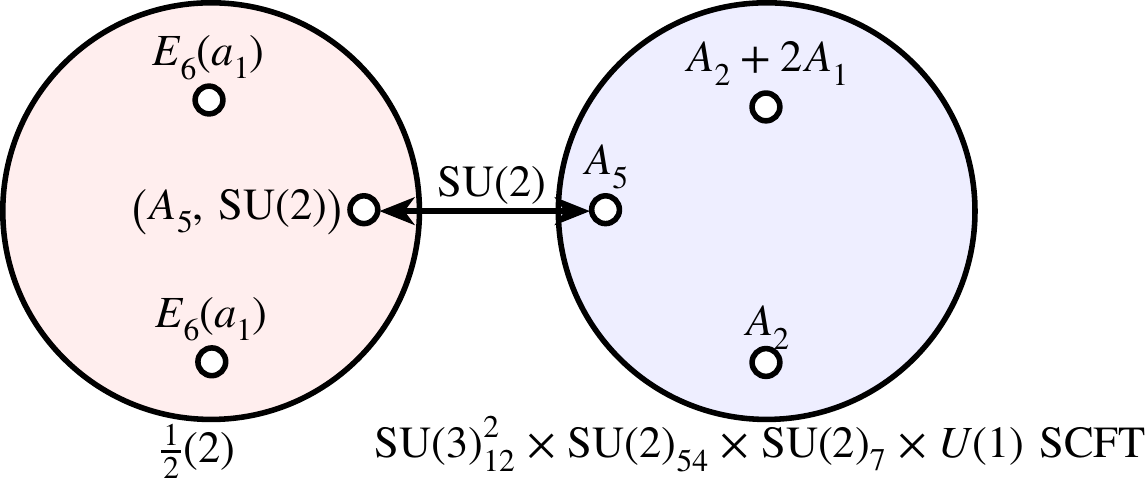}
\end{displaymath}
These give two, apparently distinct, realizations of the ${SU(3)}_{12}^2\times {SU(2)}_{54}\times {SU(2)}_7\times U(1)$ SCFT.

\subsubsection{$n=1$}\label{_5}
With three copies of the ${(E_8)}_{12}$ SCFT, we have
\begin{displaymath}
 \includegraphics[width=240pt]{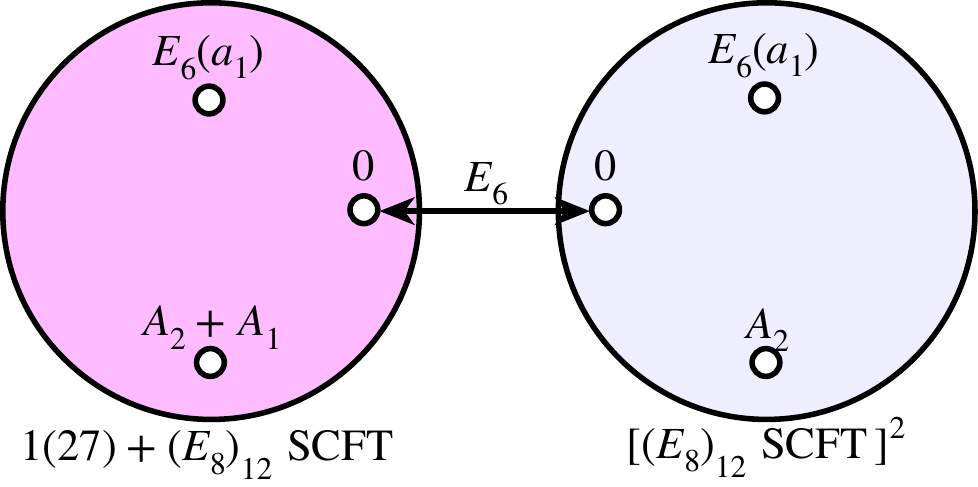}
\end{displaymath}
dual to
\begin{displaymath}
 \includegraphics[width=240pt]{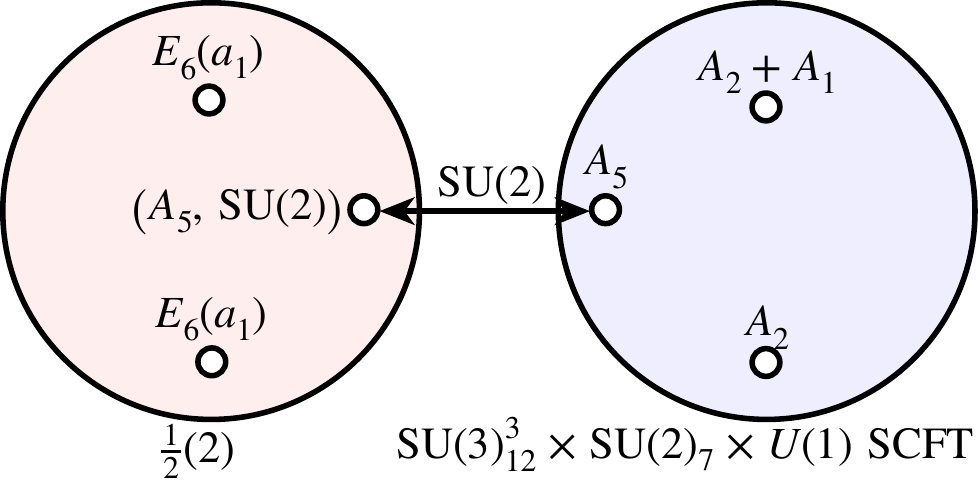}
\end{displaymath}

\subsubsection{$n=0$}\label{_6}
Finally, the $E_6$ gauging of four copies of the ${(E_8)}_{12}$ SCFT,

\begin{displaymath}
 \includegraphics[width=240pt]{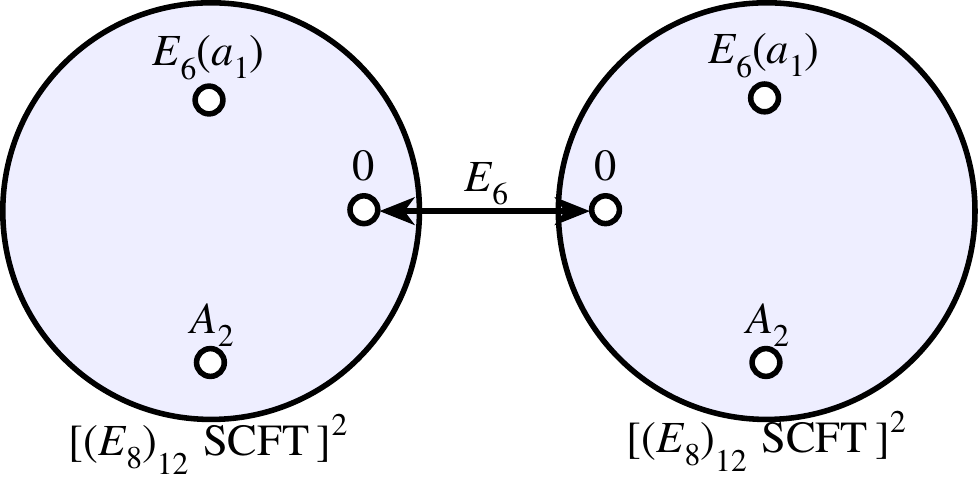}
\end{displaymath}
is dual to
\begin{displaymath}
 \includegraphics[width=240pt]{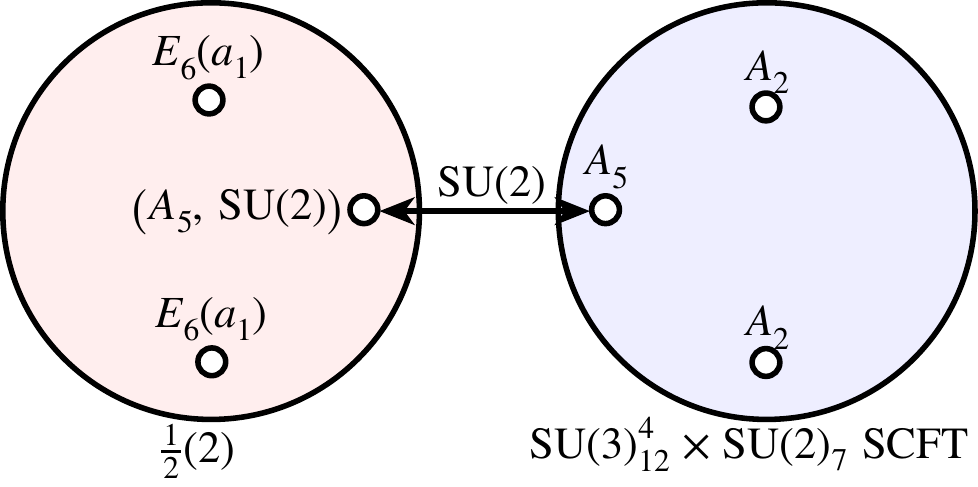}
\end{displaymath}

\subsection{Connections with F-theory}\label{connections_with_f_theory}

Placing $n$ D3-branes at a IV${}^*$, III${}^*$ or II${}^*$ singularity in F-Theory yields an $\mathcal{N}=2$ superconformal field theory on the world-volume of the D3-branes \cite{Dasgupta:1996ij,Banks:1996nj}. For $n=1$ these are, respectively, the $(E_6)_{6}$, $(E_7)_{8}$ and $(E_8)_{12}$ superconformal field theories of Minahan and Nemenschansky \cite{Minahan:1996cj}. For higher $n$, the properties of these SCFTs were computed in \cite{Aharony:2007dj}. The results may be summarized as follows

\bigskip
{
\renewcommand{\arraystretch}{2.55}
\noindent
\begin{tabular}{|c|c|c|c|c|c|}
\hline
\mbox{\shortstack{F-Theory\\singularity}}&Flavour symmetry&\mbox{\shortstack{Graded Coulomb\\branch dimensions}}&$(n_h,n_v)$\\
\hline
IV${}^*$&$(E_6)_{6n}\times SU(2)_{(n-1)(3n+1)}$&$n_{3l}=1,\quad l=1,2,...,n$&$\bigl(3n^2+14n-1, n(3n+2)\bigr)$ \\
\hline
III${}^*$&$(E_7)_{8n}\times SU(2)_{(n-1)(4n+1)}$&$n_{4l}=1,\quad l=1,2,...,n$&$\bigl(4n^2+21n-1, n(4n+3)\bigr)$ \\
\hline
II${}^*$&$(E_8)_{12n}\times SU(2)_{(n-1)(6n+1)}$&$n_{6l}=1,\quad l=1,2,...,n$&$\bigl(6n^2+35n-1, n(6n+5)\bigr)$ \\
\hline
\end{tabular}
}

In \cite{Gaiotto:2012uq}, Gaiotto and Razamat proposed a realization of these $(n\geq2)$ SCFTs as a mixed fixture, with one free hypermultiplet, in the $A_{N-1}$ theory, for $N=3n,4n$ and $6n$, respectively.

\bigskip
{
\noindent
\begin{tabular}{|l|c|l|l|}
\hline
Theory&Fixture&Manifest flavour symmetry&Enhanced to\\
\hline 
$\text{IV}^*$&$\begin{matrix}\includegraphics[width=77pt]{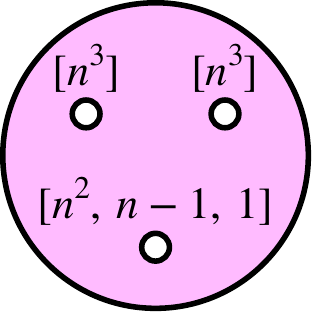}\end{matrix}$&${SU(3)}_{6n}^2\times {SU(2)}_{6n}\times U(1)^2$&${(E_6)}_{6n}\times {SU(2)}_{k} +\tfrac{1}{2}(2)$\\
\hline
$\text{III}^*$&$\begin{matrix}\includegraphics[width=77pt]{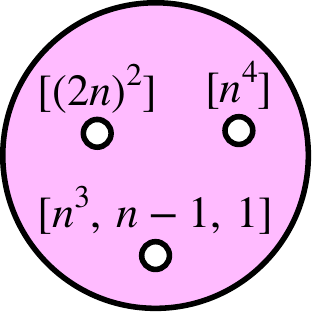}\end{matrix}$&${SU(2)}_{8n}\times {SU(4)}_{8n}\times{SU(3)}_{8n}\times U(1)^2$&${(E_7)}_{8n}\times {SU(2)}_{k} +\tfrac{1}{2}(2)$\\
\hline
$\text{II}^*$&$\begin{matrix}\includegraphics[width=77pt]{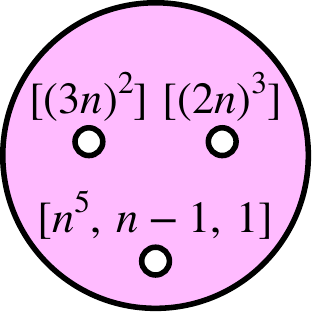}\end{matrix}$&${SU(2)}_{12n}\times {SU(3)}_{12n}\times{SU(5)}_{12n}\times U(1)^2$&${(E_8)}_{12n}\times {SU(2)}_{k} +\tfrac{1}{2}(2)$\\
\hline
\end{tabular}
}

\bigskip
For $n=2$, the $SU(2)$ flavour symmetry is manifest, and one readily verifies that it has the predicted level (given that the hypermultiplet transforms as $\tfrac{1}{2}(2)$ under the $SU(2)$). But, for $n\geq 3$, only the $U(1)$ Cartan is manifest and it is not easy to determine the level of the $SU(2)$.

We have, of course, numerous realizations of the $n=1$ theories. But we also find examples of the higher-$n$ theories
\begin{itemize}
\item We find the $n=2$ IV${}^*$ SCFT as one of our fixtures in \S\ref{interacting_fixtures_with_one_irregular_puncture} and as part of a product SCFT in \hyperlink{IntFixture39}{fixture 39} of \S\ref{interacting_fixtures_with_enhanced_global_symmetry}. It also appeared as an interacting fixture in the $D_4$ theory in \cite{Chacaltana:2011ze}.
\item We find the $n=2$ III${}^*$ SCFT as mixed \hyperlink{MixedFixture4}{fixture 4 }in \S\ref{mixed_fixtures} and as part of a product SCFT in \hyperlink{IntFixture6}{fixture 6} of \S\ref{interacting_fixtures_with_enhanced_global_symmetry}.
\item We find the $n=2$ II${}^*$ SCFT as interacting \hyperlink{IntFixture2}{fixture 2} in \S\ref{interacting_fixtures_with_enhanced_global_symmetry}.
\item We find the $n=3$ III${}^*$ SCFT as interacting \hyperlink{IntFixture7}{fixture 7} in \S\ref{interacting_fixtures_with_enhanced_global_symmetry}.
\end{itemize}

\noindent
In particular, the latter gives a nice check of the $SU(2)$ level for $n=3$.

Further examples can be found in the $\mathbb{Z}_2$-twisted sector. Notably, the fixtures
\begin{displaymath}
\begin{matrix} \includegraphics[width=77pt]{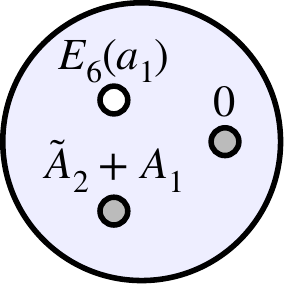}\end{matrix}
\quad\text{and}\quad
\begin{matrix} \includegraphics[width=77pt]{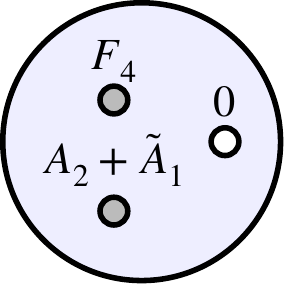}\end{matrix}
\end{displaymath}
provide realizations, respectively, of the $n=3,4$ IV${}^*$ SCFTs. Again, the $SU(2)$ levels agree with the predictions of \cite{Aharony:2007dj}. Together with the above examples, these exhaust all the IV${}^*$, III${}^*$ and II${}^*$ theories with nonzero graded Coulomb branch dimensions in degrees $\leq 12$.

\section{Isomorphic Theories}\label{isomorphic_theories}

In our table of interacting fixtures with enhanced global symmetry, we find several SCFTs which seem to be realized in more than one way. Most of these isomorphisms can be checked by various dualities. Some, however, cannot and we list them below.

\begin{align*}
\begin{matrix} \includegraphics[width=77pt]{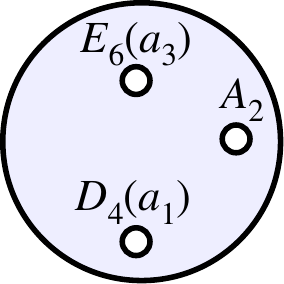}\end{matrix} \quad\simeq\quad
\begin{matrix} \includegraphics[width=77pt]{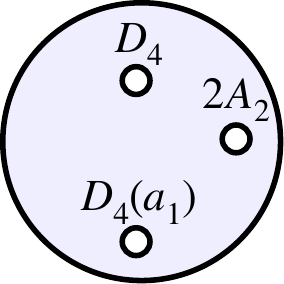}\end{matrix}&\quad\simeq\quad
{Spin(8)}_{12}^2 \times {U(1)}^2\, \text{SCFT}\\
\begin{matrix} \includegraphics[width=77pt]{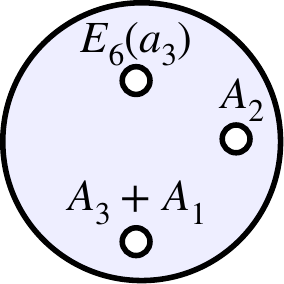}\end{matrix} \quad\simeq\quad
\begin{matrix} \includegraphics[width=77pt]{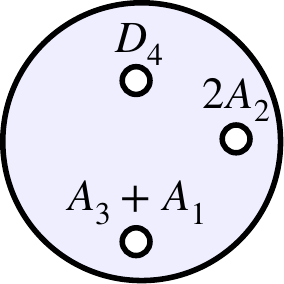}\end{matrix}&\quad\simeq\quad
{Spin(7)}_{12}^2\times {SU(2)}_{9}\times U(1)\,\text{SCFT}\\
\begin{matrix} \includegraphics[width=77pt]{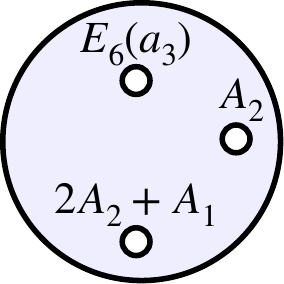}\end{matrix} \quad\simeq\quad
\begin{matrix} \includegraphics[width=77pt]{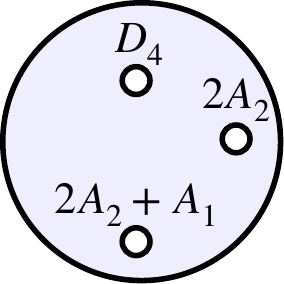}\end{matrix}&\quad\simeq\quad
{(G_2)}_{12}^2\times {SU(2)}_{26}\,\text{SCFT}\\
\begin{matrix} \includegraphics[width=77pt]{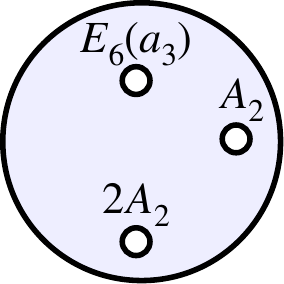}\end{matrix} \quad\simeq\quad
\begin{matrix} \includegraphics[width=77pt]{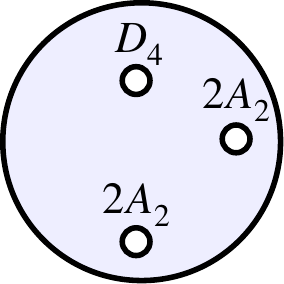}\end{matrix}&\quad\simeq\quad
{(G_2)}_{12}^3\,\text{SCFT}\\
\begin{matrix} \includegraphics[width=77pt]{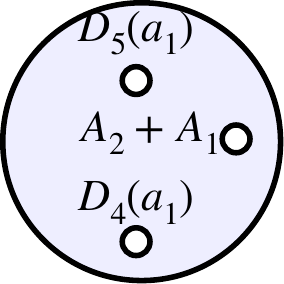}\end{matrix} \quad\simeq\quad
\begin{matrix} \includegraphics[width=77pt]{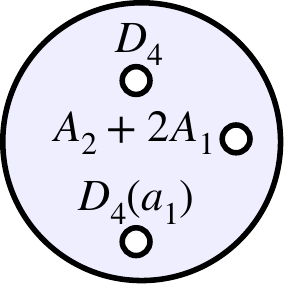}\end{matrix}&\quad\simeq\quad
{SU(3)}_{12}\times {SU(2)}_{18}^3 \times {U(1)}^3\, \text{SCFT}\\
\begin{matrix} \includegraphics[width=77pt]{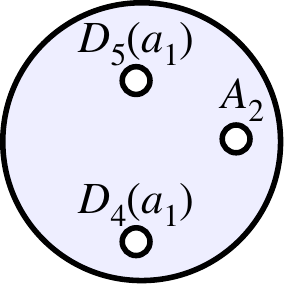}\end{matrix} \quad\simeq\quad
\begin{matrix} \includegraphics[width=77pt]{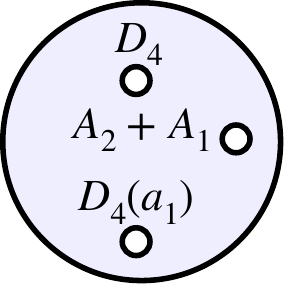}\end{matrix}&\quad\simeq\quad
{SU(3)}_{12}^2 \times {U(1)}^5\, \text{SCFT}\\
\begin{matrix} \includegraphics[width=77pt]{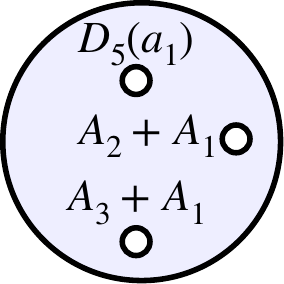}\end{matrix} \quad\simeq\quad
\begin{matrix} \includegraphics[width=77pt]{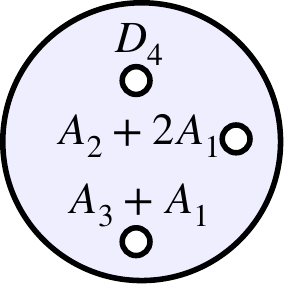}\end{matrix}&\quad\simeq\quad
{SU(3)}_{12}\times {SU(2)}_{36} \times {SU(2)}_{18} \times {SU(2)}_{9} \times {U(1)}^2\, \text{SCFT}\\
\begin{matrix} \includegraphics[width=77pt]{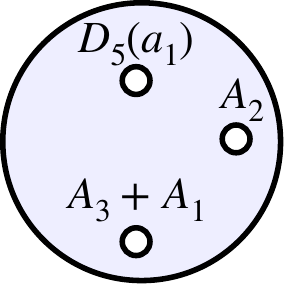}\end{matrix} \quad\simeq\quad
\begin{matrix} \includegraphics[width=77pt]{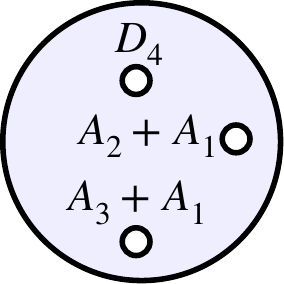}\end{matrix}&\quad\simeq\quad
{SU(3)}_{12}^2\times {SU(2)}_{9}\times U(1)^3\,\text{SCFT}\\
\begin{matrix} \includegraphics[width=77pt]{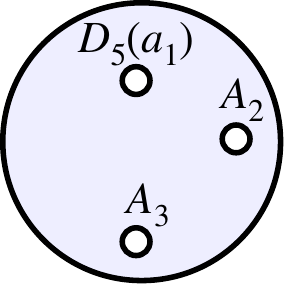}\end{matrix} \quad\simeq\quad
\begin{matrix} \includegraphics[width=77pt]{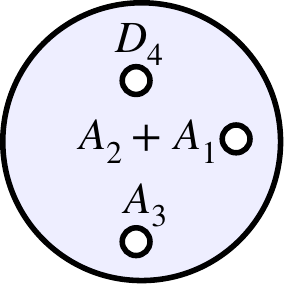}\end{matrix}&\quad\simeq\quad
{SU(3)}_{12}^2\times {Sp(2)}_{10}\times U(1)^3\,\text{SCFT}
\end{align*}

It would be nice to check these conjectured isomorphisms by comparing the expansions of the superconformal indices for these pairs of fixtures to higher order in $\tau$.

\section*{Acknowledgements}\label{Acknowledgements}
\addcontentsline{toc}{section}{Acknowledgements}

We would like to thank S.~Katz, D.~Morrison, A.~Neitzke, R.~Plesser and Y.~Tachikawa for helpful discussions. J.~D.~and O.~C.~would like to thank the Aspen Center for Physics (supported, in part, by the National Science Foundation under Grant PHY-1066293) for their hospitality when this work was initiated. O.~C.~would further like to thank the Simons Foundation for partial support in Aspen. The work of J.~D.~and A.~T.~was supported in part by the National Science Foundation under Grant PHY-1316033. The work of O.~C.~was supported in part by the INCT-Matem\'atica and the ICTP-SAIFR in Brazil through a Capes postdoctoral fellowship.

\begin{appendices}
\section{Bala-Carter Labels}\label{appendix_bala_carter_notation}

In the twisted and untwisted sectors of the $A$ and $D$ series, punctures were in one-to-one correspondence with certain classes of partitions \cite{Gaiotto:2009we,Tachikawa:2009rb,Chacaltana:2012zy,Chacaltana:2012ch}. The partition denotes how the fundamental representation (vector representation, in the case of $\mathfrak{so}(N)$) of $\mathfrak{g}$ \footnote{For untwisted (twisted) punctures in the $A$ and $D$ series, $\mathfrak{g}$ is of type $A$ $(B)$ and $D$ ($C$), respectively.} decomposes into representations of the corresponding (Nahm) $\mathfrak{su}(2)$. Moreover, one can also read off the centralizer, $\mathfrak{f}$, of $\mathfrak{su}(2)$ inside $\mathfrak{g}$, as well as the decomposition of the fundamental representation of $\mathfrak{g}$ under $\mathfrak{su}(2) \times \mathfrak{f}$, from the partition (see (2.7) in \cite{Chacaltana:2012zy}). The decomposition under $\mathfrak{su}(2)\times \mathfrak{f}$ for each puncture is precisely the information needed to compute the flavour group levels in \S\ref{flavour_groups}, as well as the expansion of the superconformal index in \S\ref{global_symmetries_and_the_superconformal_index}. In what follows, we will explain how these decompositions are obtained for the punctures in the $\mathfrak{e}_6$ theory.

In contrast to classical $\mathfrak{g}$, nilpotent orbits in the exceptional Lie algebras, which label our punctures, are not naturally classified by partitions. Here, we recall the classification of Bala and Carter \cite{BalaCarter1,BalaCarter2}, following the exposition in \cite{CollingwoodMcGovern}. Their theorem states that there is a one-to-one correspondence between nilpotent orbits in $\mathfrak{g}$ and (conjugacy classes of) pairs $(\mathfrak{l},O^{\mathfrak{\l}})$ where $\mathfrak{l}$ is a Levi subalgebra \footnote{A Levi subalgebra $\mathfrak{h}\subset\mathfrak{l}\subset\mathfrak{g}$ is a reductive subalgebra, $\mathfrak{l}$, containing the Cartan subalgebra, $\mathfrak{h}$, of $\mathfrak{g}$. See section 3.8 of \cite{CollingwoodMcGovern} for an introduction.} of $\mathfrak{g}$ and $O^{\mathfrak{\l}}$ is a \emph{distinguished} \footnote{A nilpotent orbit, $\mathcal{O}$, in $\mathfrak{g}$ is \emph{distinguished} if and only if the only Levi subalgebra of $\mathfrak{g}$, containing $\mathcal{O}$, is $\mathfrak{g}$ itself.} nilpotent orbit in $\mathfrak{l}$. By the Jacobson-Morozov theorem, any representative $X$ of $O^{\mathfrak{l}}$ embeds in a standard triple \footnote{Any $\mathfrak{su}(2)$ subalgebra of $\mathfrak{g}$ is spanned by a \emph{standard triple} $\{H,X,Y\}$ of nonzero elements of $\mathfrak{g}$ satisfying the bracket relations $[H,X]=2X, [H,Y]=-2Y$, and $[X,Y]=H$.} $\{H,X,Y\} \subset \mathfrak{l}$, where $H \in \mathfrak{h}$. $\mathfrak{l}$ then has a decomposition into $ad_H$-eigenspaces
\begin{align*} 
\mathfrak{l}=\bigoplus_{k\in\mathbb{Z}} \mathfrak{l}_k
\end{align*}
where $\mathfrak{l}_k=\{x \in \mathfrak{l} \ | \ [H,x]=kx\}$. Let $\mathfrak{l}'\equiv \mathfrak{l}_0$ and $\mathfrak{u}'\equiv \oplus_{0<k\in\mathbb{Z}} \mathfrak{l}_{k}$. Then, $\mathfrak{p}=\mathfrak{l}'+\mathfrak{u}'$ is a parabolic subalgebra of $\mathfrak{l}$, with explicit Levi decomposition into a Levi subalgebra $\mathfrak{l}'$ and the nilradical  $\mathfrak{u}'$ of $\mathfrak{p}$. (Notice that the Cartan of $\mathfrak{l}$ is contained in $\mathfrak{l}'$, so $\text{rank}(\mathfrak{l}')=\text{rank}(\mathfrak{l})$.) 

A nilpotent orbit in $\mathfrak{g}$ is then given the label $X_N(a_i)$, called the \emph{Bala-Carter label}, where $X_N$ is the Cartan type of the semisimple part of $\mathfrak{l}$, and $i$ is the number of simple roots in $\mathfrak{l}'$. The case $i=0$ is denoted just by $X_N$, and corresponds to the principal orbit in $\mathfrak{l}$, which is always distinguished.

There are 16 conjugacy classes of Levi subalgebras of $E_6$. These are specified by their semisimple parts: $0$, $A_1$, $2A_1$, $3A_1$, $A_2$, $A_2+A_1$, $2A_2$, $A_2$, $A_2+2A_1$, $A_3+A_1$, $D_4$, $A_4$, $A_4+A_1$, $A_5$, $D_5$, and $E_6$. Here, $kA_N$ denotes the direct sum of $k$ copies of $A_N$. The label $0$ denotes the Cartan subalgebra, for which the only distinguished orbit is the zero orbit. For $\mathfrak{l}$ of classical type, distinguished orbits in $\mathfrak{l}$ are easily specified in terms of their partition: for $\mathfrak{l}$ of type $A$, the only distinguished orbit is the principal orbit (which, for $A_{N-1}$, has partition $[N]$ ), while for $\mathfrak{l}$ of type $B,C,D$, distinguished orbits are those for which the partition has no repeated parts. It was found by Bala and Carter that, for $\mathfrak{l}$ of type $G_2,F_4, E_6, E_7,$ and $E_8$, there are $2,4,3,6$, and $11$ distinguished orbits, respectively. 

The distinguished orbits in the Levi subalgebras listed above give rise to 21 nilpotent orbits in $\mathfrak{e}_6$. We list these \hyperlink{decomptable}{in the table below}, along with the centralizer, $\mathfrak{f}$, and the decomposition of the 27 and 78 of $\mathfrak{e}_6$ under $\mathfrak{su}(2) \times \mathfrak{f}$ \footnote{The decomposition of the 27 determines a projection matrix, which can be used to obtain the decompositions of higher-dimensional representations. We list a projection matrix for each puncture in Appendix \ref{projection_matrices}. The decomposition of the 78 determines the levels of the flavor groups, as described in \S\ref{flavour_groups}.}. But, before that, let us give a few examples of how to obtain the decomposition of the 27 for various embeddings.

First, consider $\mathfrak{l}=D_4$. In this case there are two distinguished orbits, with partitions [7,1] and [5,3], corresponding to nilpotent orbits $D_4$ and $D_4(a_1)$, respectively, in $\mathfrak{e}_6$. The first has centralizer $\mathfrak{su}(3)$ and the second, $\mathfrak{u}(1)^2$. We can obtain the decomposition of the 27 for each of these by embedding $\mathfrak{su}(2)$ in the $\mathfrak{so}(8)$ factor in $\mathfrak{so}(8) \times \mathfrak{u}(1)^2 \subset \mathfrak{so}(10)\times\mathfrak{u}(1) \subset \mathfrak{e}_6.$ The 27 of $\mathfrak{e}_6$ decomposes under $\mathfrak{so}(10)\times \mathfrak{u}(1)$ as 
\begin{align*}
\mathfrak{e}_6&\supset \mathfrak{so}(10)\times \mathfrak{u}(1) \\
27&=1_{-4}+10_2+16_{-1}
\end{align*}
The 10 and 16 of $\mathfrak{so}(10)$ decompose under $\mathfrak{so}(8)\times \mathfrak{u}(1)$ as 
\begin{align*}
\mathfrak{so}(10)&\supset\mathfrak{so}(8)\times \mathfrak{u}(1)\\
10&=1_2+1_{-2}+(8_v)_0 \\
16&=(8_s)_{1}+(8_c)_{-1}
\end{align*}
so we have
\begin{align*}
\mathfrak{e}_6&\supset\mathfrak{so}(8)\times\mathfrak{u}(1)\times\mathfrak{u}(1) \\
27&=1_{0,-4}+1_{2,2}+1_{-2,2}+(8_v)_{0,2}+(8_s)_{1,-1}+(8_c)_{-1,-1}
\end{align*}
For $D_4(a_1)$, we embed $\mathfrak{su}(2)$ in $\mathfrak{so}(8)$ by taking
\begin{align*}
\mathfrak{so}(8)&\supset\mathfrak{su}(2) \\
8_{v,s,c}&=5+3\\
\end{align*}
which gives
\begin{align*}
\mathfrak{e}_6&\supset\mathfrak{su}(2)\times\mathfrak{u}(1)\times\mathfrak{u}(1)\\
27&=1_{0,-4}+1_{2,2}+1_{-2,2}+3_{0,2}+3_{1,-1}+3_{-1,-1}+5_{0,2}+5_{1,-1}+5_{-1,-1}
\end{align*}
For $D_4$, we embed $\mathfrak{su}(2)$ in $\mathfrak{so}(8)$ by taking
\begin{align*}
\mathfrak{so}(8)&\supset\mathfrak{su}(2) \\
8_{v,s,c}&=7+1\\
\end{align*}
which gives
\begin{align*}
\mathfrak{e}_6&\supset\mathfrak{su}(2)\times\mathfrak{u}(1)\times\mathfrak{u}(1)\\
27&=1_{0,-4}+1_{2,2}+1_{-2,2}+1_{0,2}+1_{1,-1}+1_{-1,-1}+7_{0,2}+7_{1,-1}+7_{-1,-1}
\end{align*}
For this embedding, the $\mathfrak{u}(1)^2$ centralizer enhances to $\mathfrak{su}(3)$. To see this, we can make a change of basis so that the two $\mathfrak{u}(1)$ charges are given in terms of the old ones by
\begin{align*}
q_1'&=\frac{1}{2}(q_1+q_2) \\
q_2'&=\frac{1}{2}(q_1-q_2) \\
\end{align*}
Then the decomposition becomes
\begin{align*}
\mathfrak{e}_6&\supset\mathfrak{su}(2)\times\mathfrak{u}(1)\times\mathfrak{u}(1)\\
27&=1_{-2,2}+1_{2,0}+1_{0,-2}+1_{1,-1}+1_{0,1}+1_{-1,0}+7_{1,-1}+7_{0,1}+7_{-1,0}
\end{align*}
where we recognize these $\mathfrak{u}(1)^2$ charges as the weights (in the Dynkin basis) of the 6 and $\overline{3}$ of $\mathfrak{su}(3)$. Thus, the decomposition of the 27 is given by
\begin{align*}
\mathfrak{e}_6&\supset\mathfrak{su}(2)\times\mathfrak{su}(3)\\
27&=(1,6)+(7,\overline{3})
\end{align*}

Now, consider $\mathfrak{l}=E_6$. There are three distinguished orbits in $\mathfrak{e}_6$, giving rise to nilpotent orbits $E_6, E_6(a_1),$ and $E_6(a_3)$. The decomposition of the 27 for each of these can be obtained by taking the principal embedding of $\mathfrak{su}(2)$ inside the maximal subalgebras $\mathfrak{f}_4, \ \mathfrak{sp}(4),$ and $\mathfrak{su}(3)$ of $\mathfrak{e}_6$ \footnote{One might wonder about the other maximal subalgebras of $\mathfrak{e}_6$. One finds that the principal embedding of $\mathfrak{su}(2)$ in $\mathfrak{su}(2) \times \mathfrak{su}(6)$ or $\mathfrak{su}(3)\times\mathfrak{g}_2$ again gives $E_6(a_3)$, in $\mathfrak{g}_2$ gives $E_6(a_1)$, in $\mathfrak{so}(10)\times \mathfrak{u}(1)$ gives $D_5$, and in $\mathfrak{su}(3)^3$ gives $D_4$.}, respectively. We work out the decomposition for $E_6$ (the principal nilpotent orbit in $\mathfrak{e}_6$); the decompositions for $E_6(a_1)$ and $E_6(a_3)$ follow the same steps. 

The 27 of $\mathfrak{e}_6$ decomposes under $\mathfrak{f}_4$ as
\begin{align*}
\mathfrak{e}_6&\supset\mathfrak{f}_4\\
27&=1+26
\end{align*}
The principal embedding of $\mathfrak{su}(2)$ in $\mathfrak{f}_4$ is given by taking
\begin{align*}
\mathfrak{f}_4&\supset\mathfrak{su}(2)\\
26&=9+17
\end{align*}
so the decomposition of the 27 for $E_6$ is given by
\begin{align*}
\mathfrak{e}_6&\supset\mathfrak{su}(2)\\
27&=1+9+17
\end{align*}

To see which distinguished orbit corresponds to which $E_6(a_i)$, we need to count the  number of simple roots in $\mathfrak{l}'$. To do that, we make recourse to the decomposition of the 78.
\begin{itemize}
\item For the first case (embedding via $\mathfrak{f}_4$), the 78 decomposes as $3+9+11+15+17+23$. So $\dim(\mathfrak{l}')=\dim(\mathfrak{g}_0)=6$, which is also equal to $\text{rank}(\mathfrak{l}')=\text{rank}(\mathfrak{e}_6)$. Thus $\mathfrak{l}'$ is just the Cartan subalgebra and this is the principal embedding, $E_6$.
\item For the embedding via $\mathfrak{sp}(4)$, the 78 decomposes as $3+5+7+9+2(11)+15+17$, so we have $\dim(\mathfrak{l}')=8$ and $\mathfrak{l}'$ must contain precisely one positive (hence, simple) root. Thus, this is $E_6(a_1)$.
\item Finally, for the embedding via $\mathfrak{su}(3)$, the 78 decomposes as $3(3)+3(5) +2(7)+2(9)+2(11)$, so $\dim(\mathfrak{l}')=12$ and $\mathfrak{l}'$ contains three simple roots. Hence, this is $E_6(a_3)$.
\end{itemize}

We conclude this appendix with a summary of the nilpotent orbits in $\mathfrak{e}_6$ and the corresponding decompositions of the 27 and the 78 under $\mathfrak{su}(2)\times \mathfrak{f}$.

{
\renewcommand{\arraystretch}{1.5}
\htarget{decomptable}
\begin{longtable}{|l|c|l|l|}
\hline
Bala-Carter&$\mathfrak{f}$&$27$&$78$\\
\hline 
\endhead
$0$&$\mathfrak{e}_6$&$(1;27)$&$(1;78)$\\
\hline
$A_1$&$\mathfrak{su}(6)$&$(1;\overline{15})+(2;6)$&$(1;35)+(2;20)+(3;1)$\\
\hline
$2A_1$&$\mathfrak{so}(7)\times \mathfrak{u}(1)$&${\begin{aligned}(1;&7_2+1_{-4})\\&+(2;8_{-1})+(3;1_2)\end{aligned}}$&${\begin{aligned}(1;&1_0+21_0)\\&+(2;8_3+8_{-3})+(3;7_0+1_0)\end{aligned}}$\\
\hline
$3A_1$&$\mathfrak{su}(3)\times \mathfrak{su}(2)$&${\begin{aligned}(1;&\overline{6},1)+(2;3,2)+(3;3,1)\end{aligned}}$&${\begin{aligned}(1;&8,1)+(1;1,3)+(2;8,2)\\&+(3;1,1)+(3;8,1)+(4;1,2)\end{aligned}}$\\
\hline
$A_2$&$\mathfrak{su}(3)\times \mathfrak{su}(3)$&${\begin{aligned}(1;&3,\overline{3})+(3;1,3)+(3;\overline{3},1)\end{aligned}}$&${\begin{aligned}(1;&8,1)+(1;1,8)+(3;1,1)\\&+(3;3,3)+(3;\overline{3},\overline{3})+(5;1,1)\end{aligned}}$\\
\hline
$A_2+A_1$&$\mathfrak{su}(3)\times \mathfrak{u}(1)$&${\begin{aligned}&(1;3_{2})+(2;3_{-1}+1_{1})\\ &+(3;\overline{3}_0+1_{-2})+(4;1_{1})\end{aligned}}$&${\begin{aligned}&(1;8_0+1_0)+(2;3_1+\overline{3}_{-1}+1_{-3}+1_3)\\& +(3;3_{-2}+\overline{3}_2+1_0+1_0)\\&+(4;3_1+\overline{3}_{-1})+(5;1_0)\end{aligned}}$\\
\hline
$2A_2$&$\mathfrak{g}_2$&$(1;1)+(3;7)+(5;1)$&$(1;14)+(3;7+1)+(5;7+1)$\\
\hline
$A_2+2A_1$&$\mathfrak{su}(2)\times \mathfrak{u}(1)$&${\begin{aligned}(1;&1_{2}+1_{-4})+(2;4_{-1})\\&+(3;3_{2})+(4;2_{-1})\end{aligned}}$&${\begin{aligned}(1;&1_0+3_0)+(2;4_3+4_{-3})\\&+(3;1_0+3_0+5_0)+(4;2_3+2_{-3})\\&+(5;3_0)\end{aligned}}$\\
\hline
$A_3$&$\mathfrak{sp}(2)\times \mathfrak{u}(1)$&${\begin{aligned}(1;&5_{-2}+1_{4})+(4;4_{1})\\&+(5;1_{-2})\end{aligned}}$&${\begin{aligned}(1;&10_0+1_0)+(3;1_0)\\&+(4;4_3+4_{-3})+(5;5_0)+(7;1_0)\end{aligned}}$\\
\hline
$2A_2+A_1$&$\mathfrak{su}(2)$&${\begin{aligned}(1;1)&+(2;2)+(3;3)\\&+(4;2)+(5;1)\end{aligned}}$&${\begin{aligned}(1;&3)+(2;4+2)+(3;3+1+1)\\&+(4;2+2)+(5;3+1)+(6;2)\end{aligned}}$\\
\hline
$A_3+A_1$&$\mathfrak{su}(2)\times \mathfrak{u}(1)$&${\begin{aligned}(1;&1_{4}+1_{-2})+(2;2_{-2})\\&+(3;1_{1})+(4;2_{1})\\&+(5;1_{1}+1_{-2})\end{aligned}}$&${\begin{aligned}(1;&1_0+3_0)+(2;2)_0\\&+(3;1_3+1_{-3}+1_0+1_0)\\&+(4;2_3+2_{-3}+2_0)\\&+(5;1_3 + 1_0 + 1_{-3})\\&+(6;2)_0+(7;1_0)\end{aligned}}$\\
\hline
$D_4(a1)$&$\mathfrak{u}(1)\times \mathfrak{u}(1)$&${\begin{aligned}&1_{2,2}+1_{0,-4}+1_{-2,2}\\ &+3_{1,-1}+3_{0,2}+3_{-1,-1}\\ &+5_{1,-1}+5_{0,2}+5_{-1,-1}\end{aligned}}$&${\begin{aligned}&1_{0,0}+1_{0,0}+3_{0,0}\\ &+3_{2,0}+3_{1,3}+3_{1,-3}+3_{0,0}\\ &+3_{-2,0}+3_{-1,-3}+3_{-1,3}+3_{0,0}\\ &+5_{2,0}+5_{1,3}+5_{1,-3}+5_{0,0}\\ &+5_{-2,0}+5_{-1,-3}+5_{-1,3}\\ &+7_{0,0}+7_{0,0}\end{aligned}}$\\
\hline
$A_4$&$\mathfrak{su}(2)\times \mathfrak{u}(1)$&${\begin{aligned}(1;&2_{-5})+(3;1_{-2})\\&+(5;2_{1}+1_{4})+(7;1_{-2})\end{aligned}}$&${\begin{aligned}(1;&3_0+1_0)+(3;2_3+2_{-3}+1_0)\\&+(5;1_6+1_0+1_{-6})\\ &+(7;2_3+2_{-3}+1_0)+(9;1_0)\end{aligned}}$\\
\hline
$D_4$&$\mathfrak{su}(3)$&$(1;6)+(7;\overline{3})$&$(1;8)+(3;1)+(7;8)+(11;1)$\\
\hline
$A_4+A_1$&$\mathfrak{u}(1)$&${\begin{aligned}2_{-5}&+3_{-2}+4_{1}\\&+5_{4}+6_{1}+7_{-2}\end{aligned}}$&${\begin{aligned}&1_0+2_3+2_{-3}+3_0+3_0+4_3+4_{-3}\\ &+5_6+5_0+5_{-6}+6_{-3}+6_3\\&+7_0+8_{-3}+8_3+9_0\end{aligned}}$\\
\hline
$D_5(a1)$&$\mathfrak{u}(1)$&${\begin{aligned}1_{-4}&+2_{-1}+3_2\\&+6_{-1}+7_2+8_{-1}\end{aligned}}$&${\begin{aligned}&1_0+2_3+2_{-3}+3_0+3_0+5_0\\&+6_3+6_{-3}+7_0+7_0\\&+8_3+8_{-3}+9_0+11_0\end{aligned}}$\\
\hline
$A_5$&$\mathfrak{su}(2)$&$(1;1)+(5;1)+(6;2)+(9;1)$&${\begin{aligned}&(1;3)+(3;1)+(4;2)+(5;1)+(6;2)\\&+(7;1)+(9;1)+(10;2)+(11;1)\end{aligned}}$\\
\hline
$E_6(a3)$&$-$&$1+5+5+7+9$&${\begin{aligned}&3+3+3+5+5+5\\ &+7+7+9+9+11+11\end{aligned}}$\\
\hline
$D_5$&$\mathfrak{u}(1)$&$1_2+1_{-4}+5_{-1}+9_2+11_{-1}$&${\begin{aligned}&1_0+3_0+5_3+5_{-3}+7_0+9_0\\ &+11_3+11_0+11_{-3}+15_0\end{aligned}}$\\
\hline
$E_6(a1)$&$-$&$5+9+13$&$3+5+7+9+11+11+15+17$\\
\hline
$E_6$&$-$&$1+9+17$&$3+9+11+15+17+23$\\
\hline
\end{longtable}
}

\section{Projection Matrices}\label{projection_matrices}

Our classification of interacting and mixed fixtures using the superconformal index, carried out in section \ref{global_symmetries_and_the_superconformal_index}, required that we know the decomposition of a number of higher-dimensional $\mathfrak{e}_6$ representations (and not just the 27 and the 78) under $\mathfrak{su}(2) \times \mathfrak{f}$. These are trivial to obtain using LieART \cite{Feger:2012bs}, provided we know a projection matrix for each embedding \cite{Slansky:1981yr, Feger:2012bs}. 

From the decomposition of the $27$, listed in the table above, one obtains a projection matrix simply by defining a $ 6 \times \text{rk }(\mathfrak{su}(2) \times \mathfrak{f}$) matrix, $M$, such that the LieART command 

\bigskip
{\colorbox[rgb]{1.00,0.93,1.00}{\tt In\char91\char49\char93\char61\, Project[M,WeightSystem[Irrep[E6][1,0,0,0,0,0]]]\qquad\qquad}}
\bigskip

\noindent
gives the corresponding $\mathfrak{su}(2) \times \mathfrak{f}$ weights. This projection matrix can then be used to obtain the decomposition of any $\mathfrak{e}_6$ irrep under $\mathfrak{su}(2) \times \mathfrak{f}$.

Below, we list a projection matrix for each embedding, following the conventions of LieART.

{\begin{longtable}{|l|c|l|}
\hline
Bala-Carter&$\mathfrak{f}$&Projection Matrix\\
\hline
\endhead
$A_1$&$\mathfrak{su}(6)$&$\begin{gathered}\\ \begin{pmatrix} -1&-2&-3&-2&-1&-2 \\ 1&0&0&0&0&0 \\ 0&1&0&0&0&0 \\0&0&1&0&0&0 \\ 0&0&0&1&0&0 \\ 0&0&0&0&1&0 \end{pmatrix}\\ {}\end{gathered}$\\
\hline
$2A_1$&$\mathfrak{so}(7)\times \mathfrak{u}(1)$&$\begin{gathered}\\ \begin{pmatrix} 2&3&4&3&2&2 \\ 0&0&0&0&0&1 \\ 0&0&1&0&0&0 \\ 0&1&0&1&0&0 \\ 2&1&0&-1&-2&0\end{pmatrix}\\ {}\end{gathered}$\\
\hline
$3A_1$&$\mathfrak{su}(3)\times \mathfrak{su}(2)$&$\begin{gathered}\\ \begin{pmatrix} 2&3&4&3&2&1 \\ 1&2&1&0&0&1\\ 0&0&1&2&1&1 \\ 0&1&2&1&0&1\end{pmatrix}\\ {}\end{gathered}$\\
\hline
$A_2$&$\mathfrak{su}(3)\times \mathfrak{su}(3)$&$\begin{gathered}\\ \begin{pmatrix} 2&2&0&0&0&0 \\ 0&0&0&1&0&0 \\ 0&0&0&0&1&0 \\ 0&0&0&0&0&1 \\ -1&-2&-3&-2&-1&-2 \end{pmatrix}\\ {}\end{gathered}$\\
\hline
$A_2+A_1$&$\mathfrak{su}(3)\times \mathfrak{u}(1)$&$\begin{gathered}\\ \begin{pmatrix} 3&5&7&5&3&3 \\ 0&0&0&1&0&0\\0&1&1&0&0&1\\ 1&1&-1&-1&-1&-1 \end{pmatrix}\\ {}\end{gathered}$\\
\hline
$2A_2$&$\mathfrak{g}_2$&$\begin{gathered}\\ \begin{pmatrix} 4&6&8&6&4&4 \\ 0&1&0&1&0&1 \\ 0&0&1&0&0&0 \end{pmatrix}\\ {}\end{gathered}$\\
\hline
$A_2+2A_1$&$\mathfrak{su}(2)\times \mathfrak{u}(1)$&$\begin{gathered}\\ \begin{pmatrix} 3&4&6&4&3&4 \\ 1&4&6&4&1&2 \\ -1&-2&0&2&1&0 \end{pmatrix}\\ {}\end{gathered}$\\
\hline
$A_3$&$\mathfrak{sp}(2)\times \mathfrak{u}(1)$&$\begin{gathered}\\ \begin{pmatrix} 4&7&10&7&4&6 \\ 0&1&0&1&0&0 \\ 0&0&1&0&0&0 \\ -2&-1&0&1&2&0 \end{pmatrix}\\ {}\end{gathered}$\\
\hline
$2A_2+A_1$&$\mathfrak{su}(2)$&$\begin{gathered}\\ \begin{pmatrix} 4&6&9&6&4&5 \\ 0&2&3&2&0&1\end{pmatrix}\\ {}\end{gathered}$\\
\hline
$A_3+A_1$&$\mathfrak{su}(2)\times \mathfrak{u}(1)$&$\begin{gathered}\\ \begin{pmatrix} 4&8&11&8&4&5 \\ 0&0&1&0&0&1 \\ -2&-1&0&1&2&0 \end{pmatrix}\\ {}\end{gathered}$\\
\hline
$D_4(a1)$&$\mathfrak{u}(1)\times \mathfrak{u}(1)$&$\begin{gathered}\\ \begin{pmatrix} 4&8&10&8&4&6 \\ 1&1&0&-1&-1&0\\ -1&1&0&-1&1&0 \end{pmatrix}\\ {}\end{gathered}$\\
\hline
$A_4$&$\mathfrak{su}(2)\times \mathfrak{u}(1)$&$\begin{gathered}\\ \begin{pmatrix}6&10&12&10&6&6 \\ 0&1&1&0&0&0 \\ -2&-1&-3&-2&2&0 \end{pmatrix}\\ {}\end{gathered}$\\
\hline
$D_4$&$\mathfrak{su}(3)$&$\begin{gathered}\\ \begin{pmatrix} 6&10&16&10&6&10 \\ 0&0&1&2&1&0\\ 1&2&1&0&0&0\end{pmatrix}\\ {}\end{gathered}$\\
\hline
$A_4+A_1$&$\mathfrak{u}(1)$&$\begin{gathered}\\ \begin{pmatrix} 6&10&12&10&6&7 \\ -2&-4&-6&-2&2&-3\end{pmatrix}\\ {}\end{gathered}$\\
\hline
$D_5(a1)$&$\mathfrak{u}(1)$&$\begin{gathered}\\ \begin{pmatrix} 7&12&18&12&7&10 \\ -1&-2&0&2&1&0\end{pmatrix}\\ {}\end{gathered}$\\
\hline
$A_5$&$\mathfrak{su}(2)$&$\begin{gathered}\\ \begin{pmatrix} 8&14&19&14&8&10 \\ 0&0&1&0&0&0 \end{pmatrix}\\ {}\end{gathered}$\\
\hline
$E_6(a3)$&$-$&$\begin{gathered}\\ \begin{pmatrix} 8&14&18&14&8&8\end{pmatrix}\\ {}\end{gathered}$\\
\hline
$D_5$&$\mathfrak{u}(1)$&$\begin{gathered}\\ \begin{pmatrix} 10&18&24&18&10&10 \\ -1 & -2 & 0 & 2 & 1 & 0\end{pmatrix}\\ {}\end{gathered}$\\
\hline
$E_6(a1)$&$-$&$\begin{gathered}\\ \begin{pmatrix} 12&22&30&22&12&16\end{pmatrix}\\ {}\end{gathered}$\\
\hline
$E_6$&$-$&$\begin{gathered}\\ \begin{pmatrix} 16&30&42&30&16&22\end{pmatrix}\\ {}\end{gathered}$\\
\hline
\end{longtable}

}

As an example, let's work out the decomposition of the $51975$ for the orbit $2A_2$. Running LieART, we obtain the decomposition with the following two lines of code:

\bigskip
\noindent
{\colorbox[rgb]{1.00,0.93,1.00}{\tt In\char91\char49\char93\char61\, ProjectionMatrix\char91E\char54\char44ProductAlgebra\char91SU\char50\char44G\char50\char93\char93\char61 $\begin{pmatrix} 4&6&8&6&4&4 \\ 0&1&0&1&0&1 \\ 0&0&1&0&0&0 \end{pmatrix}$ \char59\,\,\quad}}

\noindent
{\colorbox[rgb]{1.00,0.93,1.00}{\tt In\char91\char50\char93\char61\, DecomposeIrrep\char91Irrep\char91E\char54\char93\char91\char49\char44\char48\char44\char49\char44\char48\char44\char48\char44\char48\char93\char44ProductAlgebra\char91SU\char50\char44G\char50\char93\char93\,\,\,\qquad}}

\noindent
{\colorbox[rgb]{1.00,0.93,1.00}{\tt Out\char91\char50\char93\char61\vtop{\vskip-2ex\hskip-1em\hsize=32em$\begin{aligned}
&(1,1)+14(3,1)+10(5,1)+13(1,7)+13(7,1)+25(3,7)+5(9,1)+34(5,7)\\
&+4(11,1)+25(7,7)+9(1,14)+17(9,7)+16(3,14)+6(11,7)+22(5,14)\\
&+2(13,7)+15(7,14)+10(9,14)+3(11,14)+(13,14)+6(1,27)+25(3,27)\\
&+23(5,27)+21(7,27)+9(9,27)+4(11,27)+5(1,64)+12(3,64)+13(5,64)\\
&+9(7,64)+4(9,64)+(11,64)+4(1,77)+6(3,77)+2(3,77')+8(5,77)\\
&+(5,77')+4(7,77)+(7,77')+2(9,77)+(3,182)+(1,189)+2(3,189)\\
&+2(5,189)+(7,189)
\end{aligned}
$}\,\quad
}}

\bigskip
This works for all of the orbits above, except for $D_4(a_1)$, as the LieART command ``{\tt DecomposeIrrep}'' does not seem to work when the target subalgebra has more than one $\mathfrak{u}(1)$ factor. In this case, getting the decomposition is only slightly more complicated. For example, we obtain the decomposition of the $27$ of $E_6$ as follows:

\bigskip
{\colorbox[rgb]{1.00,0.93,1.00}{\tt In\char91\char49\char93\char61\, ProjectionMatrix\char91D\char53\char44ProductAlgebra\char91D\char52\char44U\char49\char93\char93 \char61 $\begin{pmatrix} 1&0&0&0&0 \\ 0&1&0&0&0\\0&0&1&1&0 \\ 0&0&1&0&1 \\ 0&0&0&1&1 \end{pmatrix}$ \char59}}

{\colorbox[rgb]{1.00,0.93,1.00}{\tt In\char91\char50\char93\char61\, ProjectionMatrix\char91D\char52\char44ProductAlgebra\char91A\char49\char93\char93 \char61  $\begin{pmatrix} 4&6&4&4 \end{pmatrix}$\char59\qquad\quad\quad\,}}

{\colorbox[rgb]{1.00,0.93,1.00}{\tt In\char91\char51\char93\char61\, DecomposeIrrep\char91\, DecomposeIrrep\char91\,\,  DecomposeIrrep\char91\,\,\,\,\,\qquad\qquad\quad}}

{\colorbox[rgb]{1.00,0.93,1.00}{\tt\qquad\quad Irrep\char91E\char54\char93\char91\char49\char44\,\char48\char44\,\char48\char44\,\char48\char44\,\char48\char44\,\char48\char93\char44\, ProductAlgebra\char91D\char53\char44\, U\char49\char93\char93\char44\qquad\qquad\quad\,\,}}

{\colorbox[rgb]{1.00,0.93,1.00}{\tt\qquad\quad  ProductAlgebra\char91D\char52\char44\, U\char49\char93\char44\,\char49\char93\char44\, ProductAlgebra\char91A\char49\char93\char44\, \char49\char93\qquad\qquad\quad\,\,}}

{\colorbox[rgb]{1.00,0.93,1.00}{\tt Out\char91\char51\char93\char61\, \char40\char49\char41\char40\char50\char41\char40\char50\char41\char43\char40\char49\char41\char40\char48\char41\char40\char45\char52\char41\char43\char40\char49\char41\char40\char45\char50\char41\char40\char50\char41\char43\char40\char51\char41\char40\char49\char41\char40\char45\char49\char41\char43\char40\char51\char41\char40\char48\char41\char40\char50\char41\,\,\qquad\quad}}

{\colorbox[rgb]{1.00,0.93,1.00}{\tt\qquad\qquad\char43\char40\char51\char41\char40\char45\char49\char41\char40\char45\char49\char41\char43\char40\char53\char41\char40\char49\char41\char40\char45\char49\char41\char43\char40\char53\char41\char40\char48\char41\char40\char50\char41\char43\char40\char53\char41\char40\char45\char49\char41\char40\char45\char49\char41\qquad\qquad\qquad\quad}}

\end{appendices}

\bibliographystyle{utphys}
\bibliography{ref}

\end{document}